\documentclass[%
reprint,
footinbib,
 amsmath,amssymb,
 aps,
prl,
]{revtex4-2}

\usepackage{graphicx}
\usepackage{dcolumn}
\usepackage{bm}
\usepackage{hyperref}

\usepackage{caption}
\usepackage{subcaption}

\usepackage{ragged2e}

\makeatletter
\long\def\@makecaption#1#2{%
  \vskip\abovecaptionskip
  \small
  \justifying
  #1.~#2\par
  \vskip\belowcaptionskip}
\makeatother

\usepackage{dsfont}
\usepackage{amsmath}
\usepackage{bbm}
\usepackage{physics}
\usepackage{comment}
\usepackage{orcidlink} 
\usepackage{amsthm}
\usepackage[T1]{fontenc}

\newcommand{\exv}{\mathbb{E}}
\newcommand{\ignore}[1]{}

\usepackage{pifont}  

\begin{document}

\preprint{APS/123-QED}

\title{Non-stabilizerness and U(1) symmetry in chaotic many-body quantum systems}

\author{Daniele Iannotti\orcidlink{0009-0009-0738-5998}$^{1,2}$}
\email{d.iannotti@ssmeridionale.it}
\author{Angelo Russotto\orcidlink{0009-0009-4303-9523}$^{3,4}$}
\email{arussott@sissa.it}
\author{Barbara Jasser\orcidlink{0009-0004-5657-2806}$^{1,2}$}
\email{b.jasser@ssmeridionale.it}
\author{Jovan Odavić\orcidlink{0000-0003-2729-8284}$^{2,5}$}
\email{jovan.odavic@unina.it}
\author{Alioscia Hamma\orcidlink{0000-0003-0662-719X}$^{1,2,5}$}
\email{alioscia.hamma@unina.it }

\affiliation{$^1$Scuola Superiore Meridionale, Largo S. Marcellino 10, 80138 Napoli, Italy}
\affiliation{$^2$INFN Sezione di Napoli, via Cintia, 80126 Napoli, Italy}
\affiliation{$^3$SISSA, via Bonomea 265, 34136 Trieste, Italy}
\affiliation{$^4$INFN Sezione di Trieste, via Valerio 2, 34127 Trieste, Italy}
\affiliation{$^5$Dipartimento di Fisica `Ettore Pancini', Universit\`a degli Studi di Napoli Federico II, Via Cintia 80126, Napoli, Italy}


\begin{abstract}
We present exact, closed-form results for the non-stabilizerness of random pure
states subject to a $U(1)$ symmetry constraint. Using stabilizer entropy as our non-stabilizerness monotone, we derive the average and the variance for $U(1)$-constrained Haar random states. 
We show that the presence of a conserved charge leads to a substantial suppression of non-stabilizerness (magic) compared to the unconstrained case, and identify a qualitative difference between entanglement and magic response. 
In the thermodynamic limit, stabilizer entropy exhibits a different leading-order scaling close to a vanishing relative charge density, implying that magic is more robust to charge density fluctuations than entanglement entropy. 
We test our analytical predictions against midspectrum
eigenstates of two chaotic many-body systems with conserved $U(1)$ charge: the complex-fermion Sachdev--Ye--Kitaev (cSYK) model and a Heisenberg XXZ chain with next-to-nearest-neighbour couplings and conserved magnetization. 
We find an excellent agreement for the non-local cSYK model and systematic deviations for the local XXZ chain, highlighting the role of interaction locality.
\end{abstract}

\maketitle

\textit{Introduction--} Eigenstates in the bulk of the spectrum of quantum chaotic Hamiltonians are widely expected to resemble random pure states in their entanglement properties~\cite{Vidmar_2017,Vidmar_2017_q,Lu_2019,Bianchi_Hackl_Kieburg_Rigol_Vidmar_2022}. For systems without conservation laws, this expectation is made precise by the Page formula~\cite{Page_1993}, which predicts near-maximal entanglement entropy for typical random states. A common lore of both classical and quantum statistical mechanics holds that, within the sector or subspace of a conserved quantity, everything is ergodic (if there are no further symmetries). However, the incorporation of global symmetry constraints (both Abelian~\cite{Huang_2019,Bianchi_Dona_2019} and non-Abelian~\cite{Patil_2023,chakraborty2025}) has refined this picture considerably, producing constrained Page curves that more faithfully capture the structure of physical many-body systems and bear the signature of the symmetry involved~\cite{Morampudi_2020,Huang_2021,Murciano_2022,Kliczkowski_2023,Yauk_2024,Rodriguez-Nieva_Jonay_Khemani_2024, Langlett_Jonay_Khemani_Rodriguez-Nieva_2025,Russotto_Ares_Calabrese_2025}.

\begin{figure}[h!]
    \centering

    \begin{subfigure}{\columnwidth}
        \centering
        \includegraphics[width=\linewidth]{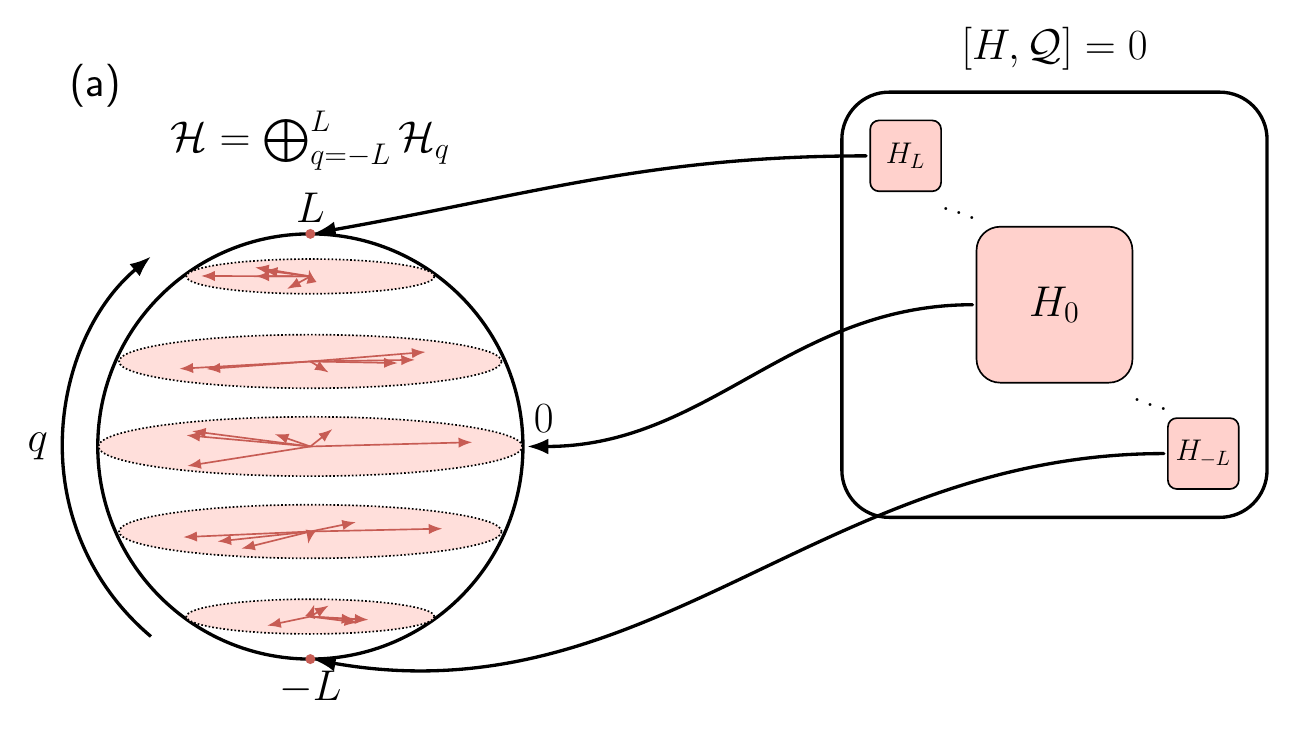}
        \label{fig:schematic}
    \end{subfigure}

     \vspace{-0.75em} 

    \begin{subfigure}{\columnwidth}
        \centering
        \includegraphics[width=\linewidth]{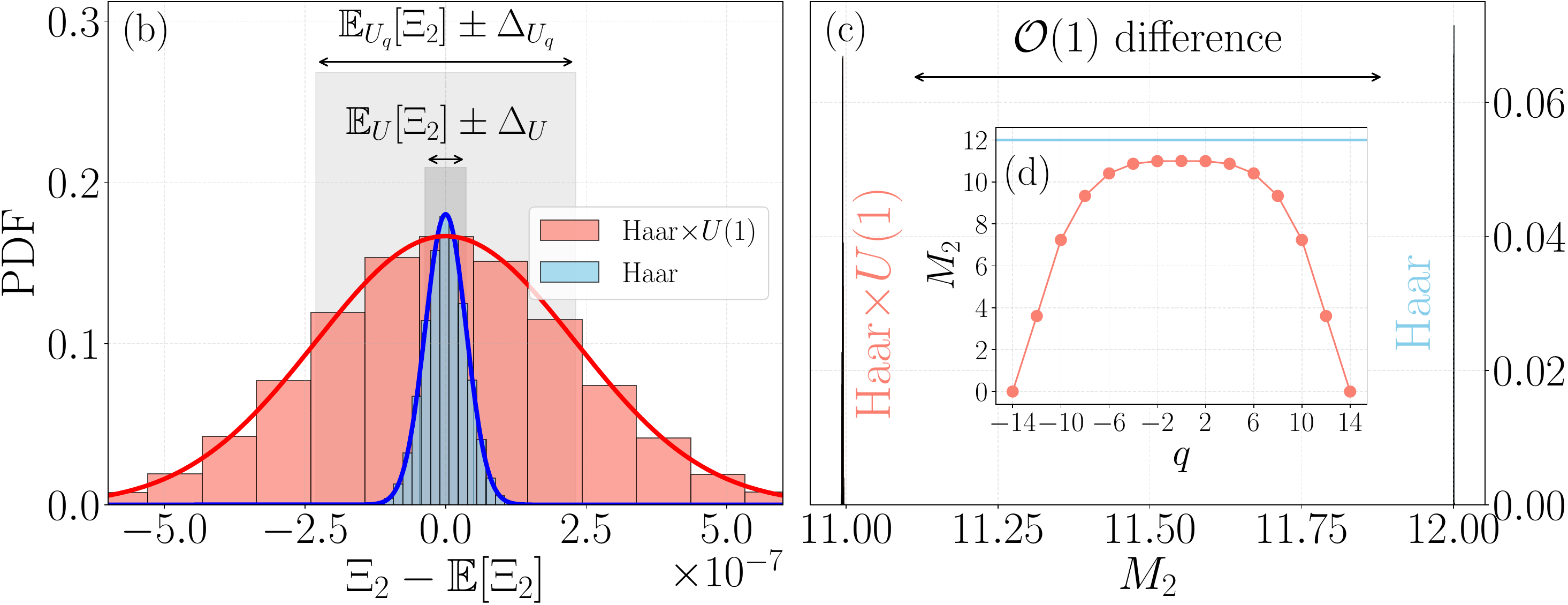}
        \label{fig:histogram}
    \end{subfigure}

     \vspace{-1.0em} 

    \caption{
    (a)~Schematic of the random-state ensemble: states are sampled uniformly within a fixed $U(1)$ charge sector $q$, modeling eigenstates of $U(1)$-symmetric chaotic Hamiltonians. (b),(c)~Analytical predictions (curves) versus numerical data for the order-2 stabilizer purity~(b) and stabilizer entropy~(c) of Haar and Haar$\times U(1)$ ensembles ($L=14$, $q=0$, $2.5\times10^5$ samples). Closed-form expressions for the $U(1)$ constrained ensemble mean $\exv_{U_q}[\Xi_2]$ and the variance $\Delta_{U_q}^2$ of stabilizer purity are among our main results. The unconstrained ensemble results are previously known. (d)~Average stabilizer entropy versus charge sector $q$ for Haar$\times U(1) $ (red) and Haar (blue).
    }
    \vspace{-0.75em}
    \label{fig:featured_image}
\end{figure}

While the effects of symmetries on entanglement and related quantities have been the subject of extensive and rich investigation~\cite{Calabrese_2021,10.21468/SciPostPhys.17.5.127, santra2025quantumresourcesnonabelianlattice, liu2024unitarydesignsrandomsymmetric,Majidy_2023,Majidy_2023_nat}, far less attention has been devoted to how symmetry-associated conserved charges shape non-stabilizerness. Non-stabilizerness (or \textit{magic}), defined as the distance of a state from the efficiently simulable stabilizer states~\cite{PhysRevA.71.022316}, is a necessary resource for quantum computation and quantum chaos that lies beyond entanglement, making the
latter only one facet of quantum complexity~\cite{Piemontese_Roscilde_Hamma_2023, Zhou_Yang_Hamma_Chamon_2020, Leone_Oliviero_Zhou_Hamma_2021, True_Hamma_2022, Hinsche_Ioannou_Nietner_Haferkamp_Quek_Hangleiter_Seifert_Eisert_Sweke_2023, Bejan_McLauchlan_Beri_2024}.

In commonly encountered local spin Hamiltonians, both in static and dynamical settings, the entanglement and magic of eigenstates have been found to be intimately linked, and their interplay has been extensively studied~\cite{Oliviero_Leone_Hamma_2022, Lami_Collura_2023,Rattacaso_Leone_Oliviero_Hamma_2023, Odavic_Viscardi_Hamma_2025, Viscardi_Dalmonte_Hamma_Tirrito_2026,Turkeshi_2025_spread,1jzy-sk9r}. However, recent studies of magic in strongly interacting non-local systems such as the Sachdev--Ye--Kitaev model~\cite{Jasser_Odavic_Hamma_2025, bera2025non,Russomanno_2025} indicate that, at finite size, magic (property of the total state) can be considerably more sensitive to symmetry constraints than entanglement~\cite{Jasser_Odavic_Hamma_2025}, pointing to structure in Hamiltonian eigenstates that entanglement-based probes alone miss.

In this Letter, we present exact, closed-form results that quantify the imprint of a $U(1)$ symmetry on the non-stabilizerness of pure random states. As our measure of magic, we employ the \textit{stabilizer entropy}~\cite{leone2022StabilizerRenyiEntropy, Leone_Bittel_2024,
Bittel_Leone_2025}, an efficiently computable, bona fide quantifier that
provides rigorous lower bounds on several other well-known magic monotones,
including stabilizer nullity~\cite{Beverland_Campbell_Howard_Kliuchnikov_2020}, min-relative entropy of magic~\cite{Liu_Winter_2022}, and robustness
of magic~\cite{Howard_Campbell_2017}. Interestingly, stabilizer entropy also admits a clear quantum mechanical
interpretation as the deficit of partial incompatibility of quantum measurements
which are at the heart of Stern--Gerlach experiments~\cite{iannotti2026van}.

We demonstrate the practical utility of our closed-form results by testing them
against the stabilizer entropy of midspectrum eigenstates in two chaotic
many-body systems with an explicitly conserved $U(1)$ charge: the
zero-dimensional complex-fermion Sachdev--Ye--Kitaev (cSYK) model and a
one-dimensional Heisenberg XXZ chain with next-to-nearest-neighbour couplings.
The cSYK model (disordered and non-local) serves as a paradigmatic toy model
for exotic strongly correlated phases such as strange metals, with deep
connections to black-hole thermodynamics and holographic
duality~\cite{Chowdhury_Georges_Parcollet_Sachdev_2022,
jha_introduction_2025}. Across the different charge sectors of this model, we
find excellent agreement between the numerical stabilizer entropy and our
analytical predictions for constrained random pure state ensembles. The XXZ
chain, a typical locally interacting chaotic
Hamiltonian~\cite{langlett2025entanglement}, does not display the same level of
agreement, highlighting the role that local interactions play in shaping
deviations from the random-state prediction.

Our main results include the ensemble average and variance of the stabilizer
purity, and consequently the stabilizer entropy (see below for
definitions), for $U(1)$ symmetric random pure states, shown schematically in Fig.~\ref{fig:featured_image}.
Most importantly, we derive asymptotic behaviour in the thermodynamic limit and identify a qualitative difference between the entanglement and magic response of constrained random pure states that
persists in the thermodynamic limit: the stabilizer entropy exhibits a different
leading-order scaling in the limit of vanishing relative charge quantum number,
implying that the magic of random pure states is more robust to conserved charge
density fluctuations than the entanglement entropy.

\textit{$U(1)$ symmetric random state ensemble--} We now describe the ensemble
of $U(1)$-symmetric Haar random states, initially introduced
in~\cite{Huang_2019,Huang_2021}, whose entanglement properties and
symmetry resolution have been extensively characterized in subsequent
works~\cite{Bianchi_Dona_2019,Murciano_2022,Lau_2022,ghasemi2024symmetryresolvedrelativeentropyrandom, Vallini2025vvq}. Consider a system of $L$ qubits. We are interested in charges $\mathcal{Q}$, which generate a $U(1)$ group acting on the global Hilbert space $\mathcal{H} = (\mathbb{C}^2)^{\otimes L}$ of dimension $d=2^L$. For concreteness, we take $\mathcal{Q}$ to be the global $z-$magnetization  
\vspace{-0.1cm}
\begin{equation}
\label{eq:defQz}
    \mathcal{Q} = \sum_{j=1}^L \sigma^z_j,
\end{equation}
\vspace{-0.25cm}

\noindent where $\sigma^{z}_j$ are Pauli matrices on the $j$-th site.
The ensemble definition is independent of the choice of the particular charge, while the dependence of the non-stabilizerness on this choice is discussed later.

The total Hilbert space of the system can be decomposed as $\mathcal{H} = \bigoplus_{q=-L}^{L} \mathcal{H}(q)$, where $\mathcal{H}(q)$ is the eigenspace with charge value $q$ and dimension $d_q=\binom{L}{\frac{L+q}{2}}$, as illustrated in Fig.~\ref{fig:featured_image}. By definition, any state $\ket{\psi(q)} \in \mathcal{H}(q)$ is an eigenstate of $\mathcal{Q}$, which means that the corresponding density matrix $\ket{\psi(q)} \bra{\psi(q)} \equiv \psi(q)$ satisfy $[\mathcal{Q},\psi(q)] = 0$. The ensemble of $U(1)$-symmetric Haar random states, with fixed value of the charge $q$, consists of density matrices $\psi_{U_q}\equiv U_q\ket{q} \bra{q}U_q^{\dagger}$, where $U_q$ is a $d_q\times d_q$ Haar random unitary matrix acting on the charge $q$ sector $\mathcal{H}(q)$, i.e. $U_q \in U(\mathcal{H}(q))$, and $\ket{q}$ is just a reference state with the only constraint of having a definite value of charge.

The projector onto a fixed charge sector $\mathcal{H}(q)$ admits an integral
representation as a direct consequence of the Peter--Weyl
theorem~\cite{harrow2013church,wigner2012group}. For the compact Abelian $U(1)$
group and the charge $\mathcal{Q}$ defined in
Eq.~\eqref{eq:defQz}~\cite{Goldstein_2018,Xavier_2018,Bonsignori_2019,Bonsignori_Calabrese_2020}, the projector reads
\vspace{-0.1cm}
\begin{equation}
\label{eq:repr_proj}
    \Pi_q = \int _{0}^{2\pi}\frac{d\varphi}{2 \pi} e^{- i \varphi q}
    \bigotimes_{j=1}^L e^{i \varphi \sigma_j^z}.
\end{equation}
The projector factorizes manifestly over individual qubits and together with the integral representation yields substantial analytical simplifications. 

\textit{Non-stabilizerness via Stabilizer Entropy (SE) --} For an $L$-qubit system, a natural operator basis is given by tensor products of the single-qubit Pauli matrices $\{ \mathbb{I}, \sigma^x, \sigma^y, \sigma^z$\}. The Clifford group is the normalizer of the Pauli group within the unitary group, i.e., it maps Pauli operators to Pauli operators. Non-stabilizerness is the resource associated with operations that lie outside the Clifford group. Gates in the Clifford group admit efficient fault-tolerant implementations, for instance via transversal constructions in quantum error-correcting codes~\cite{Campbell_Browne_2010}, whereas non-Clifford gates generally lack this property and therefore constitute a major obstacle to fault tolerance~\cite{Eastin_Knill_2009}. Besides their important role in fault-tolerant quantum computation, circuits composed of elements of the Clifford group can be efficiently simulated on a classical computer~\cite{Gottesman_1998}, whereas the introduction of nonstabilizerness increases the computational cost of such simulations~\cite{aaronson_improved_2004}. 

Stabilizer states are pure states obtained by applying Clifford unitaries ($S$,
$H$, and $\mathrm{CNOT}$ gates) to the reference state $\vert 0
\rangle^{\otimes L}$. A particularly convenient non-stabilizerness monotone is the \textit{stabilizer (R\'{e}nyi) entropy}
(SE)~\citep{leone2022StabilizerRenyiEntropy}, defined as
\begin{equation}
M_{\alpha}(\ket{\psi}):=\frac{1}{1-\alpha}\log_2\Xi_{\alpha}(\ket{\psi})\,,
\label{eq:SRE}
\end{equation}
where the stabilizer purities (SP) are given
by
\begin{equation}
\Xi_{\alpha}(\ket{\psi}):=\frac{1}{2^L}\sum_{P\in\mathbb{P}_{L}}
|\langle\psi|P|\psi\rangle|^{2\alpha}\,,\label{eq:SRE_SP}
\end{equation}
with $\ket{\psi}$ a pure quantum state, $\alpha$ the R\'{e}nyi index (set to
$\alpha = 2$ throughout~\cite{Leone_Bittel_2024}), and $\mathbb{P}_L$ the set
of Pauli strings of length $L$ built from the identity and the three Pauli
matrices. Compared to other non-stabilizerness monotones, the SE is
experimentally measurable~\cite{Oliviero_Leone_Hamma_Lloyd_2022,
Ahmad_Esposito_Stasino_Odavic_Cosenza_Sarno_Mastrovito_Viscardi_Cusumano_Tafuri},
efficiently computable via matrix product state (MPS)
representations~\cite{Haug_Piroli_2023,
Haug_Piroli_2023QUANTUM,Lami_Collura_2023,Tarabunga_Tirrito_Banuls_Dalmonte_2024}
(ultimately limited by the bond dimension), and amenable to exact evaluation for
small- to intermediate-sized systems ($L\leq
20$)~\cite{Huang_Li_Zhong_2026, Xiao_Ryu_2026,
Sierant_Valles-Muns_Garcia-Saez_2026}.

\textit{SE for $U(1)$ symmetric random state ensemble-- } The main results of this Letter concerns the mean and variance of the SE in the $U(1)$-symmetric Haar ensemble. We derive these for the conserved charge defined by Eq.~\eqref{eq:defQz}. The result extends to any other Pauli projection frame, i.e.\ for $\mathcal{Q}^{(\alpha)} =
\sum_{j=1}^L \sigma^{(\alpha)}_j$ with $\alpha = x, y, z$, by Clifford invariance of $\Xi_{\alpha}(\ket{\psi})$ and left/right invariance of the Haar measure.

We compute analytically the average $2$-SP defined in Eq.~\eqref{eq:SRE_SP}, which provides a lower bound to average $2-$SE~\cite{leone2022StabilizerRenyiEntropy} via Jensen inequality, for any fixed sector with charge $q$. To perform this computation we make use of the result, proven in~\cite{cepollaro2024stabilizer}, which allows to compute Haar averages over a generic subspace, i.e. the charge sector $\mathcal{H}(q)$ in our case. Such an average can be computed as a Haar average over the global Hilbert space and then projected onto the subspace. Specifically, for the $k-$folded pure state, we have   
\begin{equation}\label{eq:lemma}
        \exv_{U_q}[\psi_{U_q} ^{\otimes k}]=c(d,d_q)\Pi_q^{\otimes k}\exv_{U}[\psi_{U} ^{\otimes k}]\,,
    \end{equation}
where $\Pi_q$ is given by Eq.~\eqref{eq:repr_proj}, $U \in U(\mathcal{H})$ is an Haar random unitary and the prefactor scales as $c(d,d_q) \simeq \left(d/d_q\right)^k$ for large system sizes $L$~\cite{SI}.

Using the result in Eq.~\eqref{eq:lemma} and the representation from Eq.~\eqref{eq:repr_proj} for the projector $\Pi_q$, we obtain an exact formula for  the  
average $2$-SP for any finite number of qubits $L$ and charge $q$ \cite{SI}, 
whose asymptotic limit $L, q \to +\infty$ at fixed charge density $s = q/L$, yields a leading behaviour 
\begin{equation}
    \label{eq:asympt_res}
    -\log_2 \exv_{U_q}[\Xi_2(\psi_{U_q})] = m(s)\, L + g(s) + O(L^{-1}),
\end{equation}
where $m(s)$ and $g(s)$ can be written in terms of elementary functions~\cite{SI}. Moreover, $\Xi_2(\psi_{U_q})$ exhibits Lévy concentration~\cite{zhu2016CliffordGroupFails} with respect to the Haar $U(1)$-symmetric measure: as we prove in the Supplemental Material~\cite{SI}, for any $\epsilon > 0$,
\begin{equation}
\label{eq:Levyconc}
    \text{Prob}\left( \abs{ \Xi_2(\psi_{U_q}) - \exv_{U_q}[\Xi_2(\psi_{U_q})]} \geq \epsilon \right) \leq 2 e^{-\frac{d_q \epsilon^2}{9 \pi^3 \eta^2}},
\end{equation}
with $\eta = 5.4$, implying a variance loosely bounded by $O(d_q^{-1})$. Since $d_{q = sL}$ grows exponentially in $L$ for any finite charge density $s \neq \pm 1$ (the only exception being the trivial sector $q = \pm L$ where $d_q = 1$), Eq.~\eqref{eq:asympt_res} yields the typical value of $\Xi_2(\psi_{U_q})$ in the thermodynamic limit with exponentially suppressed fluctuations; see the Supplemental Material~\cite{SI} for the exact result of the variance. Due to this high concentration of measure, the stabilizer entropy $M_2$ itself concentrates around its mean, with larger tails toward the extremal values (see~\cite{jplh-zl35} for an example).

Next, we highlight here some relevant properties, especially in comparison with the known result for Haar random states~\cite{leone2022StabilizerRenyiEntropy,zhu2016CliffordGroupFails, Iannotti_Esposito_Venuti_Hamma_2025}
\begin{equation}
\label{eq:HaarM2}
    -\log_2 \exv_U[ \Xi_2(\psi_{U})] \underset{L\to \infty}{=} L -2.
\end{equation}
The volume-law coefficient $m(s)$ in Eq.~\eqref{eq:asympt_res} is always lower than the Haar one, i.e. $m(s) \leq 1$. This bound is saturated for $s = 0$, i.e. the charge sector with largest dimension, in which the linear growth of $U(1)$ symmetric Haar random states is the same as the Haar ensemble. As the charge density $s$ increase, $m(s)$ monotonically decreases until, for $s = 1$, $m(1) = 0$ . The constant term $g(s)$ is instead strictly negative and lower than the Haar value, $g(s)<-2$ with maximal value reached for $s = 0$, in which $g(0) = -3$. Furthermore, for charge density $s$ close to the sector $s =0$ from Eq.~\eqref{eq:asympt_res} it follows
\begin{equation}
\label{eq:limM2s}
    \mathbb{E}_{U_s}[M_2(\psi_{U_s})] \underset{\substack{L\to\infty\\ s\to 0}}{\simeq} L-3 + \frac{L-3}{\log2}s^4 + O(s^6).
\end{equation}
Therefore, at $s = 0$ we obtain
\begin{equation}
\label{eq:limM0}
    \lim_{L \to \infty} \left(\,\mathbb{E}_{U}[M_2(\psi_U)]-\mathbb{E}_{U_0}[M_2(\psi_{U_0})] \,\right)= 1.
\end{equation}
Note that, although the difference is of order $1$, the two ensemble predictions are distinguishable as the two variances are exponentially small in $L$, as demonstrated in Fig.~\ref{fig:featured_image}. Since $m(s)<1$ for $s\neq 0$, the difference in Eq.~\eqref{eq:limM0} for generic charge density is always divergent as $(1-m(s)) L$. The behaviour in Eq.~\eqref{eq:limM2s} is also peculiar in terms of the dependence of the charge density $s$, as the quadratic corrections are vanishing. This effect is not present, for example, in the average Von Neumann entanglement entropy of a subsystem for the same ensemble. In that case, at leading order in the subsystem size $L_A$~\cite{Bianchi_Dona_2019,Bianchi_Hackl_Kieburg_Rigol_Vidmar_2022}
\begin{equation}
\mathbb{E}_{U_s}[S_{A}(\text{Tr}_B[\psi_{U_s}])] \underset{s\to 0}{\simeq} L_A \log2 - 2 L_A s^2 + O(s^4).
\end{equation}

The results above hold for charges defined as sums of only one type of Pauli operators. More generally, consider $\mathcal{Q}_{\vec{n}} = \sum_{j=1}^L \vec{n} \cdot \vec{\sigma}_j$ with $\vec{\sigma} = \{\sigma^x, \sigma^y, \sigma^z\}$ and $\norm{\vec{n}}_2 = 1$. The same approach yields the exact average $2$-SP over Haar-random states $\psi^{(\vec{n})}_{U_q}$ in the $U(1)$ symmetry sector of $\mathcal{Q}_{\vec{n}}$ with charge $q$; the explicit finite-size result is given in the Supplemental Material~\cite{SI}.

In the thermodynamic limit at fixed charge density $s = q/L$, the asymptotic behaviour is expected to mirror Eq.~\eqref{eq:asympt_res}, though the functions $m_{\vec{n}}(s)$ and $g_{\vec{n}}(s)$ are not known in closed form for arbitrary $\vec{n}$. Nevertheless, in the sector $s = 0$ we can show that~\cite{SI}
\begin{equation}
    -\log_2 \exv[\Xi_2(\psi^{(\vec{n})}_{U_0})] \underset{L\to +\infty}{=} \begin{cases}
        L - 3, \qquad \vec{n} = \hat{x}, \hat{y}, \hat{z}, \\
        L - 2, \qquad \text{otherwise}.
    \end{cases}
\end{equation}
This reveals a sharp dichotomy: unless the conserved charge aligns exactly with one of the axes defining the stabilizer polytope (i.e.\ the basis in which the $2$-SP is computed), the average $2$-SP approaches that of globally Haar-random states up to corrections that vanish with $L$.

\begin{figure}[t!]
    \centering
    \includegraphics[width=0.9\columnwidth]{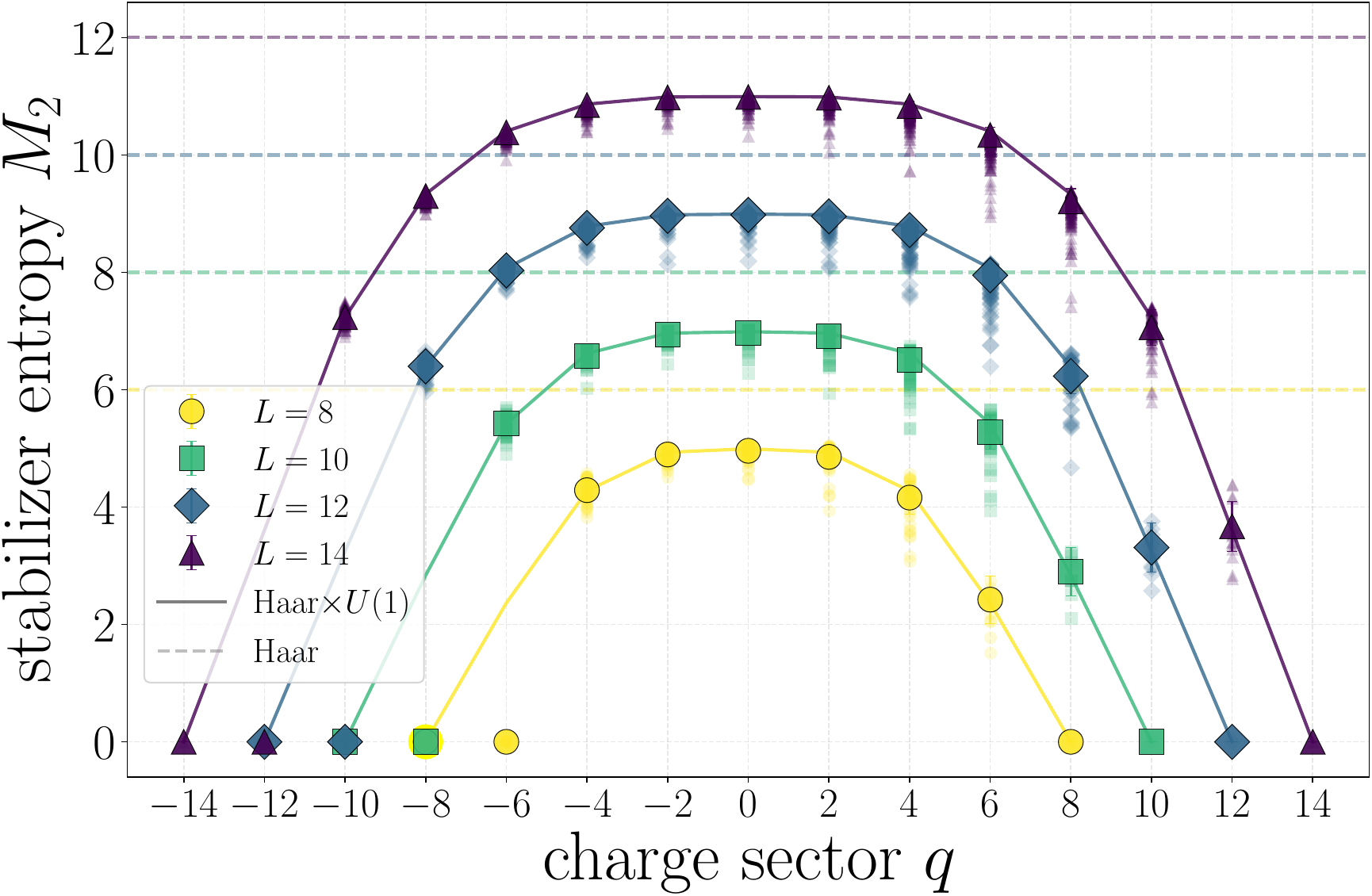}
    \caption{Disorder-averaged stabilizer entropy of cSYK eigenstates as a function of the charge sector $q$ for different system sizes $L$. Each data point is averaged over disorder realizations (for more detail see SM~\cite{SI}) and all eigenstates within the sector, regardless of energy density, since all of states exhibit chaotic behaviour in both entanglement and non-stabilizerness content~\cite{Liu_Chen_Balents_2018, Jasser_Odavic_Hamma_2025}. Faded points show single-realization data. Solid lines denote our analytical prediction for constrained Haar-random states, while dashed lines correspond to the unconstrained Haar ensemble. The two differ by an $\mathcal{O}(1)$ offset. Numerical results agree with the theory across all sectors, except $q = -L +2$ where the Hamiltonian is trivial (see main text). }
    \label{fig:cSYK}
\end{figure}

\textit{Non-local chaotic model (cSYK).---} To probe the effects of $U(1)$ symmetry on quantum state complexity, we study the complex Sachdev-Ye-Kitaev (cSYK) model~\cite{sachdev1993gapless, KITP2015, chowdhury2022sachdev}. Consider a Hilbert space $\mathcal{H}$ of dimension $d = 2^L$ describing $L$ complex fermionic modes, governed by the Hamiltonian
\begin{equation}
H_{\mathrm{cSYK}} = \frac{1}{(2L)^{3/2}} 
\sum_{i,j,k,l=1}^{L} 
J_{ij;kl}\, c_i^{\dagger} c_j^{\dagger} c_k c_l,
\label{eq:SYK}
\end{equation}
where $c_i$ and $c_i^{\dagger}$ satisfy the canonical anticommutation relations $\{c_i, c_j^{\dagger}\} = \delta_{ij}$, $\{c_i, c_j\} = \{c_i^{\dagger}, c_j^{\dagger}\} = 0$. The couplings \( J_{ij;kl} \) are complex Gaussian random variables with $\mathbb{E}[J_{ij;kl}] = 0, \; \mathbb{E}[|J_{ij;kl}|^2] = 1,$
subject to the antisymmetry conditions $J_{ij;kl} = - J_{ji;kl} = - J_{ij;lk} = J_{kl;ij}$, which ensure Hermiticity of the Hamiltonian. These all-to-all four-fermion interactions give rise to a strongly correlated zero-dimensional quantum system that is maximally chaotic in the large-$L$ limit, with Hamiltonian eigenstates that are highly delocalized. 

A defining feature of this model is the conservation of the total fermionic number. Let us introduce the operator $ \mathcal{Q}= 2\sum_{i=1}^{L} c_i^{\dagger} c_i - L$, which captures the total number of occupied fermionic modes in the system. 
By construction, the Hamiltonian in Eq.~\eqref{eq:SYK} conserves the total fermionic number, since the elementary interaction term 
\( c_i^{\dagger} c_j^{\dagger} c_k c_l \) creates and annihilates two fermions simultaneously. 
This implies that the charge operator commutes with the Hamiltonian, $[H_{\mathrm{cSYK}}, \mathcal{Q}] = 0$. 
Hence, \( \mathcal{Q} \) generates a \( U(1) \) symmetry corresponding to global phase rotations of the fermionic operators; $c_i \mapsto e^{\mathrm{i}\theta} c_i, c_i^{\dagger} \mapsto e^{-\mathrm{i}\theta} c_i^{\dagger}$, under which the Hamiltonian remains invariant. 
The quantum chaos properties mentioned above are sensitive to the presence of the $U(1)$ symmetry, see~\cite{SI}. In particular, one must restrict to fixed-charge sectors in order to probe correlations between eigenvectors with the same conserved charge. The impact of this sector-based decomposition is also clearly visible in the behavior of $M_2$. As shown in Fig.~\ref{fig:cSYK}, the average value of $M_2$ changes significantly across different charge sectors $q$. Interestingly, a similar behavior can be observed as a function of the energy density \cite{private}. The numerical data confirm that while the system approaches the theoretical expectations for a $U(1)$-symmetric Haar ensemble, even for $q=0$ it does not reach the usual Haar value; see SM~\cite{SI} for further numerical evidence. 
Notice that, for any system size L, the sector at charge $q = -L+2$ corresponds to a single fermionic excitation.
Owing to the quartic structure of the interaction term in Eq.~\eqref{eq:SYK}, the Hamiltonian acts trivially in this sector and the spectrum is fully degenerate. As a consequence, the diagonalization routine selects computational basis states as orthonormal basis within this degenerate subspace. This vanishing value is not a fundamental property of the sector itself, in principle, any orthonormal superposition within the degenerate subspace would also be a valid eigenbasis, and such choices could lead to nonzero values of $M_2$.

\begin{figure}[t!]
    \centering
    \includegraphics[width=\columnwidth]{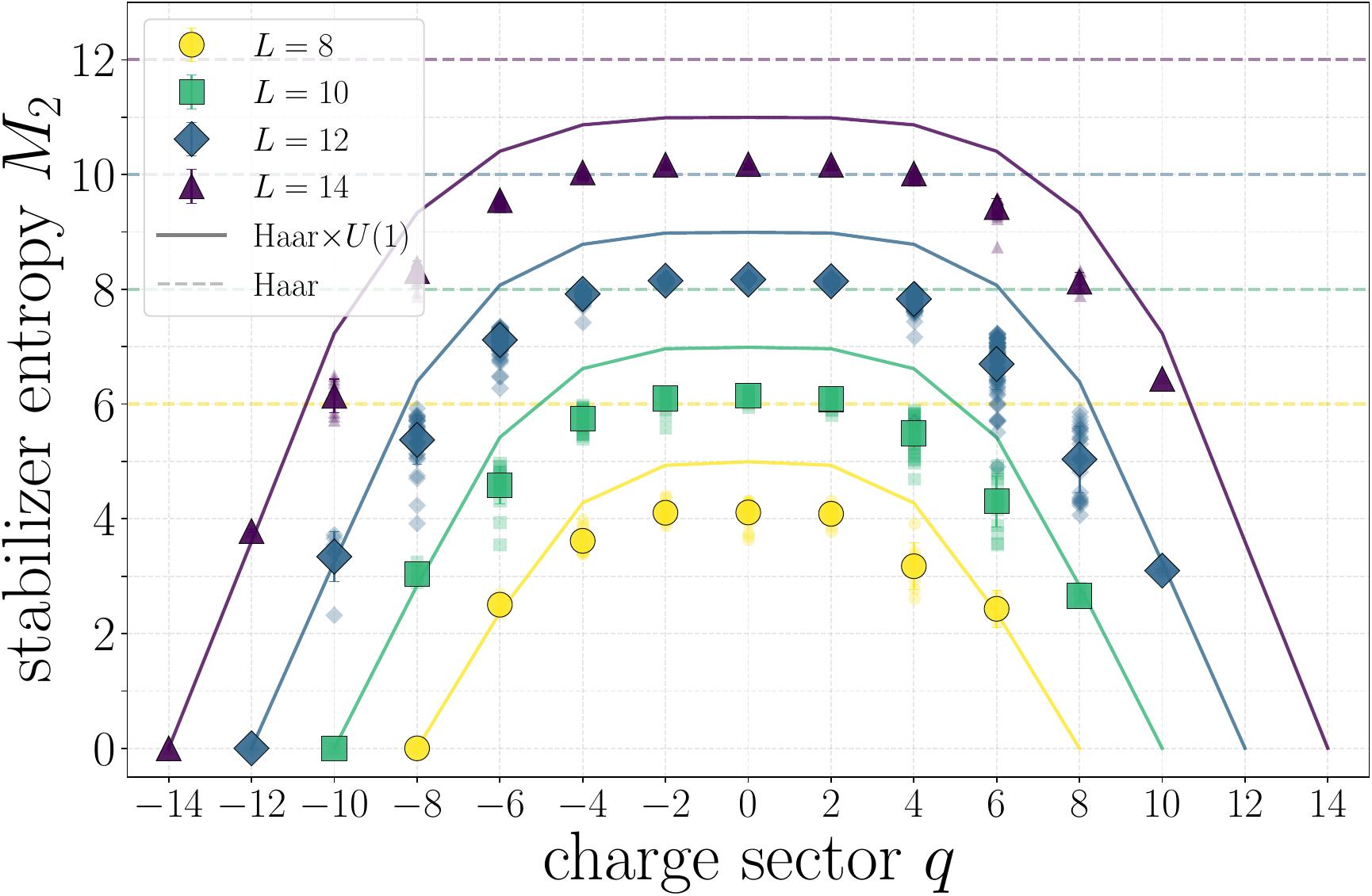}
    \caption{Stabilizer entropy of XXZ-NNN$\times U(1)$ eigenstates as a function of the magnetization sector $q$ for different system sizes $L$. Each data point is averaged over all eigenstates (mid-spectum states) within the sector in the energy density window $-0.25 < E/L < 0.25$; missing data at large positive $q$ reflect the absence of states with those magnetization quantum numbers in the selected energy interval.  Solid lines denote our analytical prediction for constrained Haar-random states, while dashed lines correspond to the unconstrained Haar ensemble.   }
    \label{fig:XXZ_main}
\end{figure}

\textit{Local chaotic model (XXZ-NNN) -- } 
When interactions are local, chaotic Hamiltonians respond differently to magic probes such as SPs in Eq.~\eqref{eq:SRE_SP}.
\noindent
Following Ref.~\cite{langlett2025entanglement}, we consider the non-integrable spin-1/2 Hamiltonian
\begin{equation}
\begin{split}
    H_{\rm XXZ-NNN \hspace{0.05cm}x \hspace{0.05cm} U(1)} &= \sum\limits_{j = 1}^{L} J_1 \, \left( \sigma^x_j \sigma^x_{j + 1}+ \sigma^y_{j} \sigma^y_{j + 1} \right) \\
    &+ \Delta \, \sigma^z_j \sigma^z_{j + 1}+ J_2 \,\sigma^z_j \sigma^z_{j+1} \sigma^z_{j + 2} \\
    &+ h_b \, (\sigma^z_1 - \sigma^z_j), \label{eq:Hamiltonian_XXZ-NNN1}
\end{split}
\end{equation}
with open boundary conditions (OBC). This Hamiltonian conserves energy and admits the total magnetization $\mathcal{Q} = \sum_{j = 1}^{L} \sigma^z_{j}$ as a scalar charge, i.e.\ $[H_{\rm XXZ-NNN\hspace{0.05cm}x \hspace{0.05cm} U(1)}, \mathcal{Q} ] = 0$. When the next-to-nearest-neighbor (NNN) coupling and boundary fields are set to zero ($J_2 = 0$, $h_b = 0$), the model reduces to an integrable one, which depending on the remaining parameter choices, either via a free-fermion mapping or Bethe Ansatz techniques. The OBC break translational symmetry, while the non-vanishing boundary fields break inversion symmetry, leaving the model with no additional symmetries and rendering it as an example of a maximally chaotic locally interacting system. To break the $U(1)$ scalar symmetry associated with magnetization conservation, one introduces a transverse field $ h_x \sum_{j = 1}^{L} \sigma^x_j$.

In Fig.~\ref{fig:XXZ_main}, we show the behaviour of the stabilizer entropy across different magnetization sectors for the local XXZ+NNN spin chain. A persistent mismatch between the analytical and numerical results is visible, when restricting to mid-spectrum eigenstates, for all system sizes considered. This should be in contrast with the analogous analysis of the cSYK model in Fig.~\ref{fig:cSYK}, where the agreement is exact. 
Both models are chaotic according to standard diagnostics of quantum complexity, and the eigenstates we target lie in the chaotic bulk; the key difference is the range of the interaction couplings. This points to locality of interactions playing a fundamental role: local constraints induce additional structure in the eigenstates that is absent in typical states, whether symmetry-constrained or otherwise. 
We note that similar conclusions can be drawn from the entanglement entropy response between the cases, making stabilizer entropy consistent with these findings~\cite{Langlett_Jonay_Khemani_Rodriguez-Nieva_2025,langlett2025entanglement,Russotto_Ares_Calabrese_2025, cao2025}.

\textit{Conclusions and Discussion--} In this work, we provided a complete
characterization of the stabilizer entropy for constrained pure random states in the presence of a conserved $U(1)$ charge. Our results reveal a distinct signature of global symmetries in the magic of quantum states qualitatively different from the corresponding entanglement
entropy response. These differences persist in the thermodynamic limit and are independent of the choice of Pauli
projection frame. We tested our analytical expressions against both local and non-local finite-size chaotic many-body Hamiltonians, finding excellent agreement when the interaction range is strictly non-local. We further clarify the subtleties of using participation entropies as a proxy for
non-stabilizerness~\cite{tirrito2025universal} (see Supplemental Material), showing that the stabilizer entropy captures a nontrivial response to symmetry constraints that participation-based proxies can obscure.

 Random pure states already play a central role in a variety of practical applications ranging from quantum information protocols to the study of quantum chaos and complexity~\cite{Hayden_2004, Pirandola_2020, Arute_2019, Neill_2018,yang2026}. Beyond this settings our work opens several avenues for future research. In particular, our findings can be directly applied to the late-time behavior of quantum circuits that explicitly preserve $U(1)$ symmetry~\cite{Rakovszky_2018,Khemani_2018} and to Floquet circuits~\cite{Friedman_2019,Jonay_2024}. More broadly, the constrained-state methodology developed here could be extended to Clifford$+T$ circuits, opening a path toward symmetry-resolved studies of the onset of quantum chaos~\cite{sonya2025nonstabilizerness, leone2025noncliffordcostrandomunitaries, Haug_Aolita_Kim_2025,Leone_2024,Piroli_2025}. 
On the many-body side, our results carry over naturally to quantum quench settings, where symmetry constraints play a central role~\cite{Odavic_Viscardi_Hamma_2025}. 
A further application lies in granular SYK models. Such models represent arrays or lattices of SYK quantum dots that provide a tractable framework for modeling strange metallic behavior in higher dimensions, and our analytical predictions for the magic content of individual dots could serve as building blocks (see Ref.~\cite{Altland_Bagrets_Kamenev_2019, chowdhury2022sachdev} and references therein). Furthermore, it would be interesting to explore the consequence of our findings in the context of the coupled and thermofield double SYK states~\cite{Sun_Zhang_2026, Zhang_Zhou_Sun_2026}. 

Complementary and concurrent works study related questions from different
angles: Ref.~\cite{Turkeshi_2025} examines the typical Pauli spectrum for
eigenstates of chaotic many-body Hamiltonians,
Ref.~\cite{tirrito2025universal} establishes connections with participation
entropies in $U(1)$-symmetric systems, and Ref.~\cite{liu2022many} studies
non-stabilizerness using alternative measures, while Ref.~\cite{PhysRevB.111.L081102} studies a one-dimensional $U(1)$ lattice gauge theory including matter fields.
As showed in Ref.~\cite{hoshino2025stabilizerrenyientropyconformal,hoshino2025stabilizerrenyientropyencodes}, topological defects yield a universal size-independent term in SE for boundary CFTs, similar to the size-independent term found in Eq.~\eqref{eq:limM2s}.
We leave the extension to
non-Abelian symmetries~\cite{Calabrese_2021,10.21468/SciPostPhys.17.5.127, santra2025quantumresourcesnonabelianlattice, liu2024unitarydesignsrandomsymmetric,Majidy_2023,Majidy_2023_nat} for future work, as well as the study of \textit{Operator Stabilizer Entropy}~\cite{p7xt-s9nz}. The relationship between constraints (like gauge-structures) and Stabilizer Entropy in  topological states could be studied also in the sensitivity to the entanglement spectrum \cite{PhysRevB.88.125117} as this is connected to stabilizer entropy through anti-flatness quantities \cite{PhysRevA.109.L040401}.

\textit{Code and Data Availability--} The code and the data for our simulations will be publicly shared at publication.
 
\section{Acknowledgements} 
A.R. and D.I. would like to acknowledge the Les Houches Summer School 2025 on Exact Solvability and Quantum Information, which facilitated work on this subject. D.I. would also like to thank Xhek Turkeshi for extensive discussions on this and related subjects. A.R. thanks Filiberto Ares for helpful discussions. A.R. acknowledges support from the European Research Council under the Advanced Grant no. 101199196 (MOSE). J.O. and A.H. acknowledge support from  PNRR MUR project PE0000023-NQSTI and the PNRR MUR project CN
00000013-ICSC.

\bibliography{refs_arXiv_v2}

@article{Vidmar_2017,
   title={Entanglement Entropy of Eigenstates of Quantum Chaotic Hamiltonians},
   volume={119},
   ISSN={1079-7114},
   url={http://dx.doi.org/10.1103/PhysRevLett.119.220603},
   DOI={10.1103/physrevlett.119.220603},
   number={22},
   journal={Physical Review Letters},
   publisher={American Physical Society (APS)},
   author={Vidmar, Lev and Rigol, Marcos},
   year={2017},
   month=nov }

@article{Vidmar_2017_q,
   title={Entanglement Entropy of Eigenstates of Quadratic Fermionic Hamiltonians},
   volume={119},
   ISSN={1079-7114},
   url={http://dx.doi.org/10.1103/PhysRevLett.119.020601},
   DOI={10.1103/physrevlett.119.020601},
   number={2},
   journal={Physical Review Letters},
   publisher={American Physical Society (APS)},
   author={Vidmar, Lev and Hackl, Lucas and Bianchi, Eugenio and Rigol, Marcos},
   year={2017},
   month=jul }

@article{Lu_2019,
   title={Renyi entropy of chaotic eigenstates},
   volume={99},
   ISSN={2470-0053},
   url={http://dx.doi.org/10.1103/PhysRevE.99.032111},
   DOI={10.1103/physreve.99.032111},
   number={3},
   journal={Physical Review E},
   publisher={American Physical Society (APS)},
   author={Lu, Tsung-Cheng and Grover, Tarun},
   year={2019},
   month=mar }

@article{Bianchi_Hackl_Kieburg_Rigol_Vidmar_2022, title={Volume-Law Entanglement Entropy of Typical Pure Quantum States}, volume={3}, DOI={10.1103/PRXQuantum.3.030201}, number={3}, journal={PRX Quantum}, publisher={American Physical Society}, author={Bianchi, Eugenio and Hackl, Lucas and Kieburg, Mario and Rigol, Marcos and Vidmar, Lev}, year={2022}, month=july, pages={030201} }

@article{Page_1993, title={Average entropy of a subsystem}, volume={71}, DOI={10.1103/PhysRevLett.71.1291}, number={9}, journal={Physical Review Letters}, publisher={American Physical Society}, author={Page, Don N.}, year={1993}, month=aug, pages={1291–1294} }

@article{Bianchi_Dona_2019, title={Typical entanglement entropy in the presence of a center: Page curve and its variance}, volume={100}, DOI={10.1103/PhysRevD.100.105010}, number={10}, journal={Physical Review D}, publisher={American Physical Society}, author={Bianchi, Eugenio and Donà, Pietro}, year={2019}, month=nov, pages={105010} }

@article{Patil_2023,
   title={Average pure-state entanglement entropy in spin systems with SU(2) symmetry},
   volume={108},
   ISSN={2469-9969},
   url={http://dx.doi.org/10.1103/PhysRevB.108.245101},
   DOI={10.1103/physrevb.108.245101},
   number={24},
   journal={Physical Review B},
   publisher={American Physical Society (APS)},
   author={Patil, Rohit and Hackl, Lucas and Fagan, George R. and Rigol, Marcos},
   year={2023},
   month=dec }

@preprint{chakraborty2025,
      title={Random matrix prediction of average entanglement entropy in non-Abelian symmetry sectors}, 
      author={Anwesha Chakraborty and Lucas Hackl and Mario Kieburg},
      year={2025},
      eprint={2512.22942},
      archivePrefix={arXiv},
      primaryClass={quant-ph},
      url={https://arxiv.org/abs/2512.22942}, 
}

@article{Morampudi_2020,
   title={Universal Entanglement of Typical States in Constrained Systems},
   volume={124},
   ISSN={1079-7114},
   url={http://dx.doi.org/10.1103/PhysRevLett.124.050602},
   DOI={10.1103/physrevlett.124.050602},
   number={5},
   journal={Physical Review Letters},
   publisher={American Physical Society (APS)},
   author={Morampudi, S. C. and Chandran, A. and Laumann, C. R.},
   year={2020},
   month=feb }

@article{Huang_2021, title={Universal entanglement of mid-spectrum eigenstates of chaotic local Hamiltonians}, volume={966}, ISSN={0550-3213}, DOI={10.1016/j.nuclphysb.2021.115373}, journal={Nuclear Physics B}, author={Huang, Yichen}, year={2021}, month=may, pages={115373} }

@article{Murciano_2022,
   title={Symmetry-resolved Page curves},
   volume={106},
   ISSN={2470-0029},
   url={http://dx.doi.org/10.1103/PhysRevD.106.046015},
   DOI={10.1103/physrevd.106.046015},
   number={4},
   journal={Physical Review D},
   publisher={American Physical Society (APS)},
   author={Murciano, Sara and Calabrese, Pasquale and Piroli, Lorenzo},
   year={2022},
   month=aug }

@article{Kliczkowski_2023,
   title={Average entanglement entropy of midspectrum eigenstates of quantum-chaotic interacting Hamiltonians},
   volume={107},
   ISSN={2470-0053},
   url={http://dx.doi.org/10.1103/PhysRevE.107.064119},
   DOI={10.1103/physreve.107.064119},
   number={6},
   journal={Physical Review E},
   publisher={American Physical Society (APS)},
   author={Kliczkowski, M. and \'Swi\k{e}tek, R. and Vidmar, L. and Rigol, M.},
   year={2023},
   month=jun }

@article{Yauk_2024,
   title={Typical entanglement entropy in systems with particle-number conservation},
   volume={110},
   ISSN={2469-9969},
   url={http://dx.doi.org/10.1103/PhysRevB.110.235154},
   DOI={10.1103/physrevb.110.235154},
   number={23},
   journal={Physical Review B},
   publisher={American Physical Society (APS)},
   author={Yauk, Yale and Patil, Rohit and Zhang, Yicheng and Rigol, Marcos and Hackl, Lucas},
   year={2024},
   month=dec }

@article{Rodriguez-Nieva_Jonay_Khemani_2024, title={Quantifying Quantum Chaos through Microcanonical Distributions of Entanglement}, volume={14}, DOI={10.1103/PhysRevX.14.031014}, number={3}, journal={Physical Review X}, publisher={American Physical Society}, author={Rodriguez-Nieva, Joaquin F. and Jonay, Cheryne and Khemani, Vedika}, year={2024}, month=july, pages={031014} }

@preprint{Langlett_Jonay_Khemani_Rodriguez-Nieva_2025, title={Quantum chaos at finite temperature in local spin Hamiltonians}, url={http://arxiv.org/abs/2501.13164}, DOI={10.48550/arXiv.2501.13164}, number={arXiv:2501.13164}, publisher={arXiv}, author={Langlett, Christopher M. and Jonay, Cheryne and Khemani, Vedika and Rodriguez-Nieva, Joaquin F.}, year={2025}, month=jan }

@article{Russotto_Ares_Calabrese_2025, title={Symmetry breaking in chaotic many-body quantum systems at finite temperature}, volume={112}, DOI={10.1103/kppn-3272}, number={3}, journal={Physical Review E}, publisher={American Physical Society}, author={Russotto, Angelo and Ares, Filiberto and Calabrese, Pasquale}, year={2025}, month=sept, pages={L032101} }

@article{PhysRevA.71.022316,
  title = {Universal quantum computation with ideal Clifford gates and noisy ancillas},
  author = {Bravyi, Sergey and Kitaev, Alexei},
  journal = {Phys. Rev. A},
  volume = {71},
  issue = {2},
  pages = {022316},
  numpages = {14},
  year = {2005},
  month = {Feb},
  publisher = {American Physical Society},
  doi = {10.1103/PhysRevA.71.022316},
  url = {https://link.aps.org/doi/10.1103/PhysRevA.71.022316}
}

@preprint{tirrito2025universal,
  title={Universal Spreading of Nonstabilizerness and Quantum Transport}, 
      author={Emanuele Tirrito and Poetri Sonya Tarabunga and Devendra Singh Bhakuni and Marcello Dalmonte and Piotr Sierant and Xhek Turkeshi},
      year={2025},
      eprint={2506.12133},
      archivePrefix={arXiv},
      primaryClass={quant-ph},
      url={https://arxiv.org/abs/2506.12133}, 
}

@article{Piemontese_Roscilde_Hamma_2023, title={Entanglement complexity of the Rokhsar-Kivelson-sign wavefunctions}, volume={107}, DOI={10.1103/PhysRevB.107.134202}, number={13}, journal={Physical Review B}, publisher={American Physical Society}, author={Piemontese, Stefano and Roscilde, Tommaso and Hamma, Alioscia}, year={2023}, month=apr, pages={134202} }

@article{Zhou_Yang_Hamma_Chamon_2020, title={Single T gate in a Clifford circuit drives transition to universal entanglement spectrum statistics}, volume={9}, ISSN={2542-4653}, DOI={10.21468/SciPostPhys.9.6.087}, abstractNote={SciPost Journals Publication Detail SciPost Phys. 9, 087 (2020) Single T gate in a Clifford circuit drives transition to universal entanglement spectrum statistics}, number={6}, journal={SciPost Physics}, author={Zhou, Shiyu and Yang, Zhicheng and Hamma, Alioscia and Chamon, Claudio}, year={2020}, month=dec, pages={087}}

@article{Leone_Oliviero_Zhou_Hamma_2021, title={Quantum Chaos is Quantum}, volume={5}, DOI={10.22331/q-2021-05-04-453}, abstractNote={Lorenzo Leone, Salvatore F. E. Oliviero, You Zhou, and Alioscia Hamma,
Quantum 5, 453 (2021).
It is well known that a quantum circuit on $N$ qubits composed of Clifford gates with the addition of $k$ non Clifford gates can be simulated on a classical computer by an algorithm scaling as…}, journal={Quantum}, publisher={Verein zur Förderung des Open Access Publizierens in den Quantenwissenschaften}, author={Leone, Lorenzo and Oliviero, Salvatore F. E. and Zhou, You and Hamma, Alioscia}, year={2021}, month=may, pages={453} }

@article{True_Hamma_2022, title={Transitions in Entanglement Complexity in Random Circuits}, volume={6}, DOI={10.22331/q-2022-09-22-818}, abstractNote={Sarah True and Alioscia Hamma,
Quantum 6, 818 (2022).
Entanglement is the defining characteristic of quantum mechanics. Bipartite entanglement is characterized by the von Neumann entropy. Entanglement is not just described by a number, however;…}, journal={Quantum}, publisher={Verein zur Förderung des Open Access Publizierens in den Quantenwissenschaften}, author={True, Sarah and Hamma, Alioscia}, year={2022}, month=sept, pages={818} }

@article{Hinsche_Ioannou_Nietner_Haferkamp_Quek_Hangleiter_Seifert_Eisert_Sweke_2023, title={One $T$ Gate Makes Distribution Learning Hard}, volume={130}, DOI={10.1103/PhysRevLett.130.240602}, number={24}, journal={Physical Review Letters}, publisher={American Physical Society}, author={Hinsche, M. and Ioannou, M. and Nietner, A. and Haferkamp, J. and Quek, Y. and Hangleiter, D. and Seifert, J.-P. and Eisert, J. and Sweke, R.}, year={2023}, month=june, pages={240602} }

@article{Bejan_McLauchlan_Beri_2024, title={Dynamical Magic Transitions in Monitored Clifford+$T$ Circuits}, volume={5}, DOI={10.1103/PRXQuantum.5.030332}, number={3}, journal={PRX Quantum}, publisher={American Physical Society}, author={Bejan, Mircea and McLauchlan, Campbell and Béri, Benjamin}, year={2024}, month=aug, pages={030332} }

@article{Oliviero_Leone_Hamma_2022, title={Magic-state resource theory for the ground state of the transverse-field Ising model}, volume={106}, DOI={10.1103/PhysRevA.106.042426}, number={4}, journal={Physical Review A}, publisher={American Physical Society}, author={Oliviero, Salvatore F. E. and Leone, Lorenzo and Hamma, Alioscia}, year={2022}, month=oct, pages={042426} }

@article{Lami_Collura_2023, title={Nonstabilizerness via Perfect Pauli Sampling of Matrix Product States}, volume={131}, DOI={10.1103/PhysRevLett.131.180401}, number={18}, journal={Physical Review Letters}, publisher={American Physical Society}, author={Lami, Guglielmo and Collura, Mario}, year={2023}, month=oct, pages={180401} }

@article{Rattacaso_Leone_Oliviero_Hamma_2023, title={Stabilizer entropy dynamics after a quantum quench}, volume={108}, DOI={10.1103/PhysRevA.108.042407}, number={4}, journal={Physical Review A}, publisher={American Physical Society}, author={Rattacaso, Davide and Leone, Lorenzo and Oliviero, Salvatore F. E. and Hamma, Alioscia}, year={2023}, month=oct, pages={042407} }

@article{Odavic_Viscardi_Hamma_2025, title={Stabilizer entropy in nonintegrable quantum evolutions}, volume={112}, DOI={10.1103/y9r6-dx7p}, number={10}, journal={Physical Review B}, publisher={American Physical Society}, author={Odavić, J. and Viscardi, M. and Hamma, A.}, year={2025}, month=sept, pages={104301} }

@article{Viscardi_Dalmonte_Hamma_Tirrito_2026, title={Interplay of entanglement structures and stabilizer entropy in spin models}, volume={9}, ISSN={2666-9366}, DOI={10.21468/SciPostPhysCore.9.1.012}, abstractNote={SciPost Journals Publication Detail SciPost Phys. Core 9, 012 (2026) Interplay of entanglement structures and stabilizer entropy in spin models}, number={1}, journal={SciPost Physics Core}, author={Viscardi, Michele and Dalmonte, Marcello and Hamma, Alioscia and Tirrito, Emanuele}, year={2026}, month=feb, pages={012} }

@article{Jasser_Odavic_Hamma_2025, title={Stabilizer entropy and entanglement complexity in the Sachdev-Ye-Kitaev model}, volume={112}, DOI={10.1103/rz86-47h3}, number={17}, journal={Physical Review B}, publisher={American Physical Society}, author={Jasser, Barbara and Odavić, Jovan and Hamma, Alioscia}, year={2025}, month=nov, pages={174204} }

@article{leone2022StabilizerRenyiEntropy,
  title = {Stabilizer {{R\'enyi Entropy}}},
  author = {Leone, Lorenzo and Oliviero, Salvatore F. E. and Hamma, Alioscia},
  year = {2022},
  month = feb,
  journal = {Physical Review Letters},
  volume = {128},
  pages = {050402--050402},
  doi = {10.1103/PhysRevLett.128.050402}
}

@article{Leone_Bittel_2024, title={Stabilizer entropies are monotones for magic-state resource theory}, volume={110}, DOI={10.1103/PhysRevA.110.L040403}, number={4}, journal={Physical Review A}, publisher={American Physical Society}, author={Leone, Lorenzo and Bittel, Lennart}, year={2024}, month=oct, pages={L040403} }

@preprint{Bittel_Leone_2025, title={Operational interpretation of the Stabilizer Entropy}, url={http://arxiv.org/abs/2507.22883}, DOI={10.48550/arXiv.2507.22883}, note={arXiv:2507.22883 [quant-ph]}, number={arXiv:2507.22883}, publisher={arXiv}, author={Bittel, Lennart and Leone, Lorenzo}, year={2025} }

@article{Beverland_Campbell_Howard_Kliuchnikov_2020, title={Lower bounds on the non-Clifford resources for quantum computations}, volume={5}, ISSN={2058-9565}, DOI={10.1088/2058-9565/ab8963}, number={3}, journal={Quantum Science and Technology}, publisher={IOP Publishing}, author={Beverland, Michael and Campbell, Earl and Howard, Mark and Kliuchnikov, Vadym}, year={2020}, month=may, pages={035009} }

@article{Liu_Winter_2022, title={Many-Body Quantum Magic}, volume={3}, DOI={10.1103/PRXQuantum.3.020333}, number={2}, journal={PRX Quantum}, publisher={American Physical Society}, author={Liu, Zi-Wen and Winter, Andreas}, year={2022}, month=may, pages={020333} }

@article{Howard_Campbell_2017, title={Application of a Resource Theory for Magic States to Fault-Tolerant Quantum Computing}, volume={118}, DOI={10.1103/PhysRevLett.118.090501}, number={9}, journal={Physical Review Letters}, publisher={American Physical Society}, author={Howard, Mark and Campbell, Earl}, year={2017}, month=mar, pages={090501} }

@article{iannotti2026van,
  title={Van Hove singularities in stabilizer entropy densities},
   volume={59},
   ISSN={1751-8121},
   url={http://dx.doi.org/10.1088/1751-8121/ae3cd6},
   DOI={10.1088/1751-8121/ae3cd6},
   number={7},
   journal={Journal of Physics A: Mathematical and Theoretical},
   publisher={IOP Publishing},
   author={Iannotti, Daniele and Campos Venuti, Lorenzo and Hamma, Alioscia},
   year={2026},
   month=feb, pages={075301} }

@article{Chowdhury_Georges_Parcollet_Sachdev_2022, title={Sachdev-Ye-Kitaev models and beyond: Window into non-Fermi liquids}, volume={94}, DOI={10.1103/RevModPhys.94.035004}, number={3}, journal={Reviews of Modern Physics}, publisher={American Physical Society}, author={Chowdhury, Debanjan and Georges, Antoine and Parcollet, Olivier and Sachdev, Subir}, year={2022}, month=sept, pages={035004} }

@preprint{jha_introduction_2025,
    title = {Introduction to {Sachdev}-{Ye}-{Kitaev} {Model}: {A} {Strongly} {Correlated} {System} {Perspective}},
    shorttitle = {Introduction to {Sachdev}-{Ye}-{Kitaev} {Model}},
    url = {http://arxiv.org/abs/2507.07195},
    doi = {10.48550/arXiv.2507.07195},
    urldate = {2026-01-03},
    publisher = {arXiv},
    author = {Jha, Rishabh},
    month = jul,
    year = {2025},
    note = {arXiv:2507.07195 [hep-th]},
}

@article{langlett2025entanglement,
  title={Entanglement patterns of quantum chaotic Hamiltonians with a scalar U (1) charge},
  author={Langlett, Christopher M and Rodriguez-Nieva, Joaquin F},
  journal={Physical Review Letters},
  volume={134},
  number={23},
  pages={230402},
  year={2025},
  publisher={APS},
  url={https://doi.org/10.1103/xxlq-d1sw}
}

@article{Huang_2019, title={Universal eigenstate entanglement of chaotic local Hamiltonians}, volume={938}, ISSN={0550-3213}, DOI={10.1016/j.nuclphysb.2018.09.013}, journal={Nuclear Physics B}, author={Huang, Yichen}, year={2019}, month=jan, pages={594–604} }

@article{Lau_2022,
   title={Page curve and symmetries},
   volume={2022},
   ISSN={1029-8479},
   url={http://dx.doi.org/10.1007/JHEP10(2022)015},
   DOI={10.1007/jhep10(2022)015},
   number={10},
   journal={Journal of High Energy Physics},
   publisher={Springer Science and Business Media LLC},
   author={Lau, Pak Hang Chris and Noumi, Toshifumi and Takii, Yuhei and Tamaoka, Kotaro},
   year={2022},
   month=oct }

@preprint{ghasemi2024symmetryresolvedrelativeentropyrandom,
      title={Symmetry-Resolved Relative Entropy of Random States}, 
      author={Mostafa Ghasemi},
      year={2024},
      eprint={2411.01491},
      archivePrefix={arXiv},
      primaryClass={hep-th},
      url={https://arxiv.org/abs/2411.01491}, 
}

@preprint{Vallini2025vvq,
title={Refinements of the Eigenstate Thermalization Hypothesis under Local Rotational Invariance via Free Probability}, 
      author={Elisa Vallini and Laura Foini and Silvia Pappalardi},
      year={2025},
      eprint={2511.23217},
      archivePrefix={arXiv},
      primaryClass={cond-mat.stat-mech},
      url={https://arxiv.org/abs/2511.23217}, 
}

@preprint{harrow2013church,
  title={The Church of the Symmetric Subspace}, 
      author={Aram W. Harrow},
      year={2013},
      eprint={1308.6595},
      archivePrefix={arXiv},
      primaryClass={quant-ph},
      url={https://arxiv.org/abs/1308.6595}, 
}

@book{wigner2012group,
  title={Group theory: and its application to the quantum mechanics of atomic spectra},
  author={Wigner, Eugene},
  volume={5},
  year={2012},
  publisher={Elsevier},
  url={https://www.sciencedirect.com/bookseries/pure-and-applied-physics/vol/5/suppl/C}
}

@article{Goldstein_2018,
   title={Symmetry-Resolved Entanglement in Many-Body Systems},
   volume={120},
   ISSN={1079-7114},
   url={http://dx.doi.org/10.1103/PhysRevLett.120.200602},
   DOI={10.1103/physrevlett.120.200602},
   number={20},
   journal={Physical Review Letters},
   publisher={American Physical Society (APS)},
   author={Goldstein, Moshe and Sela, Eran},
   year={2018},
   month=may }

@article{Bonsignori_2019,
doi = {10.1088/1751-8121/ab4b77},
url = {https://doi.org/10.1088/1751-8121/ab4b77},
year = {2019},
month = {oct},
publisher = {IOP Publishing},
volume = {52},
number = {47},
pages = {475302},
author = {Bonsignori, Riccarda and Ruggiero, Paola and Calabrese, Pasquale},
title = {Symmetry resolved entanglement in free fermionic systems},
journal = {Journal of Physics A: Mathematical and Theoretical}
}

@article{Bonsignori_Calabrese_2020, title={Boundary effects on symmetry resolved entanglement}, volume={54}, ISSN={1751-8121}, DOI={10.1088/1751-8121/abcc3a}, number={1}, journal={Journal of Physics A: Mathematical and Theoretical}, publisher={IOP Publishing}, author={Bonsignori, Riccarda and Calabrese, Pasquale}, year={2020}, month=dec, pages={015005} }

@article{Campbell_Browne_2010, title={Bound States for Magic State Distillation in Fault-Tolerant Quantum Computation}, volume={104}, DOI={10.1103/PhysRevLett.104.030503}, number={3}, journal={Physical Review Letters}, publisher={American Physical Society}, author={Campbell, Earl T. and Browne, Dan E.}, year={2010}, month=jan, pages={030503} }

@article{Eastin_Knill_2009, title={Restrictions on Transversal Encoded Quantum Gate Sets}, volume={102}, DOI={10.1103/PhysRevLett.102.110502}, number={11}, journal={Physical Review Letters}, publisher={American Physical Society}, author={Eastin, Bryan and Knill, Emanuel}, year={2009}, month=mar, pages={110502} }

@preprint{Gottesman_1998, title={The {H}eisenberg Representation of Quantum Computers}, url={http://arxiv.org/abs/quant-ph/9807006}, DOI={10.48550/arXiv.quant-ph/9807006}, note={arXiv:quant-ph/9807006}, number={arXiv:quant-ph/9807006}, publisher={arXiv}, author={Gottesman, Daniel}, year={1998}, month=july }

@article{aaronson_improved_2004,
    title = {Improved simulation of stabilizer circuits},
    volume = {70},
    url = {https://link.aps.org/doi/10.1103/PhysRevA.70.052328},
    doi = {10.1103/PhysRevA.70.052328},
    number = {5},
    urldate = {2024-10-01},
    journal = {Physical Review A},
    publisher = {American Physical Society},
    author = {Aaronson, Scott and Gottesman, Daniel},
    month = nov,
    year = {2004},
    pages = {052328},
}

@article{Oliviero_Leone_Hamma_Lloyd_2022, title={Measuring magic on a quantum processor}, volume={8}, rights={2022 The Author(s)}, ISSN={2056-6387}, DOI={10.1038/s41534-022-00666-5}, number={1}, journal={npj Quantum Information}, publisher={Nature Publishing Group}, author={Oliviero, Salvatore F. E. and Leone, Lorenzo and Hamma, Alioscia and Lloyd, Seth}, year={2022}, month=dec, pages={1–8} }

@preprint{Ahmad_Esposito_Stasino_Odavic_Cosenza_Sarno_Mastrovito_Viscardi_Cusumano_Tafuri, title={Experimental demonstration of non-local magic in a superconducting quantum processor}, url={http://arxiv.org/abs/2511.15576}, DOI={10.48550/arXiv.2511.15576}, note={arXiv:2511.15576 [quant-ph]}, number={arXiv:2511.15576}, publisher={arXiv}, author={Ahmad, Halima Giovanna and Esposito, Gianluca and Stasino, Viviana and Odavic, Jovan and Cosenza, Carlo and Sarno, Alessandro and Mastrovito, Pasquale and Viscardi, Michele and Cusumano, Stefano and Tafuri, Francesco and Massarotti, Davide and Hamma, Alioscia}, year={2025}, month=nov }

@article{Haug_Piroli_2023, title={Quantifying nonstabilizerness of matrix product states}, volume={107}, DOI={10.1103/PhysRevB.107.035148}, number={3}, journal={Physical Review B}, publisher={American Physical Society}, author={Haug, Tobias and Piroli, Lorenzo}, year={2023}, month=jan, pages={035148} }

@article{Haug_Piroli_2023QUANTUM, title={Stabilizer entropies and nonstabilizerness monotones}, volume={7}, DOI={10.22331/q-2023-08-28-1092},  journal={Quantum}, publisher={Verein zur Förderung des Open Access Publizierens in den Quantenwissenschaften}, author={Haug, Tobias and Piroli, Lorenzo}, year={2023}, month=aug, pages={1092} }

@article{Tarabunga_Tirrito_Banuls_Dalmonte_2024, title={Nonstabilizerness via Matrix Product States in the Pauli Basis}, volume={133}, DOI={10.1103/PhysRevLett.133.010601}, number={1}, journal={Physical Review Letters}, publisher={American Physical Society}, author={Tarabunga, Poetri Sonya and Tirrito, Emanuele and Bañuls, Mari Carmen and Dalmonte, Marcello}, year={2024}, month=july, pages={010601} }

@preprint{Huang_Li_Zhong_2026, title={A fast and exact approach for stabilizer Rényi entropy via the XOR-FWHT algorithm}, url={http://arxiv.org/abs/2512.24685}, DOI={10.48550/arXiv.2512.24685}, note={arXiv:2512.24685 [quant-ph]}, number={arXiv:2512.24685}, publisher={arXiv}, author={Huang, Xuyang and Li, Han-Ze and Zhong, Jian-Xin}, year={2026}, month=jan }

@preprint{Xiao_Ryu_2026, title={Exponentially Accelerated Sampling of Pauli Strings for Nonstabilizerness}, url={http://arxiv.org/abs/2601.00761}, DOI={10.48550/arXiv.2601.00761}, note={arXiv:2601.00761 [quant-ph]}, number={arXiv:2601.00761}, publisher={arXiv}, author={Xiao, Zhenyu and Ryu, Shinsei}, year={2026}, month=jan }

@preprint{Sierant_Valles-Muns_Garcia-Saez_2026, title={Computing quantum magic of state vectors}, url={http://arxiv.org/abs/2601.07824}, DOI={10.48550/arXiv.2601.07824},  number={arXiv:2601.07824}, publisher={arXiv}, author={Sierant, Piotr and Vallès-Muns, Jofre and Garcia-Saez, Artur}, year={2026}, month=jan }

@article{cepollaro2024stabilizer,
  title = {Stabilizer entropy of quantum tetrahedra},
  author = {Cepollaro, Simone and Chirco, Goffredo and Cuffaro, Gianluca and Esposito, Gianluca and Hamma, Alioscia},
  journal = {Phys. Rev. D},
  volume = {109},
  issue = {12},
  pages = {126008},
  numpages = {13},
  year = {2024},
  month = {Jun},
  publisher = {American Physical Society},
  doi = {10.1103/PhysRevD.109.126008},
  url = {https://link.aps.org/doi/10.1103/PhysRevD.109.126008}
}

@preprint{cepollaro2025stabilizersub,
    title={Stabilizer Entropy of Subspaces}, 
      author={Simone Cepollaro and Gianluca Cuffaro and Matthew B. Weiss and Stefano Cusumano and Alioscia Hamma and Seth Lloyd},
      year={2025},
      eprint={2512.23013},
      archivePrefix={arXiv},
      primaryClass={quant-ph},
      url={https://arxiv.org/abs/2512.23013}, 
}

@article{doi:10.1142/S0219887825500033,
author = {Esposito, Gianluca and Cepollaro, Simone and Cappiello, Luigi and Hamma, Alioscia},
title = {Magic of discrete Lattice Gauge theories},
journal = {International Journal of Geometric Methods in Modern Physics},
volume = {22},
number = {06},
pages = {2550003},
year = {2025},
doi = {10.1142/S0219887825500033},

URL = { https://doi.org/10.1142/S0219887825500033
},
eprint = { https://doi.org/10.1142/S0219887825500033
}}

@article{zhu2016CliffordGroupFails,
  title={The Clifford group fails gracefully to be a unitary 4-design},
  author={Zhu, Huangjun and Kueng, Richard and Grassl, Markus and Gross, David},
  journal={arXiv:1609.08172},
  year={2016},
url={
https://doi.org/10.48550/arXiv.1609.08172
}
}

@article{jplh-zl35,
  title={Asymptotically independent fluctuations of stabilizer Rényi entropy and entanglement in random unitary circuits},
   volume={7},
   ISSN={2643-1564},
   url={http://dx.doi.org/10.1103/jplh-zl35},
   DOI={10.1103/jplh-zl35},
   number={4},
   journal={Physical Review Research},
   publisher={American Physical Society (APS)},
   author={Szombathy, Dominik and Valli, Angelo and Moca, Cătălin Paşcu and Farkas, Lóránt and Zaránd, Gergely},
   year={2025},
   month=oct }

@article{Liu_Chen_Balents_2018, title={Quantum entanglement of the Sachdev-Ye-Kitaev models}, volume={97}, DOI={10.1103/PhysRevB.97.245126},  number={24}, journal={Physical Review B}, publisher={American Physical Society}, author={Liu, Chunxiao and Chen, Xiao and Balents, Leon}, year={2018}, month=june, pages={245126} }

@article{sachdev1993gapless,
  title={Gapless spin-fluid ground state in a random quantum {H}eisenberg magnet},
  author={Sachdev, Subir and Ye, Jinwu},
  journal={Physical review letters},
  volume={70},
  number={21},
  pages={3339},
  year={1993},
  publisher={APS},
url={https://doi.org/10.1103/PhysRevLett.70.3339}
}

@inproceedings{KITP2015,
  author = {KITP},
  title = {Proceedings of the KITP},
  year = {2015},
  note = {\url{http://online.kitp.ucsb.edu/online/entangled15/kitaev/}, \url{http://online.kitp.ucsb.edu/online/entangled15/kitaev2/}}
}

@article{chowdhury2022sachdev,
  title={Sachdev-Ye-Kitaev models and beyond: Window into non-Fermi liquids},
  author={Chowdhury, Debanjan and Georges, Antoine and Parcollet, Olivier and Sachdev, Subir},
  journal={Reviews of Modern Physics},
  volume={94},
  number={3},
  pages={035004},
  year={2022},
  publisher={APS},
url={https://doi.org/10.1103/RevModPhys.94.035004}
}

@article{Hayden_2004,
   title={Randomizing Quantum States: Constructions and Applications},
   volume={250},
   ISSN={1432-0916},
   url={http://dx.doi.org/10.1007/s00220-004-1087-6},
   DOI={10.1007/s00220-004-1087-6},
   number={2},
   journal={Communications in Mathematical Physics},
   publisher={Springer Science and Business Media LLC},
   author={Hayden, Patrick and Leung, Debbie and Shor, Peter W. and Winter, Andreas},
   year={2004},
   month=jul, pages={371–391} }

@article{Pirandola_2020,
   title={Advances in quantum cryptography},
   volume={12},
   ISSN={1943-8206},
   url={http://dx.doi.org/10.1364/AOP.361502},
   DOI={10.1364/aop.361502},
   number={4},
   journal={Advances in Optics and Photonics},
   publisher={Optica Publishing Group},
   author={Pirandola, S. and others},
   year={2020},
   month=dec, pages={1012} }

@article{Neill_2018,
   title={A blueprint for demonstrating quantum supremacy with superconducting qubits},
   volume={360},
   ISSN={1095-9203},
   url={http://dx.doi.org/10.1126/science.aao4309},
   DOI={10.1126/science.aao4309},
   number={6385},
   journal={Science},
   publisher={American Association for the Advancement of Science (AAAS)},
   author={Neill, C. and others},
   year={2018},
   month=apr, pages={195–199} }

@article{Arute_2019,
    author = "Arute, F. and others",
    title = "{Quantum supremacy using a programmable superconducting processor}",
    eprint = "1910.11333",
    archivePrefix = "arXiv",
    primaryClass = "quant-ph",
    doi = "10.1038/s41586-019-1666-5",
    journal = "Nature",
    volume = "574",
    number = "7779",
    pages = "505--510",
    year = "2019"
}

@misc{yang2026,
      title={Probing Entanglement and Symmetries in Random States Using a Superconducting Quantum Processor}, 
      author={Jia-Nan Yang and Lata Kh Joshi and Filiberto Ares and Yihang Han and Pengfei Zhang and Pasquale Calabrese},
      year={2026},
      eprint={2601.22224},
      archivePrefix={arXiv},
      primaryClass={quant-ph},
      url={https://arxiv.org/abs/2601.22224}, 
}

@article{Rakovszky_2018,
   title={Diffusive Hydrodynamics of Out-of-Time-Ordered Correlators with Charge Conservation},
   volume={8},
   ISSN={2160-3308},
   url={http://dx.doi.org/10.1103/PhysRevX.8.031058},
   DOI={10.1103/physrevx.8.031058},
   number={3},
   journal={Physical Review X},
   publisher={American Physical Society (APS)},
   author={Rakovszky, Tibor and Pollmann, Frank and von Keyserlingk, C. W.},
   year={2018},
   month=sep }

@article{Khemani_2018,
   title={Operator Spreading and the Emergence of Dissipative Hydrodynamics under Unitary Evolution with Conservation Laws},
   volume={8},
   ISSN={2160-3308},
   url={http://dx.doi.org/10.1103/PhysRevX.8.031057},
   DOI={10.1103/physrevx.8.031057},
   number={3},
   journal={Physical Review X},
   publisher={American Physical Society (APS)},
   author={Khemani, Vedika and Vishwanath, Ashvin and Huse, David A.},
   year={2018},
   month=sep }

@article{Friedman_2019,
   title={Spectral Statistics and Many-Body Quantum Chaos with Conserved Charge},
   volume={123},
   ISSN={1079-7114},
   url={http://dx.doi.org/10.1103/PhysRevLett.123.210603},
   DOI={10.1103/physrevlett.123.210603},
   number={21},
   journal={Physical Review Letters},
   publisher={American Physical Society (APS)},
   author={Friedman, Aaron J. and Chan, Amos and De Luca, Andrea and Chalker, J. T.},
   year={2019},
   month=nov }

@article{Jonay_2024,
   title={Slow thermalization and subdiffusion in $U(1)$ conserving Floquet random circuits},
   volume={109},
   ISSN={2469-9969},
   url={http://dx.doi.org/10.1103/PhysRevB.109.024311},
   DOI={10.1103/physrevb.109.024311},
   number={2},
   journal={Physical Review B},
   publisher={American Physical Society (APS)},
   author={Jonay, Cheryne and Rodriguez-Nieva, Joaquin F. and Khemani, Vedika},
   year={2024},
   month=jan }

@article{sonya2025nonstabilizerness,
 title={A nonstabilizerness monotone from stabilizerness asymmetry},
   volume={10},
   ISSN={2058-9565},
   url={http://dx.doi.org/10.1088/2058-9565/adfd0d},
   DOI={10.1088/2058-9565/adfd0d},
   number={4},
   journal={Quantum Science and Technology},
   publisher={IOP Publishing},
   author={Sonya Tarabunga, Poetri and Frau, Martina and Haug, Tobias and Tirrito, Emanuele and Piroli, Lorenzo},
   year={2025},
   month=sep, pages={045026} }

@preprint{leone2025noncliffordcostrandomunitaries,
      title={The non-Clifford cost of random unitaries}, 
      author={Lorenzo Leone and Salvatore F. E. Oliviero and Alioscia Hamma and Jens Eisert and Lennart Bittel},
      year={2025},
      eprint={2505.10110},
      archivePrefix={arXiv},
      primaryClass={quant-ph},
      url={https://arxiv.org/abs/2505.10110}, 
}

@article{Leone_2024,
   title={Learning t-doped stabilizer states},
   volume={8},
   ISSN={2521-327X},
   url={http://dx.doi.org/10.22331/q-2024-05-27-1361},
   DOI={10.22331/q-2024-05-27-1361},
   journal={Quantum},
   publisher={Verein zur Forderung des Open Access Publizierens in den Quantenwissenschaften},
   author={Leone, Lorenzo and Oliviero, Salvatore F. E. and Hamma, Alioscia},
   year={2024},
   month=may, pages={1361} }

@article{Turkeshi_2025,
   title={Pauli spectrum and nonstabilizerness of typical quantum many-body states},
   volume={111},
   ISSN={2469-9969},
   url={http://dx.doi.org/10.1103/PhysRevB.111.054301},
   DOI={10.1103/physrevb.111.054301},
   number={5},
   journal={Physical Review B},
   publisher={American Physical Society (APS)},
   author={Turkeshi, Xhek and Dymarsky, Anatoly and Sierant, Piotr},
   year={2025},
   month=feb }

@article{liu2022many,
  title={Many-Body Quantum Magic},
   volume={3},
   ISSN={2691-3399},
   url={http://dx.doi.org/10.1103/PRXQuantum.3.020333},
   DOI={10.1103/prxquantum.3.020333},
   number={2},
   journal={PRX Quantum},
   publisher={American Physical Society (APS)},
   author={Liu, Zi-Wen and Winter, Andreas},
   year={2022},
   month=may }

@article{PhysRevB.111.L081102,
  title = {Nonstabilizerness in U(1) lattice gauge theory},
  author = {Falc\~ao, Pedro R. Nic\'acio and Tarabunga, Poetri Sonya and Frau, Martina and Tirrito, Emanuele and Zakrzewski, Jakub and Dalmonte, Marcello},
  journal = {Phys. Rev. B},
  volume = {111},
  issue = {8},
  pages = {L081102},
  numpages = {7},
  year = {2025},
  month = {Feb},
  publisher = {American Physical Society},
  doi = {10.1103/PhysRevB.111.L081102},
  url = {https://link.aps.org/doi/10.1103/PhysRevB.111.L081102}
}

@article{p7xt-s9nz,
  title = {Magic Resources of the {H}eisenberg Picture},
  author = {Dowling, Neil and Kos, Pavel and Turkeshi, Xhek},
  journal = {Phys. Rev. Lett.},
  volume = {135},
  issue = {5},
  pages = {050401},
  numpages = {10},
  year = {2025},
  month = {Jul},
  publisher = {American Physical Society},
  doi = {10.1103/p7xt-s9nz},
  url = {https://link.aps.org/doi/10.1103/p7xt-s9nz}
}

@misc{SI,
  title = {Supplemental Material},
  note  = {See attached supplemental document}
}

@book{Vershynin_2018, place={Cambridge}, series={Cambridge Series in Statistical and Probabilistic Mathematics}, title={High-Dimensional Probability: An Introduction with Applications in Data Science}, publisher={Cambridge University Press}, author={Vershynin, Roman}, year={2018}, collection={Cambridge Series in Statistical and Probabilistic Mathematics}, url={https://www.cambridge.org/core/books/highdimensional-probability/797C466DA29743D2C8213493BD2D2102}}

@Article{10.21468/SciPostPhys.17.5.127,
	title={{Non-Abelian symmetry-resolved entanglement entropy}},
	author={Eugenio Bianchi and Pietro Dona and Rishabh Kumar},
	journal={SciPost Phys.},
	volume={17},
	pages={127},
	year={2024},
	publisher={SciPost},
	doi={10.21468/SciPostPhys.17.5.127},
	url={https://scipost.org/10.21468/SciPostPhys.17.5.127},
}

@article{PhysRevB.88.125117,
  title = {Local characterization of one-dimensional topologically ordered states},
  author = {Cui, Jian and Amico, Luigi and Fan, Heng and Gu, Mile and Hamma, Alioscia and Vedral, Vlatko},
  journal = {Phys. Rev. B},
  volume = {88},
  issue = {12},
  pages = {125117},
  numpages = {7},
  year = {2013},
  month = {Sep},
  publisher = {American Physical Society},
  doi = {10.1103/PhysRevB.88.125117},
  url = {https://link.aps.org/doi/10.1103/PhysRevB.88.125117}
}

@article{Xavier_2018,
   title={Equipartition of the entanglement entropy},
   volume={98},
   ISSN={2469-9969},
   url={http://dx.doi.org/10.1103/PhysRevB.98.041106},
   DOI={10.1103/physrevb.98.041106},
   number={4},
   journal={Physical Review B},
   publisher={American Physical Society (APS)},
   author={Xavier, J. C. and Alcaraz, F. C. and Sierra, G.},
   year={2018},
   month=jul }

@misc{krawtchouk1929generalisation,
  author    = {Weisstein, Eric W.},
  title     = {Krawtchouk Polynomial},
  howpublished = {MathWorld--A Wolfram Web Resource},
  url       = {https://mathworld.wolfram.com/KrawtchoukPolynomial.html},
  note      = {Accessed: 2026-03-30}
}

@book{ledoux2001concentration,
  title={The concentration of measure phenomenon},
  author={Ledoux, Michel},
  number={89},
  year={2001},
  publisher={American Mathematical Soc.}
}

@article{PhysRevLett.113.140401,
  title = {Quantifying Coherence},
  author = {Baumgratz, T. and Cramer, M. and Plenio, M. B.},
  journal = {Phys. Rev. Lett.},
  volume = {113},
  issue = {14},
  pages = {140401},
  numpages = {5},
  year = {2014},
  month = {Sep},
  publisher = {American Physical Society},
  doi = {10.1103/PhysRevLett.113.140401},
  url = {https://link.aps.org/doi/10.1103/PhysRevLett.113.140401}
}

@article{luitz2014participation,
  title={Participation spectroscopy and entanglement Hamiltonian of quantum spin models},
  author={Luitz, David J and Laflorencie, Nicolas and Alet, Fabien},
  journal={Journal of Statistical Mechanics: Theory and Experiment},
  volume={2014},
  number={8},
  pages={P08007},
  year={2014},
  publisher={IOP Publishing and SISSA},
  url={
https://doi.org/10.1088/1742-5468/2014/08/P08007
}
}

@article{PhysRevLett.123.180601,
  title = {Multifractal Scalings Across the Many-Body Localization Transition},
  author = {Mac\'e, Nicolas and Alet, Fabien and Laflorencie, Nicolas},
  journal = {Phys. Rev. Lett.},
  volume = {123},
  issue = {18},
  pages = {180601},
  numpages = {6},
  year = {2019},
  month = {Oct},
  publisher = {American Physical Society},
  doi = {10.1103/PhysRevLett.123.180601},
  url = {https://link.aps.org/doi/10.1103/PhysRevLett.123.180601}
}

@article{PhysRevA.111.052614,
  title = {Quantum algorithms for inverse participation ratio estimation in multiqubit and multiqudit systems},
  author = {Liu, Yingjian and Sierant, Piotr and Stornati, Paolo and Lewenstein, Maciej and P\l{}odzie\ifmmode \acute{n}\else \'{n}\fi{}, Marcin},
  journal = {Phys. Rev. A},
  volume = {111},
  issue = {5},
  pages = {052614},
  numpages = {13},
  year = {2025},
  month = {May},
  publisher = {American Physical Society},
  doi = {10.1103/PhysRevA.111.052614},
  url = {https://link.aps.org/doi/10.1103/PhysRevA.111.052614}
}

@article{Sauliere:2025bpr,
    author = "Sauliere, Arman and Magni, Beatrice and Lami, Guglielmo and Turkeshi, Xhek and De Nardis, Jacopo",
    title = "{Universality in the anticoncentration of chaotic quantum circuits}",
    eprint = "2503.00119",
    archivePrefix = "arXiv",
    primaryClass = "quant-ph",
    doi = "10.1103/lkwg-4dbt",
    journal = "Phys. Rev. B",
    volume = "112",
    number = "13",
    pages = "134312",
    year = "2025"
}

@article{Iannotti_Esposito_Venuti_Hamma_2025, title={Entanglement and Stabilizer entropies of random bipartite pure quantum states}, volume={9}, DOI={10.22331/q-2025-07-21-1797}, abstractNote={Daniele Iannotti, Gianluca Esposito, Lorenzo Campos Venuti, and Alioscia Hamma,Quantum 9, 1797 (2025).
The interplay between non-stabilizerness and entanglement in random states is a very rich arena of study for the understanding of quantum advantage and complexity. In this work, we tackle th…}, journal={Quantum}, publisher={Verein zur Förderung des Open Access Publizierens in den Quantenwissenschaften}, author={Iannotti, Daniele and Esposito, Gianluca and Venuti, Lorenzo Campos and Hamma, Alioscia}, year={2025}, month=july, pages={1797}}

@preprint{cao2025,
  author        = {Sizheng Cao and Xian-Hui Ge},
  title         = {Symmetry restoration in a fast scrambling system},
  year          = {2025},
  eprint        = {2509.26176},
  archivePrefix = {arXiv},
  url           = {https://arxiv.org/abs/2509.26176}
}

@article{singh2016average,
  title={Average coherence and its typicality for random pure states},
  author={Singh, Uttam and Zhang, Lin and Pati, Arun Kumar},
  journal={Physical Review A},
  volume={93},
  number={3},
  pages={032125},
  year={2016},
  publisher={APS}
}

@preprint{santra2025quantumresourcesnonabelianlattice,
      title={Quantum Resources in Non-Abelian Lattice Gauge Theories: Nonstabilizerness, Multipartite Entanglement, and Fermionic Non-Gaussianity}, 
      author={Gopal Chandra Santra and Julius Mildenberger and Edoardo Ballini and Alberto Bottarelli and Matteo M. Wauters and Philipp Hauke},
      year={2025},
      eprint={2510.07385},
      archivePrefix={arXiv},
      primaryClass={quant-ph},
      url={https://arxiv.org/abs/2510.07385}, 
}

@preprint{liu2024unitarydesignsrandomsymmetric,
      title={Unitary Designs from Random Symmetric Quantum Circuits}, 
      author={Hanqing Liu and Austin Hulse and Iman Marvian},
      year={2024},
      eprint={2408.14463},
      archivePrefix={arXiv},
      primaryClass={quant-ph},
      url={https://arxiv.org/abs/2408.14463}, 
}

@preprint{hoshino2025stabilizerrenyientropyconformal,
      title={Stabilizer R\'enyi Entropy and Conformal Field Theory}, 
      author={Masahiro Hoshino and Masaki Oshikawa and Yuto Ashida},
      year={2025},
      eprint={2503.13599},
      archivePrefix={arXiv},
      primaryClass={quant-ph},
      url={https://arxiv.org/abs/2503.13599}, 
}

@misc{hoshino2025stabilizerrenyientropyencodes,
      title={Stabilizer R\'{e}nyi Entropy Encodes Fusion Rules of Topological Defects and Boundaries}, 
      author={Masahiro Hoshino and Yuto Ashida},
      year={2025},
      eprint={2507.10656},
      archivePrefix={arXiv},
      primaryClass={quant-ph},
      url={https://arxiv.org/abs/2507.10656}, 
}

@misc{Franchini_2014, title={On the Spontaneous Breaking of {U(N)} symmetry in invariant Matrix Models}, url={https://arxiv.org/abs/1412.6523v2}, abstractNote={Matrix Models are the most effective way to describe strongly interacting systems with many degrees of freedom. They have proven successful in describing very different settings, from nuclei spectra to conduction in mesoscopic systems, from holographic models to various aspects of mathematical physics. This success reflects the existence of a large universality class for all these systems, signaled by the Wigner-Dyson statistics for the matrix eigenvalues. These models are defined in a base invariant way and this rotational symmetry has traditionally been read to imply that they describe extended system. In this work we show that certain matrix models, which show deviations from the Wigner-Dyson distribution, can spontaneously break their rotational ($U(N)$) invariance and localize their eigenvectors on a portion of the Hilbert space. This conclusion establishes once more a direct connection between the eigenvalue and eigenvector distributions. Recognizing this loss of ergodicity discloses the power of non-perturbative techniques available for matrix models to the study of localization problems and introduces a novel spontaneous symmetry breaking mechanism. Moreover, it brings forth the overlooked role of eigenvectors in the study of matrix models and allows for the consideration of new types of observable.}, journal={arXiv.org}, author={Franchini, Fabio}, year={2014}, month=dec }

@article{bera2025non,
  title={Non-stabilizerness of Sachdev-Ye-Kitaev model},
  author={Bera, Surajit and Schir{\`o}, Marco},
  journal={SciPost Physics},
  volume={19},
  number={6},
  pages={159},
  year={2025},
  doi={10.21468/SciPostPhys.19.6.159},
  url={https://scipost.org/SciPostPhys.19.6.159}
}

@article{Calabrese_2021,
   title={Symmetry-resolved entanglement entropy in Wess-Zumino-Witten models},
   volume={2021},
   ISSN={1029-8479},
   url={http://dx.doi.org/10.1007/JHEP10(2021)067},
   DOI={10.1007/jhep10(2021)067},
   number={10},
   journal={Journal of High Energy Physics},
   publisher={Springer Science and Business Media LLC},
   author={Calabrese, Pasquale and Dubail, Jérôme and Murciano, Sara},
   year={2021},
   month=oct }

@article{Sekino_Susskind_2008, title={Fast scramblers}, volume={2008}, ISSN={1126-6708}, DOI={10.1088/1126-6708/2008/10/065}, abstractNote={We consider the problem of how fast a quantum system can scramble (thermalize) information, given that the interactions are between bounded clusters of degrees of freedom; pairwise interactions would be an example. Based on previous work, we conjecture: The most rapid scramblers take a time logarithmic in the number of degrees of freedom. Matrix quantum mechanics (systems whose degrees of freedom are n by n matrices) saturate the bound. Black holes are the fastest scramblers in nature. The conjectures are based on two sources, one from quantum information theory, and the other from the study of black holes in String Theory.}, number={10}, journal={Journal of High Energy Physics}, author={Sekino, Yasuhiro and Susskind, L.}, year={2008}, month=oct, pages={065} }

@article{Cotler, title={Black holes and random matrices}, volume={2017}, ISSN={1029-8479}, DOI={10.1007/JHEP05(2017)118}, abstractNote={We argue that the late time behavior of horizon fluctuations in large anti-de Sitter (AdS) black holes is governed by the random matrix dynamics characteristic of quantum chaotic systems. Our main tool is the Sachdev-Ye-Kitaev (SYK) model, which we use as a simple model of a black hole. We use an analytically continued partition function |Z(β + it)|2 as well as correlation functions as diagnostics. Using numerical techniques we establish random matrix behavior at late times. We determine the early time behavior exactly in a double scaling limit, giving us a plausible estimate for the crossover time to random matrix behavior. We use these ideas to formulate a conjecture about general large AdS black holes, like those dual to 4D super-Yang-Mills theory, giving a provisional estimate of the crossover time. We make some preliminary comments about challenges to understanding the late time dynamics from a bulk point of view.}, number={5}, journal={Journal of High Energy Physics}, author={Cotler, Jordan S. and Gur-Ari, Guy and Hanada, Masanori and Polchinski, Joseph and Saad, Phil and Shenker, Stephen H. and Stanford, Douglas and Streicher, Alexandre and Tezuka, Masaki}, year={2017}, month=may, pages={118} }

@preprint{Sun_Zhang_2026, title={Connecting Magic Dynamics in Thermofield Double States to Spectral Form Factors}, url={http://arxiv.org/abs/2601.12787}, DOI={10.48550/arXiv.2601.12787}, abstractNote={Under unitary evolution, chaotic quantum systems initialized in simple states rapidly develop high complexity, precluding any efficient classical description. Quantum chaos is traditionally characterized by spectral properties of the Hamiltonian, most notably through the spectral form factor, while the hardness of classical simulation within the stabilizer formalism, commonly referred to as quantum magic, can be quantified by the stabilizer Rényi entropy. In this Letter, we propose a relation between the dynamics of the stabilizer Rényi entropy for thermofield double states and the spectral form factor, based on general arguments for chaotic systems with all-to-all interactions. This relation implies that the saturation of the stabilizer Rényi entropy is governed by a first-order dynamical transition. We then demonstrate this relation explicitly in the Sachdev-Ye-Kitaev model, using an auxiliary-spin representation of the stabilizer Rényi entropy that exhibits an emergent $Z_2$ symmetry. We further find that, in the high-temperature regime of the SYK model, the transition occurs at a finite time, with the long-time phase marked by spontaneous $Z_2$ symmetry breaking. In contrast, at low temperatures, the transition is pushed to times exponentially long in the system size. Our results reveal an intriguing interplay between quantum chaos and quantum magic.}, note={arXiv:2601.12787 [quant-ph]}, number={arXiv:2601.12787}, publisher={arXiv}, author={Sun, Ning and Zhang, Pengfei}, year={2026}, month=jan }

@article{Zhang_Zhou_Sun_2026, title={Stabilizer Renyi Entropy and Its Transition in the Coupled Sachdev-Ye-Kitaev Model}, volume={136}, DOI={10.1103/5c15-4g5n}, abstractNote={Quantum entanglement and quantum magic are two distinct fundamental resources that enable quantum systems to exhibit complex phenomena beyond the capabilities of classical computer simulations. While quantum entanglement has been extensively used to characterize both equilibrium and dynamical phases, the study of quantum magic, typically quantified by the stabilizer Rényi entropy (SRE), remains largely limited to numerical simulations of moderate system sizes. In this Letter, we establish a general framework for analyzing the SRE in solvable Sachdev-Ye-Kitaev models in the large-�� limit, which enables the application of the saddle-point approximation. Applying this method to the Maldacena-Qi coupled Sachdev-Ye-Kitaev model, we identify a series of first-order transitions of the SRE as the temperature is tuned. In particular, we uncover an intrinsic transition of the SRE that cannot be detected through thermodynamic quantities. We also discuss the theoretical understanding of the SRE in both the high-temperature and low-temperature limits. Our results pave the way for studying the SRE in strongly correlated fermionic systems in the thermodynamic limit and suggest a new class of transitions for which the SRE serves as an order parameter.}, number={8}, journal={Physical Review Letters}, publisher={American Physical Society}, author={Zhang, Pengfei and Zhou, Shuyan and Sun, Ning}, year={2026}, month=feb, pages={080201} }

@article{Altland_Bagrets_Kamenev_2019, title={Quantum Criticality of Granular Sachdev-Ye-Kitaev Matter}, volume={123}, DOI={10.1103/PhysRevLett.123.106601}, abstractNote={We consider granular quantum matter defined by Sachdev-Ye-Kitaev dots coupled via random one-body hopping. Within the framework of Schwarzian field theory, we identify a zero-temperature quantum phase transition between an insulating phase at weak and a metallic phase at strong hopping. The critical hopping strength scales inversely with the number of degrees of freedom on the dots. The increase of temperature out of either phase induces a crossover into a regime of strange metallic behavior.}, number={10}, journal={Physical Review Letters}, publisher={American Physical Society}, author={Altland, Alexander and Bagrets, Dmitry and Kamenev, Alex}, year={2019}, month=sept, pages={106601} }

@article{Piroli_2025, title={A nonstabilizerness monotone from stabilizerness asymmetry}, volume={10}, ISSN={2058-9565}, DOI={10.1088/2058-9565/adfd0d}, abstractNote={We introduce a nonstabilizerness monotone which we name basis-minimized stabilizerness asymmetry (BMSA). It is based on the notion of G-asymmetry, a measure of how much a certain state deviates from being symmetric with respect to a symmetry group G. For pure states, we show that the BMSA is a strong monotone for magic-state resource theory, while it can be extended to mixed states via the convex roof construction. We discuss its relation with other magic monotones, first showing that the BMSA coincides with the recently introduced basis-minimized measurement entropy, thereby establishing the strong monotonicity of the latter. Next, we provide inequalities between the BMSA and other nonstabilizerness measures such as the robustness of magic, stabilizer extent, stabilizer rank, stabilizer fidelity and stabilizer Rényi entropy. We also prove that the stabilizer fidelity, stabilizer Rényi entropy and BMSA with index have the same asymptotic scaling with qubit number. Finally, we present numerical methods to compute the BMSA, highlighting its advantages and drawbacks compared to other nonstabilizerness measures in the context of pure many-body quantum states. We also discuss the importance of additivity and strong monotonicity for measures of nonstabilizerness in many-body physics, motivating the search for additional computable nonstabilizerness monotones.}, number={4}, journal={Quantum Science and Technology}, publisher={IOP Publishing}, author={Sonya Tarabunga, Poetri and Frau, Martina and Haug, Tobias and Tirrito, Emanuele and Piroli, Lorenzo}, year={2025}, month=sept, pages={045026} }

@article{Haug_Aolita_Kim_2025, title={Probing quantum complexity via universal saturation of stabilizer entropies}, volume={9}, DOI={10.22331/q-2025-07-21-1801}, abstractNote={Tobias Haug, Leandro Aolita, and M.S. Kim,
Quantum 9, 1801 (2025).
Nonstabilizerness or `magic’ is a key resource for quantum computing and a necessary condition for quantum advantage. Non-Clifford operations turn stabilizer states into resourceful states,…}, journal={Quantum}, publisher={Verein zur Förderung des Open Access Publizierens in den Quantenwissenschaften}, author={Haug, Tobias and Aolita, Leandro and Kim, M. S.}, year={2025}, month=july, pages={1801} }

@article{Turkeshi_2025_spread,
   title={Magic spreading in random quantum circuits},
   volume={16},
   ISSN={2041-1723},
   url={http://dx.doi.org/10.1038/s41467-025-57704-x},
   DOI={10.1038/s41467-025-57704-x},
   number={1},
   journal={Nature Communications},
   publisher={Springer Science and Business Media LLC},
   author={Turkeshi, Xhek and Tirrito, Emanuele and Sierant, Piotr},
   year={2025},
   month=mar }

@article{1jzy-sk9r,
  title = {Anticoncentration and Nonstabilizerness Spreading under Ergodic Quantum Dynamics},
  author = {Tirrito, Emanuele and Turkeshi, Xhek and Sierant, Piotr},
  journal = {Phys. Rev. Lett.},
  volume = {135},
  issue = {22},
  pages = {220401},
  numpages = {9},
  year = {2025},
  month = {Nov},
  publisher = {American Physical Society},
  doi = {10.1103/1jzy-sk9r},
  url = {https://link.aps.org/doi/10.1103/1jzy-sk9r}
}

@article{PhysRevA.109.L040401,
  title = {Quantifying nonstabilizerness through entanglement spectrum flatness},
  author = {Tirrito, Emanuele and Tarabunga, Poetri Sonya and Lami, Gugliemo and Chanda, Titas and Leone, Lorenzo and Oliviero, Salvatore F. E. and Dalmonte, Marcello and Collura, Mario and Hamma, Alioscia},
  journal = {Phys. Rev. A},
  volume = {109},
  issue = {4},
  pages = {L040401},
  numpages = {6},
  year = {2024},
  month = {Apr},
  publisher = {American Physical Society},
  doi = {10.1103/PhysRevA.109.L040401},
  url = {https://link.aps.org/doi/10.1103/PhysRevA.109.L040401}
}

@article{Russomanno_2025,
   title={Nonstabilizerness in the unitary and monitored quantum dynamics of XXZ-staggered and Sachdev-Ye-Kitaev models},
   volume={112},
   ISSN={2469-9969},
   url={http://dx.doi.org/10.1103/njgn-fksh},
   DOI={10.1103/njgn-fksh},
   number={6},
   journal={Physical Review B},
   publisher={American Physical Society (APS)},
   author={Russomanno, Angelo and Passarelli, Gianluca and Rossini, Davide and Lucignano, Procolo},
   year={2025},
   month=aug }

@article{Majidy_2023,
   title={Non-Abelian symmetry can increase entanglement entropy},
   volume={107},
   ISSN={2469-9969},
   url={http://dx.doi.org/10.1103/PhysRevB.107.045102},
   DOI={10.1103/physrevb.107.045102},
   number={4},
   journal={Physical Review B},
   publisher={American Physical Society (APS)},
   author={Majidy, Shayan and Lasek, Aleksander and Huse, David A. and Yunger Halpern, Nicole},
   year={2023},
   month=jan }

@article{Majidy_2023_nat,
   title={Noncommuting conserved charges in quantum thermodynamics and beyond},
   volume={5},
   ISSN={2522-5820},
   url={http://dx.doi.org/10.1038/s42254-023-00641-9},
   DOI={10.1038/s42254-023-00641-9},
   number={11},
   journal={Nature Reviews Physics},
   publisher={Springer Science and Business Media LLC},
   author={Majidy, Shayan and Braasch, William F. and Lasek, Aleksander and Upadhyaya, Twesh and Kalev, Amir and Yunger Halpern, Nicole},
   year={2023},
   month=oct, pages={689–698} }

@misc{private,
  author = {Lucignano, Procolo and Passarelli, Gianluca and Rossini, Davide and Russomanno, Angelo},
  title  = {},
  year   = {2024},
 note   = {Private communication}
}

\clearpage

\end{document}


\setcounter{secnumdepth}{3}
\renewcommand{\theequation}{S.\arabic{equation}}

\title{Supplemental Material for Non-stabilizerness and U(1) symmetry in chaotic many-body quantum systems}

\author{Daniele Iannotti\orcidlink{0009-0009-0738-5998}$^{1,2}$}
\email{d.iannotti@ssmeridionale.it}
\author{Angelo Russotto\orcidlink{0009-0009-4303-9523}$^{3,4}$}
\email{arussott@sissa.it}
\author{Barbara Jasser\orcidlink{0009-0004-5657-2806}$^{1,2}$}
\email{b.jasser@ssmeridionale.it}
\author{Jovan Odavić\orcidlink{0000-0003-2729-8284}$^{2,5}$}
\email{jovan.odavic@unina.it}
\author{Alioscia Hamma\orcidlink{0000-0003-0662-719X}$^{1,2,5}$}
\email{alioscia.hamma@unina.it }

\affiliation{$^1$Scuola Superiore Meridionale, Largo S. Marcellino 10, 80138 Napoli, Italy}
\affiliation{$^2$INFN Sezione di Napoli, via Cintia, 80126 Napoli, Italy}
\affiliation{$^3$SISSA, via Bonomea 265, 34136 Trieste, Italy}
\affiliation{$^4$INFN Sezione di Trieste, via Bonomea 265, 34136 Trieste, Italy}
\affiliation{$^5$Dipartimento di Fisica `Ettore Pancini', Universit\`a degli Studi di Napoli Federico II, Via Cintia 80126, Napoli, Italy}

\maketitle


\section{Average Stabilizer purity of order 2 for Haar$\times U(1)$ ensemble}
\label{app:average_SP}
\begin{figure}
    \centering
    \includegraphics[width=0.99\columnwidth]{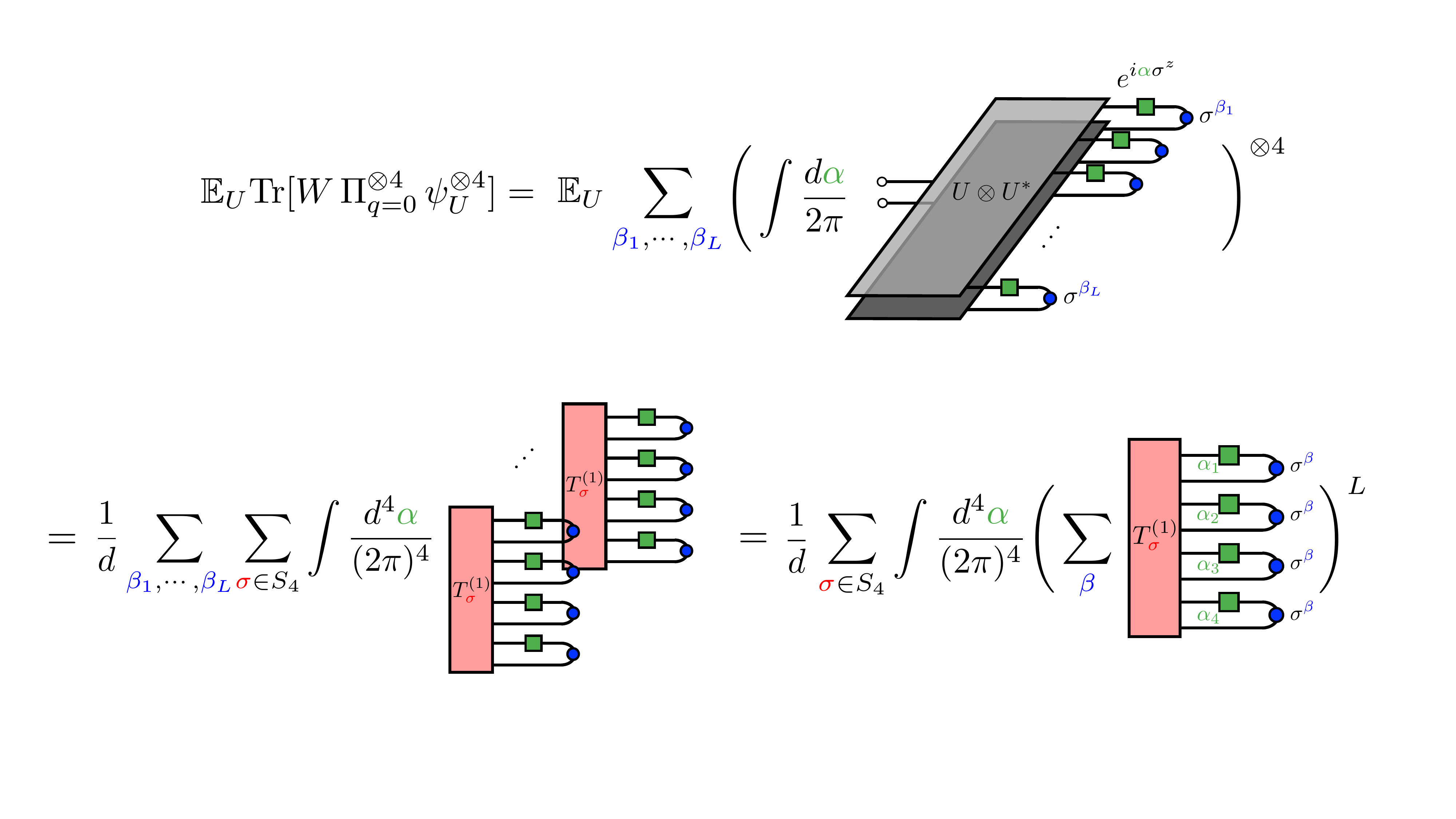}
    \caption{Graphical representation of $\exv_U \text{Tr}[W \,\Pi_{q}^{\otimes 4}\, \psi_U^{\otimes4}]$. We choose, for clarity, $q = 0$. A different value of the charge sector $q$ appears only as a phase inside the integral over $\alpha$. Left hand side and right hand side of the bottom part of the figure correspond to, respectively, Eq.~\eqref{eq:step1} and Eq.~\eqref{eq:step2}.
    }\label{fig:TN_representation}
\end{figure}
\noindent
In this section, we analytically derive the exact result of the average of Stabilizer purity (SP) of order 2 according to the  Haar$\times U(1)$ ensemble~\cite{Bianchi_Dona_2019} for any size $L$ and charge $q$. In particular, we obtain the expression
\begin{equation}
    \label{eq:finitesizeavg}
       \exv_{U_q}[\Xi_2(\psi_{U_q})]= \frac{12 \,d_{q}^2 + 8 \,d_{q} + 2^{2-L}\, h(L,q)}{4! \binom{d_q + 3}{4}},
     \end{equation}
with $h(L,q) = \sum_{k = 0}^L \binom{L}{k}\binom{L-k}{\frac{L-q}{2}}^4 \mathcal{F}(k,L,q)^4$,
where $\mathcal{F}(k,L,q) \equiv  {}_2F_1(-k,-\frac{L-q}{2},1-k+\frac{L+q}{2};-1)$ is the hypergeometric function.
This result extends to the other directional charges $\mathcal{Q}^{(x,y)}$ by Clifford invariance of $\Xi_{\alpha}(\ket{\psi})$ and left/right invariance of the Haar measure. Moreover, Eq.~\eqref{eq:finitesizeavg} is symmetric under $q \to -q$, as this follows directly from the Clifford invariance of SPs, e.g.\ $\Xi_{\alpha}(\bigotimes_j \sigma^x_j \ket{\psi}) = \Xi_{\alpha}(\ket{\psi})$. Along the same lines, an analogous result can be obtained for discrete $\mathbb{Z}_d$ symmetries~\cite{cepollaro2025stabilizersub,doi:10.1142/S0219887825500033} by employing the characteristic functions of the corresponding symmetry group. Together with the continuous U(1) case above, this completes the picture for typical Abelian symmetries.

The total Hilbert space of the system can be decomposed as $\mathcal{H} = \bigoplus_{q=-L}^{L} \mathcal{H}(q)$, where $\mathcal{H}(q)$ is the eigenspace with charge value $q$ and dimension $d_q=\binom{L}{\frac{L+q}{2}}$ and $\text{dim}(\mathcal{H})=d$. By definition, any state $\ket{\psi(q)} \in \mathcal{H}(q)$ is an eigenstate of $\mathcal{Q}$, which means that the corresponding density matrix $\ket{\psi(q)} \bra{\psi(q)} \equiv \psi(q)$ satisfy $[\mathcal{Q},\psi(q)] = 0$. The ensemble of $U(1)$-symmetric Haar random states, with fixed value of the charge $q$, consists of density matrices $\psi_{U_q}\equiv U_q\ket{q} \bra{q}U_q^{\dagger}$, where $U_q$ is a $d_q\times d_q$ Haar random unitary matrix acting on charge $q$ sector $\mathcal{H}(q)$, i.e. $U_q \in U(\mathcal{H}(q))$, and $\ket{q}$ is just a reference state with the only constraint of having a definite value of the charge $q$.
\noindent
Let us write the average in a useful manner
\begin{equation}
    \mathbb{E}_{U_q}[\Xi_2(\psi_{U_q})]=\mathbb{E}_{U_q}[\text{Tr}[W \,\psi(q)^{\otimes4}]],
\end{equation}
where $W = \sum_{P\in \mathcal{P}_L}P^{\otimes 4}/d$ and $\mathcal{P}_L$ is  the quotient group of the Pauli group containing
all +1 phases for $L$ qubits. 
We can rewrite the expectation value over random unitaries in a fixed eigenspace charge sector as an expectation value over Haar unitaries over the whole Hilbert space (see Eq.~(5) in the main text and Lemma from Ref.~\cite{cepollaro2024stabilizer})
\begin{equation}
    \mathbb{E}_{U_q}[\text{Tr}[W \,\psi(q)^{\otimes4}]] = c(d_0,d_q)  \mathbb{E}_{U}[\text{Tr}[W \,\Pi_q^{\otimes 4}\,\psi_{U}^{\otimes4}]],
\end{equation}
where $\psi_U$ is now an usual Haar random state and $\Pi_q$ a projector onto the eigenspace of $\mathcal{Q}$ with eigenvalue $q$ and the coefficient $c(d_0,d_q)$ is 
\begin{equation}
    c(d_0,d_q) = \binom{d_0 +3}{4}\binom{d_q +3}{4}^{-1}.
\end{equation}
The Haar average can now be computed
\begin{equation}
\label{eq:expect}
    \mathbb{E}_{U_q}[\text{Tr}[W \,\psi(q)^{\otimes4}]]  = \frac{1}{\binom{d_q+3}{4} 4!} \sum_{\sigma \in S_4} \text{Tr}[W \,\Pi_q^{\otimes 4}\, T_{\sigma}],
\end{equation}
using Weingarten calculus where $T_\sigma$ is the permutation matrix of 4 elements, i.e. $$T_\sigma= \sum_{i_1,i_2,i_3,i_4} \ket{i_{\sigma^{-1}(1)},i_{\sigma^{-1}(2)},i_{\sigma^{-1}(3)},i_{\sigma^{-1}(4)}}\bra{i_1,i_2,i_3,i_4}\,.$$
In order to proceed with the calculation, we use the integral representation of the projector~\cite{Goldstein_2018,Xavier_2018,Bonsignori_2019,Bonsignori_Calabrese_2020} $\Pi_q$ for $\mathcal{Q}$:
\begin{equation}
    \Pi_q = \int _{0}^{2\pi}\frac{d\alpha}{2 \pi} e^{- i \alpha q} \bigotimes_{i=1}^L e^{i \alpha \sigma_i^z}.
\end{equation}
This form is convenient since it expresses the projector as a factorized contribution over qubit.
In Fig.~\ref{fig:TN_representation} we show a graphical representation of the trace in Eq.~\eqref{eq:expect} once we use the factorized expression for the projector $\Pi_q$.
\noindent
As a first example of calculations one can do, we show how to get the dimension of the Hilbert space $d_q$\footnote{The dimension of the Hilbert space where the projector acts trivially is given by the trace since a projector respects $\Pi^2=\Pi$ and $\Pi^2 \psi= \lambda \psi$ if $\Pi \psi= \lambda \psi$. Namely $\lambda^2=\lambda$, hence $\lambda=0,1$. So the trace is counting the number of ones, that is the dimension of the subsystem.} as
\begin{equation}
\begin{split}
    d_q &= \text{Tr}[\Pi_q]= \int _{0}^{2\pi}\frac{d\alpha}{2 \pi} e^{- i \alpha q} \text{Tr} [\bigotimes_{j=1}^L e^{i \alpha \sigma_j^z} ]= \int _{0}^{2\pi}\frac{d\alpha}{2 \pi} e^{- i \alpha q} \text{Tr}^L \left[ e^{i \alpha \sigma_j^z} \right] \\
    &= \int _{0}^{2\pi}\frac{d\alpha}{2 \pi} e^{- i \alpha q} \text{Tr}^L \left[ \cos(\alpha) \mathbbm{1}+ i \sin(\alpha) \sigma_j^z \right]  = 2^L \int _{0}^{2\pi}\frac{d\alpha}{2 \pi} e^{- i \alpha q} \cos^L(\alpha)= \binom{L}{\frac{L-q}{2}} \quad \abs{q}\leq L \And L-q \text{ even}
    \label{eq:dimq}
\end{split}
\end{equation}
otherwise is zero, i.e. it holds when $q$ is an integer from $-L$ to $L$. Thus, we proceed to evaluate the following necessary integral for the average
\begin{equation}
\label{eq:step1}
\begin{split}
    \text{Tr}[W \,\Pi_q^{\otimes 4}\, T_{\sigma}] =\int_{[0,2 \pi]^{\times 4}} \frac{d\bm{\alpha}}{(2\pi)^4} e^{-i q \sum_{i=1}^4 \alpha_i}
    \text{Tr}[W \prod_{i=1}^4 \left(\bigotimes_{j_i=1}^L e^{i \alpha_i\sigma_{j_i}^{z}}\right) T_\sigma],
\end{split}  
\end{equation}
where we use $d\bm{\alpha}$ as a shorthand for $\prod_{i = 1}^4 d\alpha_i$ and the label $i$ indicates the corresponding replica.
Since both $W$ and $T_{\sigma}$ can be factorized over each qubit site (see Fig. \ref{fig:TN_representation}) we simplify the trace as
\begin{equation}
\label{eq:step2}
     \int_{[0,2 \pi]^{\times 4}}   \frac{d\bm{\alpha}}{(2\pi)^4} e^{-i q \sum_{j=0}^4 \alpha_j}  \left(\text{Tr}[W^{(1)} \left(\bigotimes_{j=1}^4 e^{i\alpha_j \sigma^{z}_j}\right) T_{\sigma}^{(1)}]\right)^L,
\end{equation}
with each term being listed in Table~\eqref{tab:S_4_perm}.

\begin{table*}[t]
\centering
\begin{tabular}{|c|c||c|c|}
\hline
\textbf{Permutation $\sigma$} & \textbf{Trace Value} 
& \textbf{Permutation $\sigma$} & \textbf{Trace Value} \\
\hline
$()$ & $8\!\left(\sum_{j=1}^4 \cos\alpha_j+\sin\alpha_j\right)$
& $(1\,3)$ & $4\cos(\alpha_1+\alpha_3)\cos(\alpha_2+\alpha_4)$ \\

$(3\,4)$ & $4\cos(\alpha_1+\alpha_2)\cos(\alpha_3+\alpha_4)$
& $(1\,3\,4)$ & $2\cos(\alpha_1+\alpha_2+\alpha_3+\alpha_4)$ \\

$(2\,3)$ & $4\cos(\alpha_2+\alpha_3)\cos(\alpha_1+\alpha_4)$
& $(1\,3)(2\,4)$ & $8\!\left(\sum_{j=1}^4 \cos\alpha_j+\sin\alpha_j\right)$ \\

$(2\,3\,4)$ & $2\cos(\alpha_1+\alpha_2+\alpha_3+\alpha_4)$
& $(1\,3\,2\,4)$ & $4\cos(\alpha_1+\alpha_2)\cos(\alpha_3+\alpha_4)$ \\

$(2\,4\,3)$ & $2\cos(\alpha_1+\alpha_2+\alpha_3+\alpha_4)$
& $(1\,4\,3\,2)$ & $4\cos(\alpha_1+\alpha_3)\cos(\alpha_2+\alpha_4)$ \\

$(2\,4)$ & $4\cos(\alpha_1+\alpha_3)\cos(\alpha_2+\alpha_4)$
& $(1\,4\,2)$ & $2\cos(\alpha_1+\alpha_2+\alpha_3+\alpha_4)$ \\

$(1\,2)$ & $4\cos(\alpha_1+\alpha_2)\cos(\alpha_3+\alpha_4)$
& $(1\,4\,3)$ & $2\cos(\alpha_1+\alpha_2+\alpha_3+\alpha_4)$ \\

$(1\,2)(3\,4)$ & $8\!\left(\sum_{j=1}^4 \cos\alpha_j+\sin\alpha_j\right)$
& $(1\,4)$ & $4\cos(\alpha_2+\alpha_3)\cos(\alpha_1+\alpha_4)$ \\

$(1\,2\,3)$ & $2\cos(\alpha_1+\alpha_2+\alpha_3+\alpha_4)$
& $(1\,4\,2\,3)$ & $4\cos(\alpha_1+\alpha_2)\cos(\alpha_3+\alpha_4)$ \\

$(1\,2\,3\,4)$ & $4\cos(\alpha_1+\alpha_3)\cos(\alpha_2+\alpha_4)$
& $(1\,4)(2\,3)$ & $8\!\left(\sum_{j=1}^4 \cos\alpha_j+\sin\alpha_j\right)$ \\

$(1\,2\,4\,3)$ & $4\cos(\alpha_2+\alpha_3)\cos(\alpha_1+\alpha_4)$
& $(1\,3\,2)$ & $ 2\cos(\prod_{j=1}^4 \alpha_j)$ \\

$(1\,2\,4)$ & $2\cos(\alpha_1+\alpha_2+\alpha_3+\alpha_4)$
& $(1\,3\, 4 \,2)$ & $ 4 \cos (\alpha_1+\alpha_4) \cos (\alpha_2+\
\alpha_3)$ \\
\hline
\end{tabular}
\caption{Values of $\mathrm{Tr}\!\left[W^{(1)}\left(\bigotimes_{j=1}^4 e^{i\alpha_j\sigma^z_j}\right)T^{(1)}_\sigma\right]$ for all permutations $\sigma\in S_4$ before integration.}
\label{tab:S_4_perm}
\end{table*}

By computing the sum over the permutation group $S_4$ we obtain
\begin{equation}
\label{eq:full_int}
    \begin{split}
         \sum_{\sigma \in S_4} \text{Tr}[W \, \Pi_q^{\otimes 4} \, T_\sigma]&= 2^{L+2} \int_{[0,2 \pi]^{\times 4}} \frac{d\bm{\alpha}}{(2\pi)^4}  e^{-i q \sum_{j=1}^4 \alpha_j} \,  [3 \, \left(2 \cos (\text{$\alpha $1}+\text{$\alpha $2}) \cos (\text{$\alpha $3}+\text{$\alpha $4})\right)^L\\
         &+2 \cos^L (\sum_{j=1}^4 \alpha_j)+2^{2L} \Big(\prod_{j=1}^{4} \cos\alpha_j + \prod_{j=1}^{4} \sin\alpha_j \Big)^L]
    \end{split}
\end{equation}
The first integral is computed in the following way 
\begin{equation}
    \begin{split}
        &\int_{[0,2 \pi]^{\times 4}} \frac{d\bm{\alpha}}{(2\pi)^4} e^{-i q \sum_{j=1}^4 \alpha_j} 2^{L+2} \, 3 \, \left(2 \cos (\text{$\alpha $1}+\text{$\alpha $2}) \cos (\text{$\alpha $3}+\text{$\alpha $4}) \right)^L \\
        &= \frac{2^{2 L+2}\, 3}{(2\pi)^4} \left(\int_0^{2\pi}d\alpha_1\, \int_0^{2\pi} d\alpha_2\, \cos^L(\alpha_1+\alpha_2) e^{-i q  (\alpha_1+\alpha_2) } \right)^2\\
        &= \frac{2^{2 L+2}\, 3}{(2\pi)^2} \left(\int_0^{2\pi}ds \cos^L(s) e^{-i q  s } \right)^2\\
        &= \frac{2^{2 L+2}\, 3}{(2\pi)^2 2^{2L}} \left( \sum_{j=0}^L \binom{L}{j} \int_0^{2\pi}ds \, e^{i(L- 2j-q) s}  \right)^2\\
        &= \begin{cases}
            & 12 \,  d_q^2 \quad |q|\leq L \And L-q \text{ even}\\
            & 0 \quad \text{otherwise}
        \end{cases},
    \end{split}
\end{equation}
where we used standard change of variables and $\int_0^{2\pi} ds\, e^{i n s}= 2 \pi$ if and only if $n=0$ otherwise is zero for integer values of $n$.
The second integral can be obtained analogously
\begin{equation}
    \begin{split}
        &\int_{[0,2 \pi]^{\times 4}} \frac{d\bm{\alpha}}{(2\pi)^4} \, e^{-i q \sum_{j=1}^4 \alpha_j} \, 2^{L+3}  \left(\cos (\text{$\alpha $1}+\text{$\alpha $2}+\text{$\alpha $3}+\text{$\alpha $4}) \right)^L\\
    &=\frac{2^{L+3}}{(2\pi)^4} \left( (2 \pi)^3 \int_0^{2\pi}d s \cos^L(s) e^{-i q  s } \right)\\
        &= \begin{cases}
            & 2^{3} d_q \quad |q|\leq L \And L-q \text{ even}\\
            & 0 \quad \text{otherwise}
        \end{cases}.
    \end{split}
\end{equation}
We are left with the evaluation of the last integral 
\begin{equation}
    \begin{split}
         \frac{2^{3L+2}}{(2\pi)^4} \int_{[0,2 \pi]^{\times 4}} d\bm{\alpha} &\, e^{-i q  \sum_{j=0}^4 \alpha_j} \Big(\prod_{j=1}^{4} \cos\alpha_j + \prod_{j=1}^{4} \sin\alpha_j \Big)^L
        =  \frac{2^{3L+2}}{(2\pi)^4} \int d\bm{\alpha} \, e^{-i q  \sum_j \alpha_j} \sum_{k=0}^{L} \binom{L}{k} \prod_{j=1}^{4} (\cos\alpha_j)^{L-k} (\sin\alpha_j)^k\\
        =  \frac{2^{3L+2}}{(2\pi)^4}  \sum_{k=0}^{L} & \binom{L}{k} \left(\int_0^{2 \pi} d\alpha (\cos\alpha)^{L-k} (\sin\alpha)^k  e^{-i q  \alpha} \right)^4 \\
    \end{split}
\end{equation}
where we use the binomial expansion and even $L$. Let us now compute the final integral
\begin{equation}
    \begin{split}
        \int_0^{\pi} d\alpha & (\cos \alpha)^{L-k} (\sin \alpha)^k e^{- i q \alpha}
        = \int_0^{\pi} d\alpha \left(\frac{e^{i \alpha} + e^{-i \alpha}}{2}\right )^{L-k} \left(\frac{e^{i \alpha} - e^{-i \alpha}}{2i}\right )^{k} e^{-i q \alpha} \\
        = \frac{(-i)^k}{2^{L}} &\sum_{n = 0}^{L-k}  \sum_{m =0}^k \binom{L-k}{n} \binom{k}{m} (-1)^m \int_{0}^{\pi} d\alpha \, e^{i \alpha (L-2n - 2m - q)},
    \end{split}
\end{equation}
where we can use the trivial identity $\int_{0}^{\pi} d\alpha \,e^{i \alpha (L-2n - 2m - q)} = \pi \,\delta_{n+m,(L-q)/2}$ to obtain
\begin{equation}
\begin{split}
    \int_0^{\pi} d\alpha & (\cos \alpha)^{L-k} (\sin \alpha)^k e^{- i q \alpha}= \frac{(-i)^k \, \pi}{2^L} \sum_{m =0}^k (-1)^m\binom{L-k}{\frac{L-q}{2}-m} \binom{k}{m}.
    \label{eq:cossin4}
\end{split}
\end{equation}
We identify these terms as Kravchuk type polynomials~\cite{krawtchouk1929generalisation}
\begin{equation}
\label{eq:generic_integral}
    J_q(a,b)\equiv \int_0^{2 \pi} \frac{d \alpha}{2 \pi} \cos^a \alpha \,\sin^b \alpha \, e^{-i q \alpha}= \frac{(-i)^b \,}{2^{a+b}} \sum_{m =0}^b (-1)^m\binom{a}{\frac{a+b+q}{2}-m} \binom{b}{m} \quad a+b+q \text{ even},
\end{equation}
yielding a compact form expression reading
\begin{equation}
\label{eq:defh_SM}
    h(L,q) \equiv 2^{4 L} \sum_{k=0}^L \binom{L}{k} J_q(L-k,k)^4 .
\end{equation}
The final result is
\begin{equation}
    \exv_{U_q}[\Xi_2(\psi_{U_q})]= \frac{12 \,d_{q}^2 + 8 \,d_{q} + 2^{2-L}\, h(L,q)}{4! \binom{d_q + 3}{4}}.
\end{equation}

\subsection{Asymptotic behaviour}
\label{app:asymptotics}

We now derive the asymptotic behaviour of the average $2-$SP in the scaling limit $L,q \to +\infty$ with finite ratio $s = q/L$. Our starting point is the finite size average
\begin{equation}
\label{eq:app_finitesizeavg}
        \exv_{U_q}[\Xi_2(\psi_{U_q})]= \frac{12 \,d_{q}^2 + 8 \,d_{q} + 2^{2-L}\, h(L,q)}{4! \binom{d_q + 3}{4}},
\end{equation}
where $h(L,q)$ is given by Eq.~\eqref{eq:defh_SM}. 
By inspection we notice that the term giving the leading contribution in the considered scaling limit is the final one, and therefore we drop the remaining terms in the numerator. We verify a posteriori that this approximation is valid.  By retracing the steps that lead to Eq.~\eqref{eq:app_finitesizeavg}, we re-express it as 
\begin{equation}
    2^{2-L} h(L,q)  = \frac{2^{3 L +2}}{(2\pi)^4}\int_{[0,2 \pi]^{\times 4}} d\bm{\alpha} \left( \prod_{i = 1}^4 \cos\alpha_i + \prod_{i = 1}^4 \sin\alpha_i\right)^L e^{- i s L \sum_i \alpha_i}.
\end{equation}
We analyze the integral performing a saddle point approximation. The integral takes the form
\begin{equation}
\label{eq:sp_int}
    \int_{[0,2 \pi]^{\times 4}} d\bm{\alpha} \exp(L \, F\left(\bm{\alpha};s\right))
    \equiv\int_{[0,2 \pi]^{\times 4}}  d\bm{\alpha} \left( \prod_{i = 1}^4 \cos\alpha_i + \prod_{i = 1}^4 \sin\alpha_i\right)^L e^{- i s L \sum_i \alpha_i} ,
\end{equation}
where we have defined
\begin{equation}
    F(\bm{\alpha};s) = \log\left( \prod_{i = 1}^4 \cos\alpha_i + \prod_{i = 1}^4 \sin\alpha_i\right) - i s \sum_{i=1}^4 \alpha_i.
\end{equation}
In the $L\to +\infty$ limit, the integral \eqref{eq:sp_int} is dominated by the contribution around the saddles of the function $F(\bm{\alpha};s)$. 
Let us first analyse the case $s = 0$. The function $F(\bm{\alpha};s = 0)$ has many local maxima with value 0, in particular for any choice of $\alpha_i \in \{0,\pi\}$ and also for any choice of $\alpha_i \in \{\pi/2,3 \pi/2\}$. Therefore, there are $32$ local maxima. Using the fact that $L$ is always even and the periodicity of $F(\bm{\alpha};s = 0)$ it is easy to show that the saddle point contribution of each one of the $32$ local maxima is equivalent.

\begin{figure}[t!]
\centering
\includegraphics[width=0.5\columnwidth]{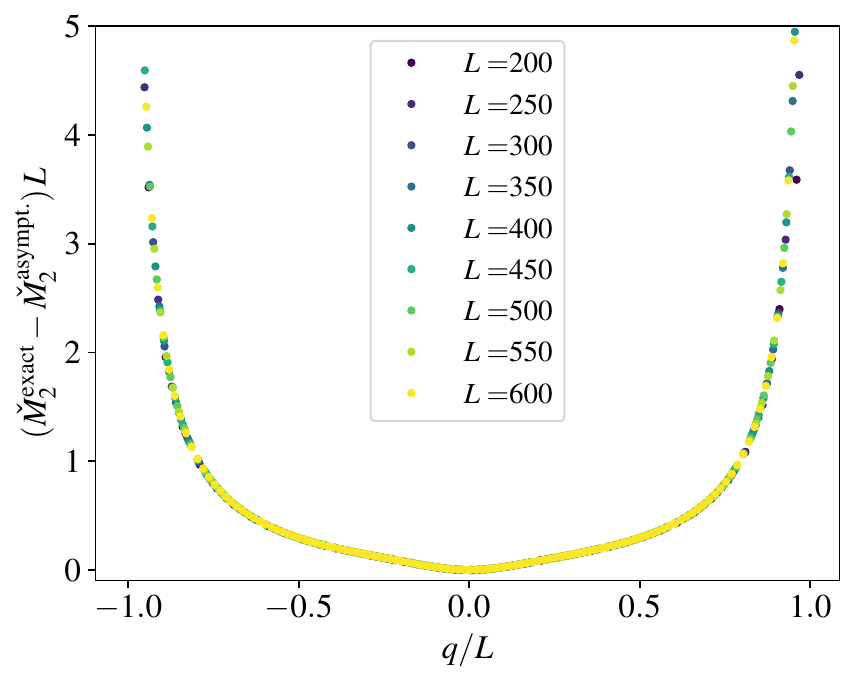}
      \caption{Numerical check of the validity of the asymptotic behaviour in Eq.~\eqref{eq:app_asymptres}. We compute numerically, as a function of the charge density $q/L$, the difference between the exact prediction $\check M_2^{\text{exact}} = -\log_2 \exv_{U_q}[\Xi_2(\psi_{U_q})]$ originating from Eq.~\eqref{eq:app_finitesizeavg} and the derived asymptotic expression Eq.~$\check M_2^{\text{asympt.}}$ \eqref{eq:app_asymptres}. We multiply the latter difference by the number of qubits $L$. By increasing $L$ we observe a clear collapse of the data, thereby verifying that the difference $\check M_2^{\text{exact}}-\check M_2^{\text{asympt.}}$ is of order $O(L^{-1})$.}
    \label{fig:M2diff}
\end{figure}

\noindent
We now show that the generic case $s \neq 0$ present similar symmetry properties allowing us to analyse only one of the possible saddles. The stationarity condition $\partial_{\alpha_i} F(\bm{\alpha};s) = 0$ is given by
\begin{equation}
\label{eq:statcond}
    \frac{-\sin \alpha_i \prod_{j\neq i} \cos \alpha_j + \cos \alpha_i \prod_{j\neq i} \sin \alpha_j}{\prod_j \cos \alpha_j + \prod_j \sin \alpha_j} = i s \qquad \forall i = 1,\cdots,4.
\end{equation}
As we show, Eq.~\eqref{eq:statcond} admits a solution in which $\alpha_i = \alpha^*$ for any $i = 1,\cdots,4$. Before determining the solution $\alpha^*$ notice that any $\bm{\alpha} = \bm{\alpha}^* + \bm{n} \pi + q \pi/2$ with $\bm{n} = \{0,1\}^4$ and $q = 0,1$, remains a solution. The saddle point contribution of each one of these $32$ saddles is always the same as can be proved explicitly by using the fact that both $L$ and $M$ are even. In summary, we can simplify the analysis by only determining the saddle point contribution of the  saddle $\bm{\alpha}^* =(\alpha^*,\alpha^*,\alpha^*,\alpha^*)$ and multiply the final result by $32$. 
 With the assumption of symmetric saddle, the condition in Eq.~\eqref{eq:statcond} significantly simplifies to
\begin{equation}
\label{eq:statcond}
    is(3+ \cos(4 \alpha^*)) + \sin(4 \alpha^*) = 0.
\end{equation}
Define now $z(s)$ such that $2\cos(4 \alpha^*) = z + z^{-1}$ and $2 i \sin (4 \alpha^*) = z - z^{-1}$, where we have left implicit the dependence on $s$ of both $\alpha^*$ and $z$. The solution of Eq.~\eqref{eq:statcond} is given by
\begin{equation}
    z(s) = \frac{3s+ \sqrt{1+8s^2}}{1-s},
\end{equation}
which, using the definition of $z(s)$, implies that 
\begin{equation}
\label{eq:saddle}
    \alpha^* = - \frac{i}{4} \ln z(s).
\end{equation}
The value of the function $F(\bm{\alpha};s)$ at the saddle is thus given by
\begin{equation}
   F(\bm{\alpha}^*;s) =  - \ln\left( \frac{3-\sqrt{1+8s^2}}{2}\right) - s \ln z(s)
\end{equation}
This result allows us to perform the saddle point approximation of Eq.~\eqref{eq:sp_int} yielding
\begin{equation}
\label{eq:final_sp}
 \int_{[0,2 \pi]^{\times 4}} d\bm{\alpha} \exp(L \, F(\bm{\alpha};s)) \simeq\\
 \simeq 32\, e^{L F(\bm{\alpha}^*;s)} \frac{(2\pi)^2}{L^2} \frac{1}{\sqrt{|\det \left(H |_{\bm{\alpha}^*} \right)|}},
\end{equation}
where $(H |_{\bm{\alpha}^*})_{ij} = \partial_{\alpha_i}\partial_{\alpha_j} F(\bm{\alpha};s)|_{\bm{\alpha^*}}$ denotes the Hessian of $F(\bm{\alpha};s)$ evaluated at the saddle $\bm{\alpha} = \bm{\alpha}^* $ and the prefactor $32$ comes from the counting of the different possible equivalent saddles. Using Eq.~\eqref{eq:saddle} we find that, defining $\xi(s) \equiv|\det \left(H |_{\bm{\alpha}^*} \right)|^{-1/2}$ we have
\begin{equation}
\label{eq:xidef}
    \xi(s) = \left[ \frac{(3+\sqrt{1+8s^2})^5}{4(1-s^2)^4(1+8s^2+3 \sqrt{1+8s^2})}\right]^{1/2}.
\end{equation}
Next we make use of the approximation Eq.~\eqref{eq:final_sp} in Eq.~\eqref{eq:app_finitesizeavg} together with standard approximations
\begin{equation}
    \binom{d_q + 3}{4} \simeq \frac{d_q^4}{4!},
\end{equation}
and
\begin{equation}
    d_q = \binom{L}{\left(\frac{1-s}{2}\right)L} \simeq\left( \frac{2}{\pi L (1-s^2)}\right)^{1/2} 2^{L H_2((1-s)/2)},
\end{equation}
with $H_2(x)$ is the entropy function $H_2(x) = -x \log_2 x - (1-x) \log_2(1-x)$. Putting all these pieces together, we arrive at the final result
\begin{multline}
\label{eq:app_asymptres}
    -\log_2 \exv_{U_q}[\Xi_2(\psi_{U_q})] \underset{\substack{L,q \to +\infty \\ s = q/L}}{=}
     \left[ 4 H_2\left(\frac{1-s}{2}\right) - \frac{1}{\ln 2} F(\bm{\alpha}^*;s)-3 \right] L 
    - \log_2(8(1-s^2)^2 \xi(s)) + O(L^{-1}),
\end{multline}
which is exactly the result stated in the main text, where we have defined 
\begin{equation}
    \begin{cases}
        m(s) =   4 H_2\left(\frac{1-s}{2}\right) - \frac{1}{\ln 2} F(\bm{\alpha}^*;s)-3  ,\\
        g(s) = - \log_2(8(1-s^2)^2 \xi(s)).
    \end{cases}
\end{equation}
In Fig.~\ref{fig:M2diff}, we compare numerically the obtained asymptotic result with the exact formula for $\check{M}_2 \equiv -\log_2 \mathbb{E}_{U_q}[\Xi_2(\psi_{U_q})]$ in Eq.~\eqref{eq:app_finitesizeavg}, confirming that the large $L$ asymptotics in Eq.~\eqref{eq:app_asymptres} is accurate up to order $O(1)$.

\subsection{Mixed local charge generalization}

We here compute a further generalization of the result in Eq.~\eqref{eq:app_finitesizeavg} using a charge defined as $\mathcal{Q}_{\Vec{n}}=\sum_{j=1}^L \Vec{n} \cdot \Vec{\sigma}_j$, such that $\norm{\Vec{n}}_2=1$. The projector onto the eigenspace of $\mathcal{Q}$ with eigenvalue $q$ is
\begin{equation}
    \Pi_q = \int _{0}^{2\pi}\frac{d\alpha}{2 \pi} e^{- i \alpha q} \bigotimes_{j=1}^L e^{i \alpha \Vec{n} \cdot \Vec{\sigma}_j}.
\end{equation}
The main property of the above calculations remains the factorization of the projector over each qubit. We can now compute as we did before the integrals
\begin{widetext}
\begin{equation}
\label{eq:exactmixcharge}
    \begin{split}
         \sum_{\sigma \in S_4} \text{Tr}[W \, \Pi_q^{\otimes 4} \, T_\sigma]&= 2^{L+2}\int \frac{d\bm{\alpha}}{(2\pi)^4}  e^{-i q \sum_{j=1}^4 \alpha_j} \,  \Bigg[2^L \, 3 \, \left( \cos(\alpha_1+\alpha_4) \cos(\alpha_2+\alpha_3) + (f(\Vec{n})-1) \prod_{j=1}^4 \sin \alpha_j  \right)^L\\
         &+2 \left(\cos(\sum_{j=1}^4 \alpha_j)+(g(\Vec{n})-1) \prod_{j=1}^4 \sin \alpha_j \right)^L+2^{2L} \left( \prod_{j=1}^4 \cos\alpha_j+ \left( \sum_{k=1}^3 n_k^4 \right)\prod_{j=1}^4 \sin\alpha_j \right)^L \Bigg]
    \end{split}
\end{equation}
\end{widetext}
with $f(\Vec{n})= (n_1 - n_2 - n_3)  (n_1 + n_2 - n_3)  (n_1 - n_2 + n_3)  (n_1 + n_2 + n_3)$ and $g(\Vec{n})=\sum_{k=1}^3 n_k^4- 6( n_1^2 n_2^2  + n_1^2 n_3^2  + n_2^2 n_3^2 )$.
Notice that when one of the components of $\Vec{n}$ is one and the other zeros we obtain the same result as before.
We derive each piece one by one.
\begin{widetext}
    \begin{equation}
    \label{eq:mix1}
    \begin{split}
       I_{1,\vec{n}}\equiv &\int_{[0,2 \pi]^{\times 4}} \frac{d\bm{\alpha}}{(2\pi)^4}  e^{-i q \sum_{j=1}^4 \alpha_j} \, 2^{2L+2} \, 3 \, \Big( \cos(\alpha_1+\alpha_4) \cos(\alpha_2+\alpha_3) + (f(\Vec{n})-1) \prod_{j=1}^4 \sin \alpha_j  \Big)^L\\
        &=2^{2L+2} \, 3  \int_{[0,2 \pi]^{\times 4}} \frac{d\bm{\alpha}}{(2\pi)^4}  e^{-i q \sum_{j=1}^4 \alpha_j} \, \sum_{k=0}^L \binom{L}{k} \cos^{L-k}(\alpha_2+\alpha_3) \cos^{L-k}(\alpha_1+\alpha_4) (f(\Vec{n})-1)^k \prod_{j=2}^4 \sin^k\alpha_j \\
        &=2^{2L+2} \, 3 \sum_{k=0}^L \binom{L}{k} (f(\Vec{n})-1)^k  \left( \int \frac{d \alpha_1 \, d \alpha_2}{(2 \pi)^2} e^{-i q (\alpha_1+ \alpha_2)} \cos^{L-k}(\alpha_1+\alpha_2) \sin^k \alpha_1 \sin^k \alpha_2 \right)^2 \\
        &= 2^{2L+2} \, 3 \sum_{k=0}^L \binom{L}{k} (f(\Vec{n})-1)^k  \left( \frac{(-1)^k}{2^{L+k}} \sum_{j=0}^k \binom{k}{j}^2 \binom{L-k}{\frac{L+q}{2}-j} \right)^2\\
        &=2^{2L+2} \, 3 \sum_{k=0}^L \binom{L}{k} (f(\Vec{n})-1)^k \frac{1}{2^{2L + 2 k}} \left(\sum_{j=0}^k \binom{k}{j}^2 \binom{L-k}{\frac{L+q}{2}-j}\right)^2
        \end{split}
    \end{equation}
\end{widetext}

\begin{widetext}
    \begin{equation}
    \label{eq:mix2}
    \begin{split}
       I_{2,\vec{n}} \equiv& 2^{L+3}\int_{[0,2 \pi]^{\times 4}} \frac{d\bm{\alpha}}{(2\pi)^4}  e^{-i q \sum_{j=1}^4 \alpha_j} \,  \left(\cos(\sum_{j=1}^4 \alpha_j)+(g(\Vec{n})-1) \prod_{j=1}^4 \sin \alpha_j \right)^L\\
        &=  2^{L+3} \int_{[0,2 \pi]^{\times 4}} \frac{d\bm{\alpha}}{(2\pi)^4}  e^{-i q \sum_{j=1}^4 \alpha_j} \, \sum_{s=0}^L \binom{L}{s} (g(\Vec{n})-1)^s \cos^{L-k}(\sum_{j=1}^4 \alpha_j) \prod_{j=1}^4 \sin^k \alpha_j \\
        &= 2^{L+3} \sum_{s=0}^L \binom{L}{s} (g(\Vec{n})-1)^s \int \frac{d\bm{\alpha}}{(2\pi)^4}  e^{-i q \sum_{j=1}^4 \alpha_j} \cos^{L-k}(\sum_{j=1}^4 \alpha_j) \prod_{j=1}^4 \sin^k \alpha_j  \\
        &= 2^{L +3} \sum_{k=0}^L \binom{L}{k} (g(\Vec{n})-1)^k \frac{1}{2^{L + 3 k}} \left(\sum_{j=0}^k \binom{k}{j}^4 \binom{L-k}{\frac{L+q}{2}-j}\right) 
        \end{split}
    \end{equation}
\end{widetext}

\begin{figure}
\centering
    \includegraphics[width=0.5\linewidth]{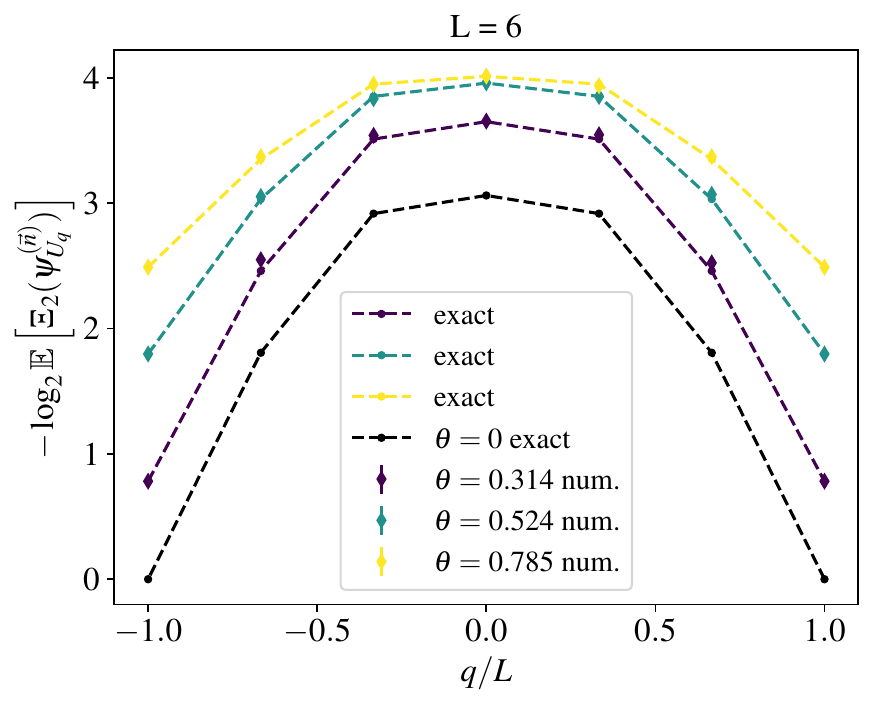}
    \caption{Comparison between analytical (dashed lines) and numerical (markers with errorbars) predictions for the average of the 2-SP for $U(1)$ symmetric Haar random state with local charge $\mathcal{Q}_{\Vec{n}}=\sum_{j=1}^L \Vec{n} \cdot \Vec{\sigma}_j$ as a function of the charge density $q/L$ and $\vec{n}=(\sin \theta,0,\cos \theta)$ ($L=6$ and 200 realizations).}
    \label{fig:avgSP_generic}
\end{figure}

\begin{widetext}
    \begin{equation}
    \label{eq:mix3}
    \begin{split}
        I_{3,\vec{n}}\equiv& \int_{[0,2 \pi]^{\times 4}} \frac{d\bm{\alpha}}{(2\pi)^4}  e^{-i q \sum_{j=1}^4 \alpha_j} \, 2^{3L+2}    \Big( \prod_{j=1}^4 \cos\alpha_j+ \left( \sum_{k=1}^3 n_k^4 \right)\prod_{j=1}^4 \sin\alpha_j \Big)^L\\
        & =\int_{[0,2 \pi]^{\times 4}} \frac{d\bm{\alpha}}{(2\pi)^4}  e^{-i q \sum_{j=1}^4 \alpha_j} \, 2^{3L+2}  \sum_{k=0}^L \binom{L}{k} \prod_{j=1}^4 \cos^{L-k}\alpha_j \left( \sum_{k=1}^3 n_k^4 \right)^k \prod_{j=1}^4 \sin^k\alpha_j \\
        &=2^{3L+2}  \sum_{k=0}^L \binom{L}{k} \left( \sum_{k=1}^3 n_k^4 \right)^k \left(\int_0^{2 \pi} \frac{d \alpha}{2\pi} e^{-i q \alpha} \cos^{L-k}\alpha   \sin^k\alpha \right)^4 \\
        &= 2^{3L+2} \frac{1}{2^{4L}} \sum_{k=0}^L \binom{L}{k} \left( \sum_{k=1}^3 n_k^4 \right)^k \left[ \sum_{m =0 }^k (-1)^m\binom{L-k}{\frac{L-q}{2}-m}\binom{k}{m} \right]^4
        \end{split}
    \end{equation}
\end{widetext}
where the last identity is from Eq. \eqref{eq:cossin4}. The final result is
\begin{equation}
\label{eq:finalresmix}
    -\log_2 \exv\left[ \Xi_2(\psi^{(\vec{n})}_{U_q})\right] = -\log_2\left(\frac{I_{1,\vec{n}}+I_{2,\vec{n}}+I_{3,\vec{n}}}{4 ! \binom{d_q+3}{4}}\right).
\end{equation}
In Fig.\ref{fig:avgSP_generic}, we show the analytical result just obtained compared with the numerical result as a function of $\vec{n}$ and $q$, where the numerical sampling for $U(1)$-constrained pure random states is described in Sect.~\ref{app:sampling}.

\subsubsection{Asymptotic behaviour for generic $\vec{n}$}
\begin{figure}

    \includegraphics[width=0.5\linewidth]{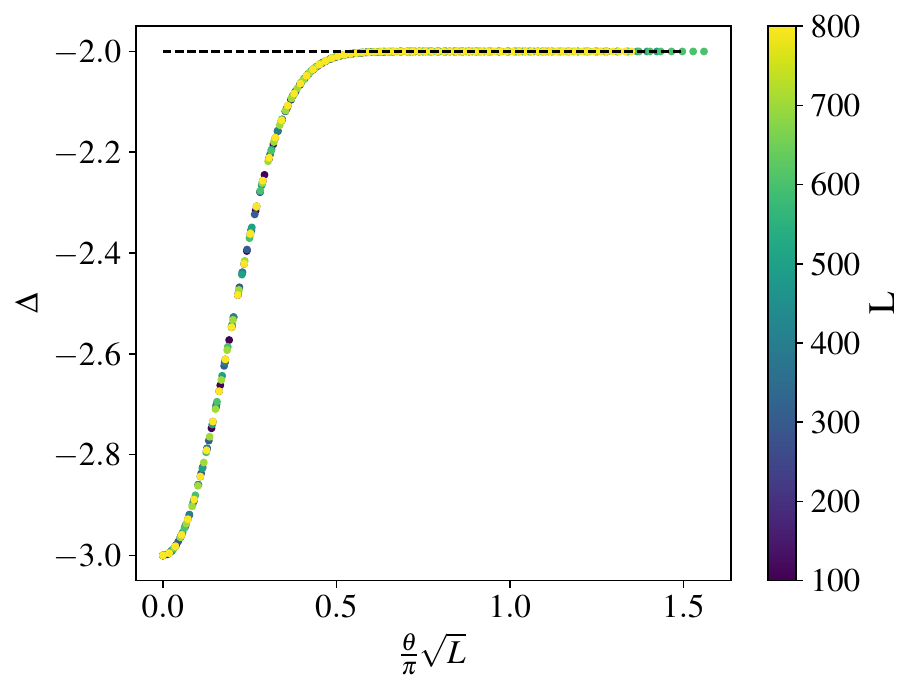}
    \caption{Numerical check of the asymptotic result in Eq.~\eqref{eq:asympttheta}. We numerically compute, using the exact formula Eq.~\eqref{eq:finalresmix}, the difference $\Delta \equiv -\log_2 \exv[\Xi_2(\psi_{U_{q=0}})]-L$ for different conserved charges parametrized by $\vec{n} = (\sin \theta,0,\cos\theta)$ and for increasing number of qubits $L$. When the angle $\theta$ is scaled as $\sqrt{L}$ the data corresponding to different number of qubits $L$ perfectly collapse. }
    \label{fig:asympttheta}
\end{figure}

In this section we analyze the asymptotic behaviour of Eq.~\eqref{eq:exactmixcharge} in the limit $L,q \to +\infty$ and fixed charge density $s = q/L$. By numerical inspection of the results in Eqs.~\eqref{eq:mix1},\eqref{eq:mix2}, and \eqref{eq:mix3}, we observe that the leading contribution to $-\log_2 \exv[\Xi(\psi^{(\vec{n})}_{U_q})]$ is given by Eq.~\eqref{eq:mix3}. Let us define $w_{\vec{n}} \equiv \sum_{k=1}^3 n_k^4$. Notice that the case $\mathcal{Q}_{\Vec{n}} = \pm \mathcal{Q}^{(x,y,z)}$ corresponds to $w_{\vec{n}} = 1$, which is analyzed in detail in Sec.~\eqref{app:asymptotics}. All other choices of $\vec{n}$ corresponds to $1/3 \leq w_{\vec{n}} <1$. Consider now only the leading term Eq.~\eqref{eq:mix3}. More specifically, we are required to compute the asymptotic behaviour of the following integral
\begin{equation}
    \int_{[0,2 \pi]^{\times 4}} d\bm{\alpha} \exp(L \, F_{\vec{n}}\left(\bm{\alpha};s\right))
    \equiv\int_{[0,2 \pi]^{\times 4}}  d\bm{\alpha} \left( \prod_{i = 1}^4 \cos\alpha_i + w_{\vec{n}} \prod_{i = 1}^4 \sin\alpha_i\right)^L e^{- i s L \sum_i \alpha_i} .
\end{equation}
Let us count the number of equivalent saddles in the case $s = 0$ analogously to Sec.~\ref{app:asymptotics}. If $w_{\vec{n}}<1$, the number of equivalent saddles is halved with respect to the case $w_{\vec{n}} = 1$. The reason is that a saddle is now obtained if and only if $\alpha_j \in \{0,\pi\}$ and not anymore for $\alpha_j \in \{\pi/2,3\pi/2\}$ leaving us with only $16$ equivalent saddles. Let us therefore define 
\begin{equation}
\label{eq:Nsaddles}
    \mathcal{N}_{\vec{n}} = \begin{cases}
        32, \qquad w_{\vec{n}} = 1, \\
        16, \qquad w_{\vec{n}} < 1.
    \end{cases}
\end{equation}
As a result of the standard saddle-point approximation, we have
\begin{equation}
\label{eq:final_sp_n}
 \int_{[0,2 \pi]^{\times 4}} d\bm{\alpha} \exp(L \, F_{\vec{n}}(\bm{\alpha};s = 0)) \simeq\mathcal{N}_{\vec{n}}\, e^{L F_{\vec{n}}(\bm{\alpha}^*;s = 0)} \frac{(2\pi)^2}{L^2} \frac{1}{\sqrt{|\det \left(H_{\vec{n}} |_{\bm{\alpha}^*} \right)|}}.
\end{equation}
By explicit computation of the symmetric saddle $\bm{\alpha}^* = \bm{0}$ we obtain
\begin{equation}
    F_{\vec{n}}(\bm{\alpha}^*;s = 0) = 0,
\end{equation}
and 
\begin{equation}
    \frac{1}{\sqrt{|\det \left(H_{\vec{n}} |_{\bm{\alpha}^*} \right)|}} = 1,  \qquad \forall \hspace{0.15cm} w_{\vec{n}}.
\end{equation}
Putting all the pieces together and using the same steps leading to Eq.~\eqref{eq:app_asymptres}, we find 
\begin{equation}
    -\log_2 \exv[\Xi_2(\psi^{(\vec{n})}_{U_{q = 0}})] \underset{\substack{L\to +\infty \\ s = 0}}{=} L -\log_2 \left(\frac{\mathcal{N}_{\vec{n}}}{4}\right).
\end{equation}
Finally, using Eq.~\eqref{eq:Nsaddles}, we obtain the result shown in the main text
\begin{equation}
\label{eq:asympttheta}
    -\log_2 \exv[\Xi_2(\psi^{(\vec{n})}_{U_{q = 0}})] \underset{\substack{L\to +\infty \\ s = 0}}{=} \begin{cases}
        L-3, \qquad \vec{n} = \hat{x},\hat{y},\hat{z} , \\
        L-2, \qquad \text{otherwise}.
    \end{cases}
\end{equation}
Next, we use the following parametrization: $\vec{n}$ as $\vec{n}=(\sin \theta,0,\cos \theta)$. The angle $\theta$ is in the interval $\theta \in [0,\pi/4]$ since the result is symmetric under $\theta \to \pi/2 -\theta$ because of the Clifford invariance of $\Xi_2$. As shown in Fig.~\ref{fig:asympttheta}, we have the following scaling behaviour
\begin{equation}
    -\log_2 \exv[\Xi_2(\psi^{(\vec{n})}_{U_{q = 0}})] = L +\mathcal{M}\left(\theta \,\sqrt{L}\right),
\end{equation}
where the scaling function $\mathcal{M}(x)$ satisfies $\mathcal{M}(0) = -3$ and $\mathcal{M}(+\infty) = -2$.

\subsection{Lévy typicality}
\label{app:Levy}
\noindent
In addition to the mean value according to the Haar$\times U(1)$ ensemble, we can state its Lévy typicality, i.e. as the dimension of the subspace $d_q$ grows, the probability of finding a state with 2-SP different from its average value decays double exponentially in the system size $L$, reading
\begin{equation}
\label{eq:Levyconc}
    \text{Prob}\left( \abs{ \Xi_2(\psi_{U_q})- \exv_{U_q}[\Xi_2(\psi_{U_q})]}\geq \epsilon \right)\leq 2 e^{-\frac{d_q \epsilon^2}{9 \pi^3 \eta^2}}.
\end{equation}
One says then that $\Xi_2$ \textit{typically} if is close to its average or simply that the random variable is \textit{typical}.
To demonstrate this, one need to know that 2-SP is Lipschitz-continuous with Lipschitz constant $\eta=5.4$~\cite{zhu2016CliffordGroupFails}
\begin{equation}
    \abs{\Xi_2(\psi)-\Xi_2(\phi)}\leq 5.4 \norm{\psi-\phi}_2,
\end{equation}
and the Lévy lemma~\cite{ledoux2001concentration}.
It states that for every Lipschitz-continuous function $f: S^{2 d-1} \rightarrow \mathbb{R}$ with Lipschitz constant $\eta$, that is,
$$
|f(x)-f(y)| \leq \eta\|x-y\|,
$$
where $\|x-y\|$ is the Euclidean norm in the surrounding space $\mathbb{R}^{2 d}$ of $S^{2 d-1}$ and $x$ is drawn randomly according to the uniform measure on the sphere $S^{2 d-1}$, one has
$$
\operatorname{Prob}\left(|f(x)-\mathrm{E}[f(x)]| \geq \epsilon\right) \leq 2 \exp \left(\frac{-d \epsilon^2}{9 \pi^3 \eta^2}\right), \quad \forall \epsilon \geq 0.
$$
Hence,
\begin{equation}
    \abs{\Xi_2(\psi(q))-\Xi_2(\phi(q))}\leq 5.4 \norm{\psi(q)-\phi(q)}_2 = 5.4 \norm{\Pi_q(\psi-\phi) \Pi_q}_2 \leq 5.4 \norm{\Pi_q}^2_{\infty} \norm{\psi-\phi}_2  =   5.4 \norm{\psi-\phi}_2
\end{equation}
where we use the Hölder inequality for the Schatten-$p$ norms and the fact that $\Pi_q$ is a projector.
Eq.~\eqref{eq:Levyconc} provides straightaway a bound on the variance, see Ref.~\cite{Vershynin_2018}. Defining $X_q \equiv \Xi_2(\psi_{U_q})$, we have
\begin{equation}
\begin{split}
    \Delta_{U_q}^2[X_q] = \mathbb{E}[(X_q-\mathbb{E}[X_q])^2] &= \int_{0}^{\infty}dt \,\text{Prob}((X_q-\exv[X_q])^2 \geq t)  \\
    &= \int_0^{\infty} ds \,2 s\, \text{Prob}(|X_q-\mathbb{E}[X_q]|\geq s) \\
    &\leq 4 \int_{0}^{\infty} ds \, s \exp\left(-\frac{d_q s^2}{9 \pi^3 \eta^2}\right),
\end{split}
\end{equation}
where in the last line we used explicitly Eq.~\eqref{eq:Levyconc}. The solution of the last integral provides the bound
\begin{equation}
\Delta_{U_q}^2[\Xi_2(\psi_{U_q})] \leq \frac{18 \pi^3 \eta^2}{d_q},
\end{equation}
which is precisely the bound $\Delta_{U_q}^2[\Xi_2(\psi_{U_q})] \leq O(d_q^{-1})$ stated in the main text.

\section{Variance of Stabilizer purity of order 2 for Haar$\times U(1)$ ensemble}
\label{app:variance_SP}
\noindent
In this section, we analytically derive the exact result for the variance for 2-SP over the ensemble of Haar random states constrained to a $U(1)$ symmetry, namely
\begin{equation}
    \Delta_{U_q}^2[\Xi_2(\psi_{U_q})] = \exv_{U_q}[\Xi_2(\psi_{U_q})^2]-\exv_{U_q}[\Xi_2(\psi_{U_q})]^2\,.
\end{equation}
At this point, we need only to evaluate $\exv_{U_q}[\Xi_2(\psi_{U_q})^2]$, which for simplicity we rewrite this average as
\begin{equation}
\begin{split}
    \exv_{U_q}[\Xi_2(\psi_{U_q})^2]&= \exv_{U_q}[\text{Tr}[W^{\otimes 2} \,\psi(q)^{\otimes8}]]= \frac{1}{\binom{d_q+7}{8} 8!} \sum_{\sigma \in S_8} \text{Tr}[W^{\otimes 2} \,\Pi_q^{\otimes 8}\, T_{\sigma}],
\end{split}
\end{equation}
and we can use the same factorization on a single site we made for the case of the mean value, i.e.
\begin{equation}
     \int_{[0,2 \pi]^{\times 8}}   \frac{d\bm{\alpha}}{(2\pi)^8} e^{-i q \sum_{i=0}^8 \alpha_i}  \left(\text{Tr}[\left(W^{(1)}\right)^{\otimes 2} \left(\bigotimes_{i=1}^8 e^{i\alpha_i \sigma^{z}_i}\right) T_{\sigma}^{(1)}]\right)^L.
\end{equation}
Although there are $8!$ terms in the sum, we can reduce it to $13$, which read
\begin{equation}
    16 \, 64^L \int_{[0,2 \pi]^{\times 8}} \frac{d\bm{\alpha}}{(2\pi)^8}  e^{-i q \sum_{j=1}^8 \alpha_j}  \Big(\prod_{j=1}^{4} \cos\alpha_j + \prod_{j=1}^{4} \sin\alpha_j \Big)^L \Big(\prod_{j=5}^{8} \cos\alpha_j + \prod_{j=5}^{8} \sin\alpha_j \Big)^L = 16 \left(\frac{h(L,q)}{2^L}\right)^2
\end{equation}
\begin{equation}
    96 \, 32^L \int_{[0,2 \pi]^{\times 8}} \frac{d\bm{\alpha}}{(2\pi)^8}  e^{-i q \sum_{j=1}^8 \alpha_j} \, \cos^L(\alpha_5+\alpha_8) \, \cos^L(\alpha_6+\alpha_7) \,  \Big(\prod_{j=1}^{4} \cos\alpha_j + \prod_{j=1}^{4} \sin\alpha_j \Big)^L = 96 \frac{d_q^2 h(L,q)}{2^L}
\end{equation}
\begin{equation}
    640 \, 16^L \int_{[0,2 \pi]^{\times 8}} \frac{d\bm{\alpha}}{(2\pi)^8}  e^{-i q \sum_{j=1}^8 \alpha_j} \, \cos^L(\alpha_5+\alpha_8+\alpha_6+\alpha_7) \,  \Big(\prod_{j=1}^{4} \cos\alpha_j + \prod_{j=1}^{4} \sin\alpha_j \Big)^L = 640 \frac{d_q h(L,q)}{2^L}
\end{equation}
\begin{equation}
\begin{split}
    256 \, 32^L \int_{[0,2 \pi]^{\times 8}} \frac{d\bm{\alpha}}{(2\pi)^8}  e^{-i q \sum_{j=1}^8 \alpha_j} &\, \Big( \sin \alpha_6 \sin \alpha_7 \sin \alpha_8 (\sin(\alpha_3+\alpha_4) \cos \alpha_1 \cos \alpha_2 \cos \alpha_5 \\
    &- \cos(\alpha_3+\alpha_4) \sin \alpha_1 \sin \alpha_2 \sin \alpha_5)\\
    &+ \cos \alpha_6 \cos \alpha_7 \cos \alpha_8 (\cos(\alpha_3+\alpha_4) \cos \alpha_1 \cos \alpha_2 \cos \alpha_5 \\
    &+ \sin(\alpha_3+\alpha_4) \sin \alpha_1 \sin \alpha_2 \sin \alpha_5) \Big)^L\\
    &= 256 \, 2^{5L}  \sum_{k=0}^{L}  \binom{L}{k} \, J_q(L-k,k)^3 \sum_{j=0}^{k} \binom{k}{j} (-1)^{k-j} \sum_{p=0}^{L-k} \binom{L-k}{p}  \\
    & \, J_q(k - j + p, L - k - p + j)  J_q(j + p, L - p - j)^3 \equiv  256 \, 2^{5L} \,\mathcal{K}_1(L,q)
\end{split}
\end{equation}
\begin{equation}
\begin{split}
    960  \, 8^L \int_{[0,2 \pi]^{\times 8}} \frac{d\bm{\alpha}}{(2\pi)^8}  e^{-i q \sum_{j=1}^8 \alpha_j} \, \cos^L(\alpha_1+\alpha_2) \, &  \Big(\cos(\alpha_2-\alpha_7)\cos(\alpha_3+\alpha_4+\alpha_5-\alpha_6)\\
    &+ \cos(\alpha_2+\alpha_7)\cos(\alpha_3+\alpha_4+\alpha_5+\alpha_6) \Big)^L\\
    &= 960 \, d_q \sum_{k=0}^{L/2} \binom{L}{2 k} \binom{2 k}{\frac{2 k-q}{2}}^2 \binom{L-2k}{\frac{L-2k+q}{2}}^2\\
    & =960\, d_q \mathcal{K}_2(L,q)
\end{split}
\end{equation}
\begin{equation}
\begin{split}
    5920  \, 4^L \int_{[0,2 \pi]^{\times 8}} \frac{d\bm{\alpha}}{(2\pi)^8}  e^{-i q \sum_{j=1}^8 \alpha_j} \, &  \Big(\cos(\alpha_1-\alpha_8)\cos(\alpha_2+\alpha_3+\alpha_4+\alpha_5+\alpha_6-\alpha_7)\\
    &+ \cos(\alpha_1+\alpha_8)\cos(\alpha_2+\alpha_3+\alpha_4+\alpha_5+\alpha_6+\alpha_7) \Big)^L\\
    &= 5920 \sum_{k=0}^{L/2} \binom{L}{2 k} \binom{2 k}{\frac{2 k-q}{2}}^2 \binom{L-2k}{\frac{L-2k+q}{2}}^2
    \\
    &= 5920 \,\mathcal{K}_2(L,q)
\end{split}
\end{equation}
\begin{equation}
     144\,16^L \int_{[0,2 \pi]^{\times 8}} \frac{d\bm{\alpha}}{(2\pi)^8}  e^{-i q \sum_{j=1}^8 \alpha_j} \, \left(\cos(\alpha_1+\alpha_8) \cos(\alpha_2+\alpha_7) \cos(\alpha_3+\alpha_6) \cos(\alpha_4+\alpha_5)\right)^L= 144\, d_q^4
\end{equation}
\begin{equation}
     3648\,8^L \int_{[0,2 \pi]^{\times 8}} \frac{d\bm{\alpha}}{(2\pi)^8}  e^{-i q \sum_{j=1}^8 \alpha_j} \, \left(\cos(\alpha_1+\alpha_8) \cos(\alpha_2+\alpha_7) \cos(\alpha_3+\alpha_6+\alpha_4+\alpha_5)\right)^L= 3648\, d_q^3
\end{equation}

\begin{figure}
    \centering
    \includegraphics[width=0.6\linewidth]{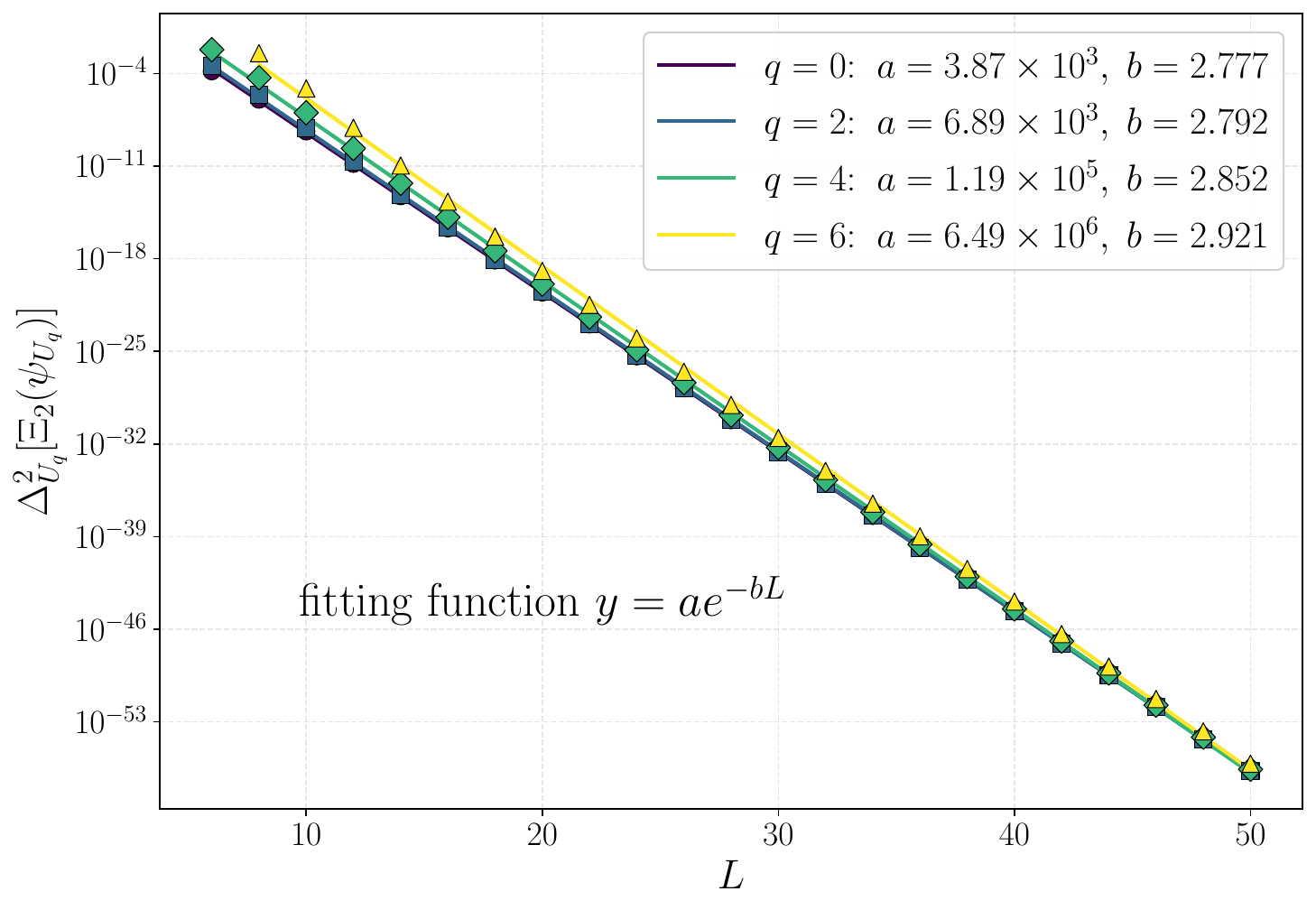}
    \caption{Analytical result of the variance of the 2-SP, $\Delta_{U_q}^2[\Xi_2(\psi_{U_q})]$, according to the Haar×$U(1)$ measure as a function of the system size, $L$, for different values of the charge, $q=0,2,4,6$. An exponential fitting $y= a \, e^{-b L}$ is shown for each value of the charge $q$.  }
    \label{fig:variance}
\end{figure}
\begin{figure}
    \centering
    \includegraphics[width=1.0\linewidth]{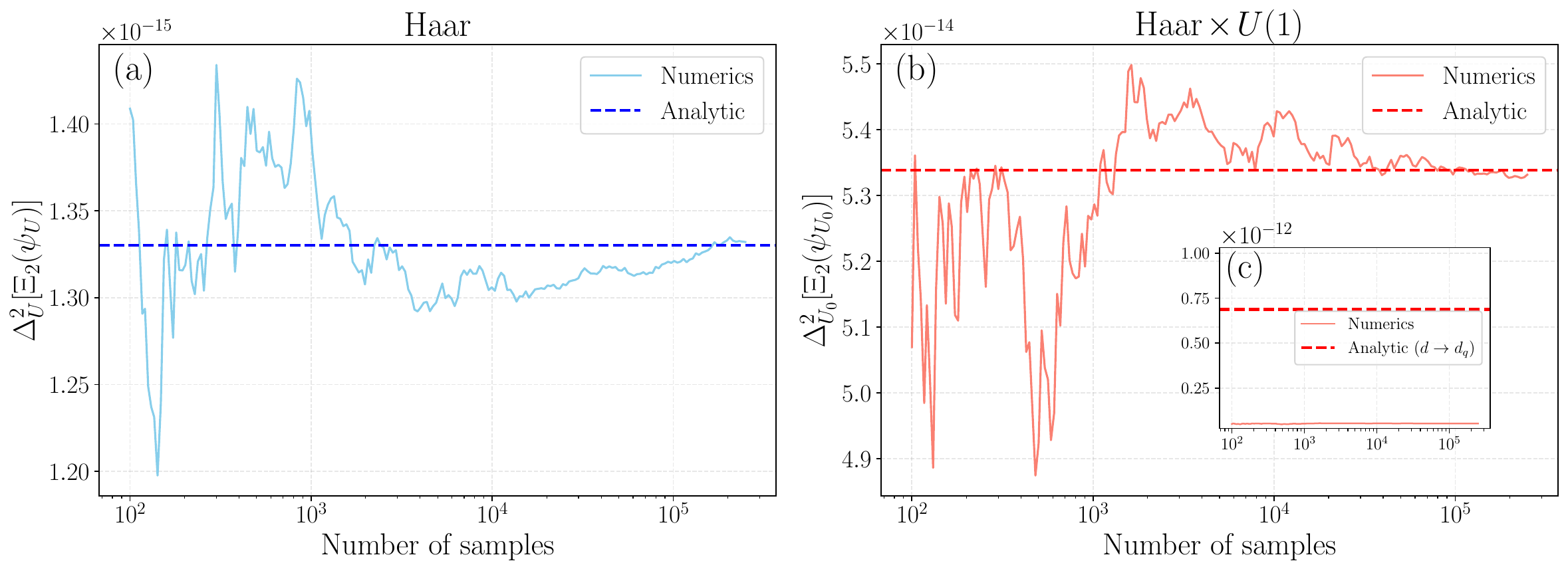}
    \caption{Numerical (empirical) variance convergence as a function of the number of sample realizations for states with $L = 14$ qubits, corresponding to a total Hilbert space dimension $d = 2^{14}$ (same data as in Fig.~1 of the main text). Panel~(a) shows the unconstrained random pure state ensemble sampled according to the Haar measure. The dashed line represents the known theoretical prediction $\Delta^2[\Xi_2] \approx \mathcal{O}(1/d^4)$; for the complete expression beyond leading order, see Eq.~(10) of Ref.~\cite{Iannotti_Esposito_Venuti_Hamma_2025}. One observes a relatively slow convergence (requiring more than $10^5$ samples) of the numerical data toward the theoretical asymptotic prediction at these scales. Panel~(b) shows the constrained random pure state ensemble sampled according to the Haar$\times U(1)$ measure with charge sector $q = 0$. 
    In panel~(c), we show that a naive dimensional replacement $d\to d_{q}=\binom{L}{\frac{L+q}{2}}$, does not yield agreement with the empirical sampling of the constrained ensemble. }
    \label{fig:variance_Samples}
\end{figure}


\begin{equation}
     17152\,4^L \int_{[0,2 \pi]^{\times 8}} \frac{d\bm{\alpha}}{(2\pi)^8}  e^{-i q \sum_{j=1}^8 \alpha_j} \, \left(\cos(\alpha_1+\alpha_8) \cos(\alpha_2+\alpha_7+\alpha_3+\alpha_6+\alpha_4+\alpha_5)\right)^L= 17152\, d_q^2
\end{equation}

\begin{equation}
\begin{split}
    1152  \, 8^L \int_{[0,2 \pi]^{\times 8}} \frac{d\bm{\alpha}}{(2\pi)^8}  e^{-i q \sum_{j=1}^8 \alpha_j} \, &  \Big(\cos(\alpha_1-\alpha_3)\cos(\alpha_2-\alpha_4-\alpha_5+\alpha_6) \cos(\alpha_7-\alpha_8)\\
    &+ \cos(\alpha_1+\alpha_3)\cos(\alpha_2+\alpha_4+\alpha_5+\alpha_6) \cos(\alpha_7+\alpha_8) \Big)^L\\
    &=  1152 \sum_{k=0}^{L/2} \binom{L}{2 k} \binom{2 k}{\frac{2 k-q}{2}}^3 \binom{L-2k}{\frac{L-2k+q}{2}}^3\\
    &= 1152 \,\mathcal{K}_3(L,q)
\end{split}
\end{equation}

\begin{equation}
\begin{split}
     1536 \,8^L \int_{[0,2 \pi]^{\times 8}} \frac{d\bm{\alpha}}{(2\pi)^8} & e^{-i q \sum_{j=1}^8 \alpha_j} \, \Big(\cos(\alpha_1+\alpha_2+\alpha_3+\alpha_4+\alpha_5) \cos\alpha_6 \cos\alpha_7 \cos\alpha_8 \\
     &+ \sin(\alpha_1+\alpha_2+\alpha_3+\alpha_4+\alpha_5) \sin\alpha_6 \sin\alpha_7 \sin\alpha_8\Big)^L= 1536\, \frac{h(L,q)}{2^L}
\end{split}
\end{equation}

\begin{equation}
     8704\,2^L \int_{[0,2 \pi]^{\times 8}} \frac{d\bm{\alpha}}{(2\pi)^8}  e^{-i q \sum_{j=1}^8 \alpha_j} \, \left(\cos(\sum_{i=1}^8\alpha_1)\right)^L= 8704\, d_q
\end{equation}

\begin{equation}
    \begin{split}
        96 \, 2^{3 L} \int_{[0,2 \pi]^{\times 8}} \frac{d\bm{\alpha}}{(2\pi)^8}  e^{-i q \sum_{j=1}^8 \alpha_j}  \, & \Big(\cos(\alpha_1+\alpha_2)\cos(\alpha_3+\alpha_4)\cos(\alpha_5
        +\alpha_6)\cos(\alpha_7+\alpha_8)\\
        &+\cos(\alpha_1-\alpha_2)\cos(\alpha_3-\alpha_4)\cos(\alpha_5-\alpha_6)\cos(\alpha_7-\alpha_8)     \\&+\sin(\alpha_1+\alpha_2)\sin(\alpha_3+\alpha_4)\sin(\alpha_5+\alpha_6)\sin(\alpha_7+\alpha_8)\\
        &+\sin(\alpha_1-\alpha_2)\sin(\alpha_3-\alpha_4)\sin(\alpha_5-\alpha_6)\sin(\alpha_7-\alpha_8)) \Big)^L \\
        &= 96\,  2^{3 L} \sum_{k=0}^L \binom{L}{k} \sum_{j=0}^k \binom{k}{j} \sum_{p=0}^{L-k} \binom{L-k}{p} \\
        & J_q(k - j, L - k - p)^4  J_q(j, p)^4 \\
    &= 96 \, 2^{3 L} \,\mathcal{K}_4(L,q)
    \end{split}
\end{equation}
where the functions $d_q$, $J_q(a,b)$ and $h(L,q)$ used in the result are those defined respectively in Eqs.~\eqref{eq:dimq}, \eqref{eq:generic_integral} and  \eqref{eq:defh_SM}. Hence, we find the non-trivial part of the computation of the variance reduces to
\begin{equation}
\begin{aligned}
\sum_{\sigma \in S_8} \text{Tr}[W^{\otimes 2} \,\Pi_q^{\otimes 8}\, T_{\sigma}] 
=&\;\frac{h(L,q)}{2^{L}}\left(96 \,d_q^2 + 640\, d_q + 1536+ 16 \frac{h(L,q)}{2^{L}}\right) \\
&+ d_q(144 \,d_q^3
+ 3648\, d_q^2
+ 17152\, d_q
+ 8704 )\\
&+ 256\,2^{5L}\mathcal{K}_1(L,q)
+ \left(960\,2^{5L} + 5920\right)\mathcal{K}_2(L,q)\\
&+ 1152\,\mathcal{K}_3(L,q)
+ 96\,2^{3L}\mathcal{K}_4(L,q).
\end{aligned}
\label{eq:avgsquareSP}
\end{equation}
In Fig.~\ref{fig:variance} we show the behavior of the full variance $\Delta_{U_q}^2[\Xi_2(\psi_{U_q})]$ as a function of the system size $L$ and the charge $q$.
Additionally, due to the intricate number of terms in Eq.~\eqref{eq:avgsquareSP}, we present a consistent comparison between the analytical and numerical results in Fig.~\ref{fig:variance_Samples}.
In the case of the known theoretical prediction for the variance over the Haar ensemble $\Delta^2[\Xi_2] \approx \mathcal{O}(1/d^4)$~\cite{zhu2016CliffordGroupFails,Iannotti_Esposito_Venuti_Hamma_2025}, one observes a relatively slow convergence (requiring more than $10^5$ samples) of the numerical data toward the theoretical asymptotic prediction at these scales. A similar analysis is shown for the Haar$\times U(1)$ measure sampled states and that a naive dimensional replacement $d\to d_{q}$, where $d_q=\binom{L}{\frac{L+q}{2}}$, does not yield agreement with the empirical sampling of the constrained ensemble. 
In fact, a discrepancy of two orders of magnitude persists. Quantities such as participation entropies fail to capture this distance (see Sect.~\ref{sec:participation}), highlighting the important subtleties inherent in the stabilizer entropy response. The practical procedure for sampling $U(1)$-constrained pure random states is described in Sect.~\ref{app:sampling}.

\section{Generating Haar states with fixed magnetization in Pauli $X,Y,Z$ frames}\label{app:sampling}

\begin{figure}[ht!]
\centering
\includegraphics[width=\columnwidth]{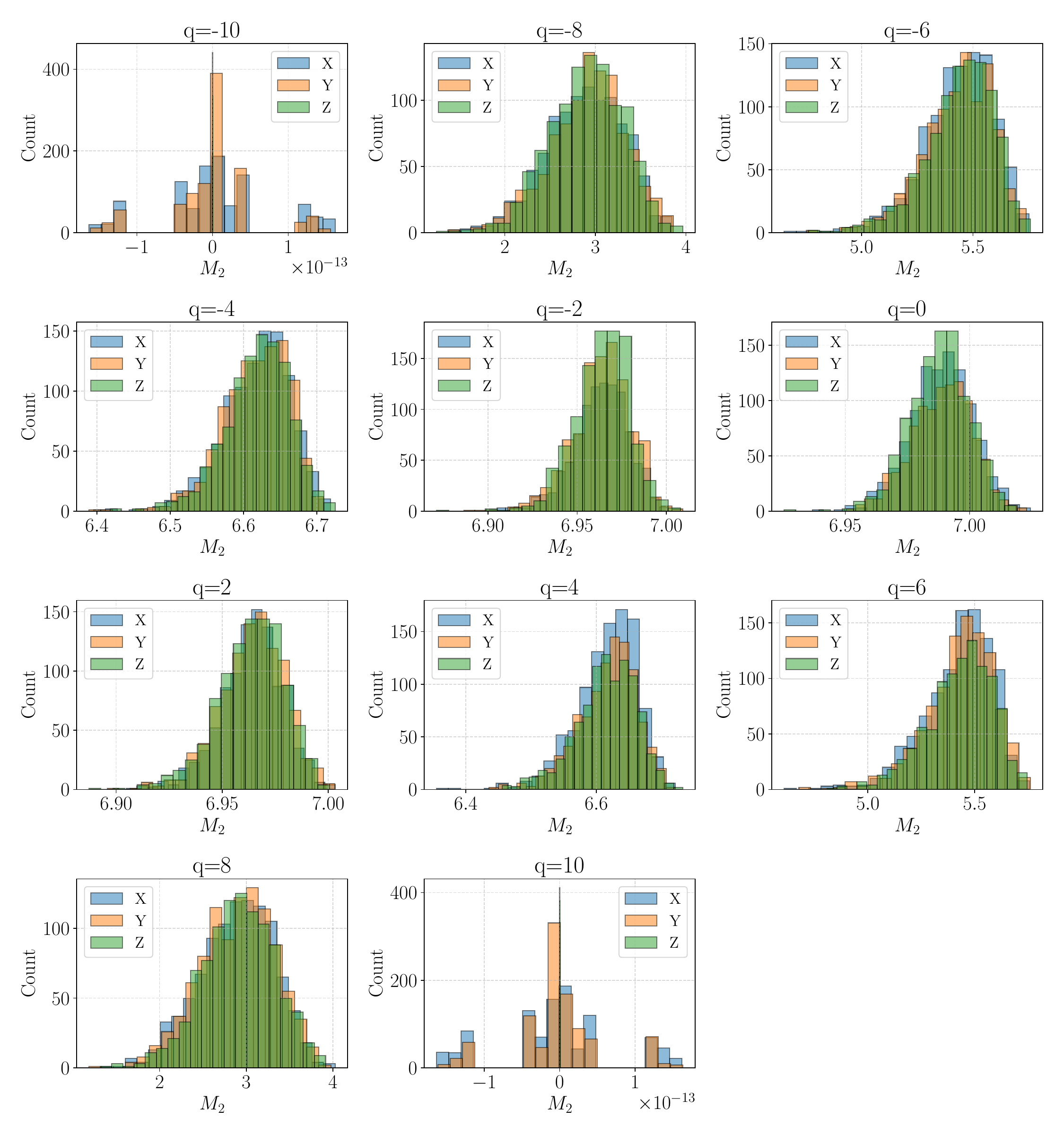}
    \caption{
Numerical checks demonstrating the response of Haar random states subjected to constraints across different Pauli matrix conservation symmetries and their universality. The system size is fixed at $L = 10$, and $1000$ instances of Haar random states with fixed magnetization $q \in \{-10, -8, \ldots, 8, 10\}$ are generated; the corresponding unnormalized histograms are shown. Appreciable skewness and excess kurtosis of the distributions can be observed for $q = \pm 4, \pm 6, \pm 8$ at finite size, which disappear in the thermodynamic limit as demonstrated in Fig.~\ref{fig:variance}. 
    } \label{fig:different_projections_supplemental}
\end{figure}

A Haar random state $\vert \psi^{\rm Haar} \rangle$ is a state drawn uniformly in the Hilbert space. Such a state is nothing more than a normalized vector with $2^L$ complex entries drawn from the normal distribution $\mathcal{N}(0,1)$. 
Imposing constrains such as magnetization requires building appropriate projectors, which we define below. The total magnetization operator along an axis $\alpha \in \{ \sigma^x,\sigma^y,\sigma^z\}$ is defined as 
\begin{equation}
    \mathcal{Q}^{(\alpha)} = \sum\limits_{j = 1}^{L} \sigma_j^{\alpha},
\end{equation}
where the $\sigma_{j}^{\alpha}$ is the Pauli matrix acting on the $j$-th qubit. The eigenvalues $q$ of $\mathcal{Q}^{(\alpha)}$ range from $-L$ to $+L$ in steps of 2. The $\sigma^z$-direction is straightforward because it is the computational basis ($\vert 0 \rangle$ and $\vert 1 \rangle$). The magnetization $q$ corresponds to the total difference between the \lq\lq up\rq\rq and \lq\lq down\rq\rq spins. The projector $\Pi^{z}_{q}$ is a diagonal matrix in the computation basis and reads
\begin{equation}
    \big[ \Pi^{z}_{q} \big]_{jj} = \begin{cases}
1, & \text{if state $\vert j \rangle$ has magnetization $m$}\\
0, & \text{otherwise}
\end{cases}
\end{equation}
To obtain a constrained state therefore we just need to act on the Haar random state 
\begin{equation}
    \vert \psi^{z-{\rm Haar}}_{q} \rangle = \Pi^{z}_{q} \vert \psi^{\rm Haar} \rangle,
\end{equation}
and appropriately normalize. Projecting along $\sigma^x$ is mathematically equivalent $\sigma^z$ if the basis is rotated. To perform the rotation we use the Hadamard $H$ gate that acts as $H \vert 0 \rangle = \vert + \rangle$ and $H \vert 1 \rangle = \vert - \rangle$  and projection is performed as 
\begin{equation}
    \vert \psi^{x-{\rm Haar}}_q \rangle = \Pi^{x}_{q} \vert \psi^{\rm Haar} \rangle =  H^{\otimes L} \Pi^{z}_{q} H^{\otimes L}  \vert \psi^{\rm Haar} \rangle,
\end{equation}
and properly normalized.
Projecting along $\sigma^y$ requires the rotation matrix
\begin{equation}
    U_{y} = R_{x} (-\pi/2) = \dfrac{1}{\sqrt{2}} \begin{pmatrix}
1 & i \\
i & 1 
\end{pmatrix},
\end{equation}
and takes the simular route
\begin{equation}
    \vert \psi^{y-{\rm Haar}}_q \rangle = \Pi^{y}_{q} \vert \psi^{\rm Haar} \rangle = U_{y}^{\otimes L} \Pi^{z}_{q} U_{y}^{\dagger, \otimes L} \vert \psi^{\rm Haar} \rangle. 
\end{equation}
In Fig.~\ref{fig:different_projections_supplemental}, we demonstrate what happens if we sample from these projected ensembles while imposing different constrains to pure random states. We observe similar response between different projection operator demonstrating universality between scalar charges projection directions for the case of SE. Formal reason on why this is the case in terms of the stabilizer entropy response can be found in the main text, and is related with the non-increasing property of SE under Clifford transformations.

As the dimension of the subspace $\mathcal{H}_q$ of the Hilbert space grows, as shown in Sect. \ref{app:Levy}, the probability density function of the 2-SP, $\Xi_2$, concentrates, resulting in a very peaked distribution around its mean value that can be conveniently modeled as a Gaussian distribution with mean and variance computed respectively in Sec.~\ref{app:average_SP} and \ref{app:variance_SP} for $d_q \to \infty$.

\section{Average Participation Entropy for the Haar$\times U(1)$ ensemble}\label{sec:participation}
\noindent
To complent our analysis on stabilizer purity and entropy we also compute the participation entropy (PE). PE characterizes the spread of a wave function in the computational basis and is defined as
\begin{equation} 
\label{eq:pe_entropy}
\mathcal{S}_k(|\psi\rangle) = 
\frac{1}{1-k}\log\!\left[
\sum_{x=0}^{2^{L}-1} |\langle x | \psi \rangle|^{2k}
\right],
\end{equation}
where $k$ is the Rényi index and 
$\mathcal{B}=\{|x\rangle\,|\,x=0,\dots,2^L-1\}$ denotes the computational basis. It is a well-established measure of coherence for pure states~\cite{PhysRevLett.113.140401,luitz2014participation,PhysRevLett.123.180601,PhysRevA.111.052614,Sauliere:2025bpr}. A formal connection between non-stabilizerness and anticoncentration for $U(1)$-symmetric states was established in~\cite{tirrito2025universal},
\begin{equation}
\label{eq:relation_sre}
M_{a}(|\psi\rangle) \leq 
\frac{a}{a-1}\mathcal{S}_{b}(|\psi\rangle),
\qquad 
a>1,\; b\leq2 .
\end{equation}
Next, we compute the average participation entropy over the Haar$\times U(1)$ ensemble and compare it with the behavior of $M_\alpha$. 
Using Jensen's inequality we obtain
\begin{equation}
\begin{split}
\mathbb{E}_{U_q}[\mathcal{S}_k(|\psi_{U_q}\rangle)]
&\ge
\frac{1}{1-k}
\log
\left[
\mathbb{E}_{U_q}
\left[
\sum_x |\langle x|\psi_{U_q}\rangle|^{2k}
\right]
\right] \\
&=
\frac{1}{1-k}
\log
\left[
\frac{k! \, d_q!}{(d_q+k-1)!}
\right].
\end{split}
\end{equation}
The expectation value appearing above can be computed explicitly,
\begin{equation}
\begin{split}
\mathbb{E}_{U_q}
\!\left[
\sum_{x} |\langle x | \psi_{U_q} \rangle|^{2k}
\right]
&=
\frac{1}{\binom{d_q+k-1}{k} k!}
\sum_{x}
\sum_{\pi\in S_k}
\Tr
\!\left[
|x\rangle\!\langle x|^{\otimes k}
\Pi_q^{\otimes k}
T_\pi
\right]
\\
&=
\frac{1}{\binom{d_q+k-1}{k}}
\sum_x |\langle x|\Pi_q|x\rangle|^{k}
\\
&=
\frac{1}{\binom{d_q+k-1}{k}}
\sum_x
\left(
\frac{1}{2\pi}
\int_0^{2\pi}
d\alpha\,
e^{i\alpha(\sum_i(-1)^{x_i}-q)}
\right)^k
\\
&=
\frac{1}{\binom{d_q+k-1}{k}}
\sum_x
\delta_{q,\sum_i(-1)^{x_i}}
\\
&=
\frac{\binom{L}{\frac{L+q}{2}}}
{\binom{d_q+k-1}{k}}
=
\frac{k! \, d_q!}{(d_q+k-1)!}.
\end{split}
\end{equation}
From the moments 
\(
\sum_x |\langle x|\psi_{U_q}\rangle|^{2k}
\)
one can reconstruct the distribution of outcome probabilities 
\(w=|\langle x|\psi_{U_q}\rangle|^2\) via a Laplace transform, yielding the Porter–Thomas distribution within the symmetry sector,
\begin{equation}
P(w)=
\frac{d_q-1}{d_q}
\left(
1-\frac{w}{d_q}
\right)^{d_q-2}.
\end{equation}
We can also compute the average Shannon participation entropy
\begin{equation}
\mathcal{S}_1(|\psi\rangle)
=
-\sum_x |\langle x|\psi\rangle|^2
\log |\langle x|\psi\rangle|^2 ,
\end{equation}
which corresponds to the relative entropy of coherence.
Since $|\psi_{U_q}\rangle$ is Haar-random within the $d_q$-dimensional subspace 
$\mathcal{H}_q\subset\mathbb{C}^d$, only $d_q$ amplitudes are non-zero. 
Using the result of~\cite{singh2016average} for the Haar ensemble we obtain
\begin{equation}
\mathbb{E}_{U_q}
[\mathcal{S}_1(|\psi_{U_q}\rangle)]
=
\sum_{p=1}^{d_q}\frac{1}{p}-1 .
\end{equation}
Although a strong relation between PE and SE exists for $U(1)$-symmetric states, the two quantities exhibit different sensitivities to the symmetry constraint when analyzed over random-state ensembles. We note that effects of the symmetry on random states we just presented and its constrained Porter-Thomas distribution could be useful in invariant Matrix Models and random matrix theory in general~\cite{Franchini_2014}.

\section{Hamiltonian systems - additional analysis}

We present supporting numerical data for the Hamiltonian model systems considered in this work: the Heisenberg XXZ next-nearest-neighbor (XXZ-NNN) model, with and without an explicit $U(1)$ symmetry, and the mixed-field Ising model (MFIM), which does not conserve magnetization. Note that the main text reports only on the former. In the final subsection, we provide additional checks on the complex Sachdev-Ye-Kitaev (cSYK) model prominently featured in the main text.

\subsection{Local models: Heisenberg XXZ next-nearest-neighbor (NNN) and the mixed-field Ising (MFIM) model}

\begin{figure}[t]
    \centering
    \includegraphics[width=\linewidth]{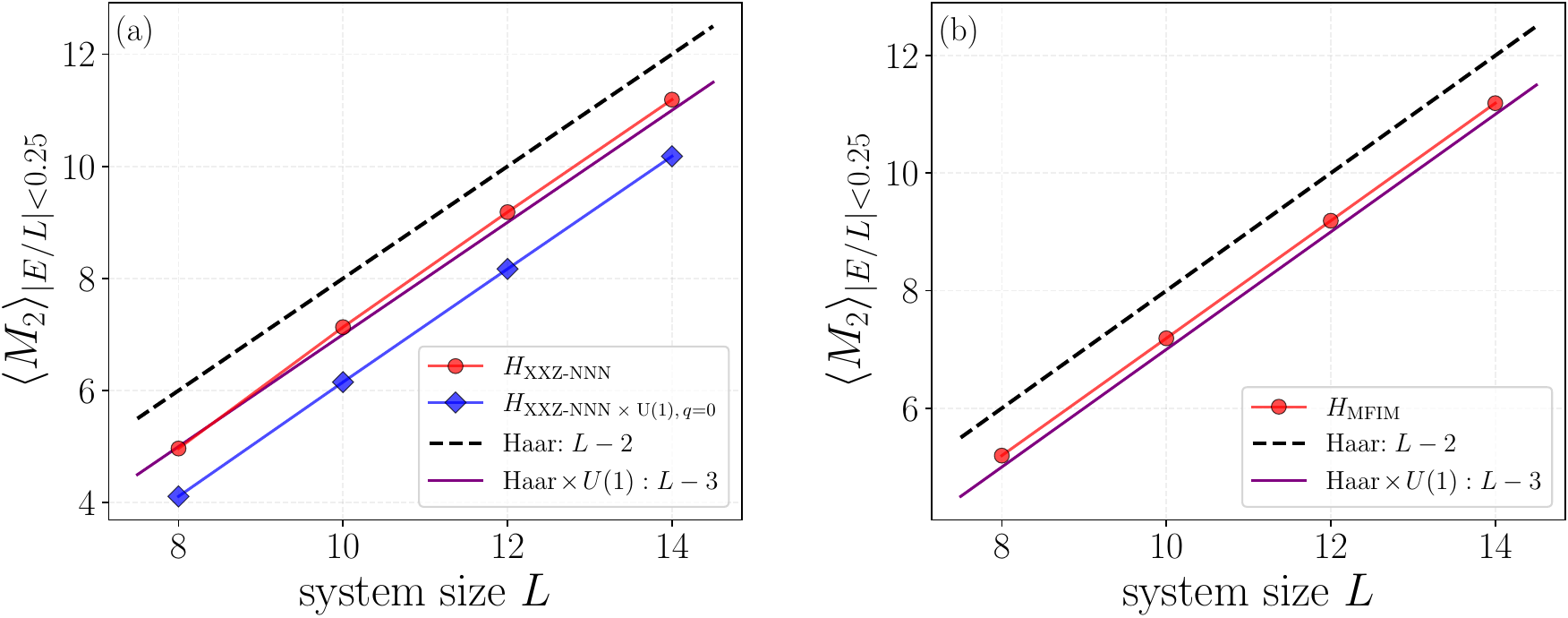}
    \caption{Average stabilizer entropy $M_2$ of mid-spectrum eigenstates (energy-density window $|E/L|<0.25$). (a) XXZ-NNN Hamiltonian (defined in the main text) for various system sizes $L$, with parameters $J_1=1$, $\Delta=0.5$, $h_b=0.25$, and either no magnetization conservation ($J_2=0.0$, $h_x=0.75$) or conserved magnetization ($h_x=0$, $J_2=0.6$). (b) MFIM Hamiltonian for various system sizes in the maximally chaotic regime set by the transverse and longitudinal fields $(g,h)=(1.1,0.35)$~\cite{Rodriguez-Nieva_Jonay_Khemani_2024, Langlett_Jonay_Khemani_Rodriguez-Nieva_2025, Russotto_Ares_Calabrese_2025}. To drive the system deep into the non-integrable regime, we apply asymmetric boundary fields $(h_1,h_L)=(1/4,-1/4)$ in the Hamiltonian in Eq.~\eqref{eq:mfim}.}
    \label{fig:XXZ_supplemental}
\end{figure}

As shown in Fig.~\ref{fig:XXZ_supplemental}(a), the average stabilizer entropy over mid-spectrum eigenstates of the XXZ-NNN model (red points) does not saturate the Haar value (dashed black line) across system sizes $L$; instead, it approaches a lower bound, revealing a noticeable deficit in the stabilizer entropy. This behavior may be viewed as a signature of an emergent $U(1)$ symmetry in the model. Although the agreement with the Haar $\times\, U(1)$ prediction (solid purple line) could in principle be attributed to interaction locality, as argued in Ref.~\cite{Rodriguez-Nieva_Jonay_Khemani_2024}, the residual gap we observe (approximately $0.2$ in absolute value and essentially constant across system sizes) rules out such an explanation. Figure~\ref{fig:XXZ_supplemental}(a) further illustrates the effect of the additional scalar charge associated with magnetization conservation on both the eigenstates and their stabilizer entropy: restricting to the $q=0$ sector (equivalent to taking a cross section of the data in Fig.~3 of the main text) produces a further suppression of the non-stabilizerness measured by $M_2$.

We also present analogous results for the MFIM spin chain, another Hamiltonian that does not conserve magnetization, defined as
\begin{equation}
    H^{\mathrm{MFIM}} = \sum_{j=1}^{L-1} \sigma^z_j \sigma^z_{j+1} + g\sum_{j=1}^{L} \sigma^x_j + h\sum_{j=1}^{L} \sigma^z_j + h_1 \sigma^z_1 + h_L \sigma^z_L,
    \label{eq:mfim}
\end{equation}
where we impose open boundary conditions (OBC). The OBC explicitly break inversion symmetry and eliminate the remaining discrete symmetries, leaving energy conservation as the sole global symmetry. Figure~\ref{fig:XXZ_supplemental}(b) shows trends consistent with those of the XXZ-NNN model, including the same ${\sim}0.2$ gap between the mid-spectrum eigenstates and the Haar $\times\, U(1)$ expectation.

\subsection{Non-local model: complex Sachdev-Ye-Kitaev (cSYK) model}\label{app:Spectral SYK}

\begin{figure}[ht]
    \centering
    \includegraphics[width=1.0\linewidth]{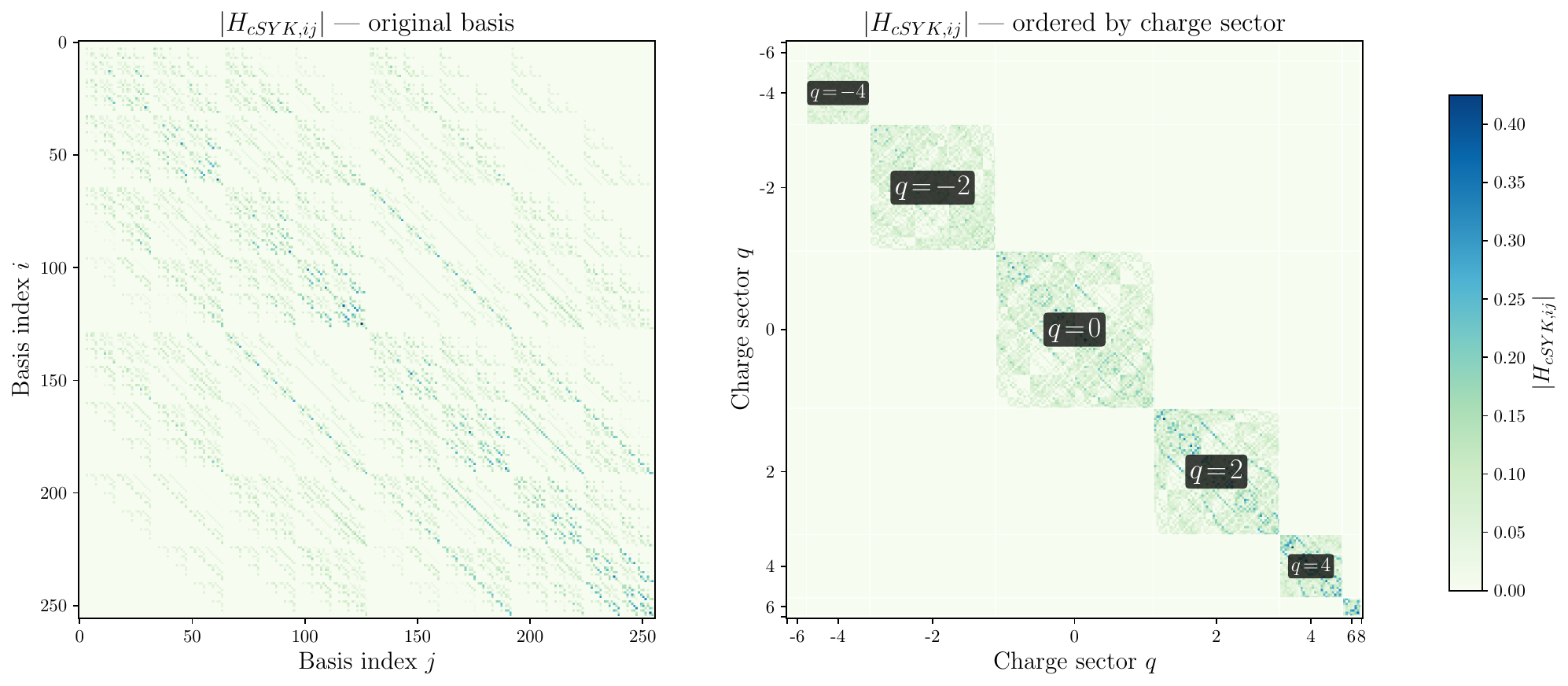}
    \caption{Magnitude of the Hamiltonian matrix elements $|H_{{\rm cSYK},ij}|$ for the complex $\mathrm{SYK}_4$ model with $L=8$ fermionic modes. \textit{Left panel:} Heatmap in the original Fock basis, where the block structure is not manifest. \textit{Right panel:} Matrix reordered by particle-number (charge) sectors $q$, revealing the block-diagonal structure imposed by the $U(1)$ symmetry. Each block corresponds to a fixed fermion-number sector, illustrating the decomposition of the Hilbert space into conserved-charge subspaces.}
    \label{fig:cSYKblocks}
\end{figure}

As discussed in the main text, the cSYK model commutes with the charge operator $\mathcal{Q}$, leading to a block-diagonal structure $\mathcal{H} = \bigoplus_{q} \mathcal{H}_q$ that we make explicit in Fig.~\ref{fig:cSYKblocks}. The charge sectors span $q \in \{-L, -L + 2, \ldots, L-2, L\}$ and can be diagonalized independently. The total spectrum is therefore the union of all eigenvalues across the symmetry-induced blocks.

The SYK model gained its prominence as an example of a fast scrambler~\cite{Sekino_Susskind_2008, sachdev1993gapless, chowdhury2022sachdev} (a feature attributed to black holes) and because its eigenvalues and eigenstates display features characteristic of quantum chaotic systems. The SYK model is zero-dimensional, and its interactions span a complete graph, making it strongly nonlocal. Moreover, the model is disordered, with couplings sampled from a normal distribution. Here we consider the variant defined in terms of complex fermions (see main text), which admits an explicit charge conservation, as opposed to the Majorana fermion version~\cite{KITP2015, Jasser_Odavic_Hamma_2025} which does not.

In this work, we consider many different disorder realizations of the cSYK model, as listed in Table~\ref{tab:statistics}. We note that most computational effort expended goes towards constructing the disordered Hamiltonian (whose individual number scales $\mathcal{O}(L^4)$) rather than in evaluating the stabilizer entropy. 

\begin{table}[ht]
    \centering
    \begin{tabular}{| c | c | c | c | c |}
    \hline
        $L$ & 8 & 10 & 12 & 14 \\
        \hline
        Realizations & 2000 & 1000 & 200 & 100 \\
        \hline
    \end{tabular}
    \caption{Number of disorder realizations considered for each system size.}
    \label{tab:statistics}
\end{table}

\begin{figure}[ht]
    \centering
    \includegraphics[width=0.8\linewidth]{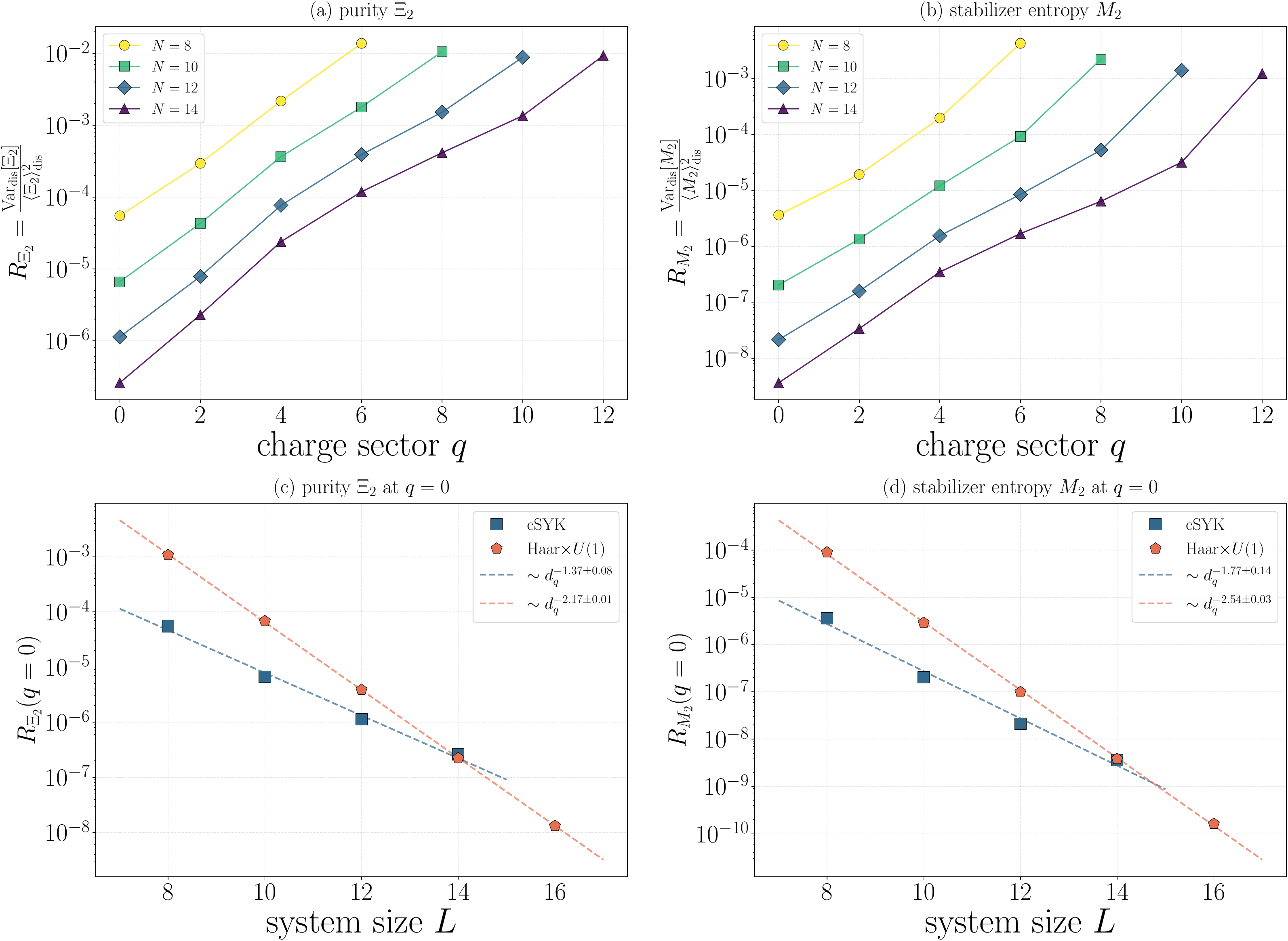}
    \caption{Relative variance signaling self-averaging with respect to disorder across different charge sectors. Panels in the left column~(\textit{(a)} and~\textit{(c)}) show the scaling of the purity with system size, while panels in the right column~(\textit{(b)} and~\textit{(d)}) show the stabilizer entropy. An  decay is observed, confirming that the quantities under study are self-averaging. The plot shows only half of the possible values of the charge sector , as the results for negative values are symmetric.}
    \label{fig:cSYKself}
\end{figure}

In certain respects, SYK models are self-averaging: that is, very few samples suffice to provide a reliable estimate of the target observable for large enough systems, while in others they are not. For example, the two-point correlators are strongly self-averaging~\cite{KITP2015}, whereas the spectral form factor at intermediate to late times~\cite{Cotler} is not, with the behavior dominated by large oscillations. This implies that self-averaging cannot be taken for granted and must be verified on a case-by-case basis, motivating the relative-variance analysis presented in Fig.~\ref{fig:cSYKself}. We observe a clear fast decay of the relative variance with system size, confirming the self-averaging property for both the stabilizer purity and the stabilizer entropy across different charge sectors.

\begin{figure}[ht]
    \centering
    \includegraphics[width=1.0\linewidth]{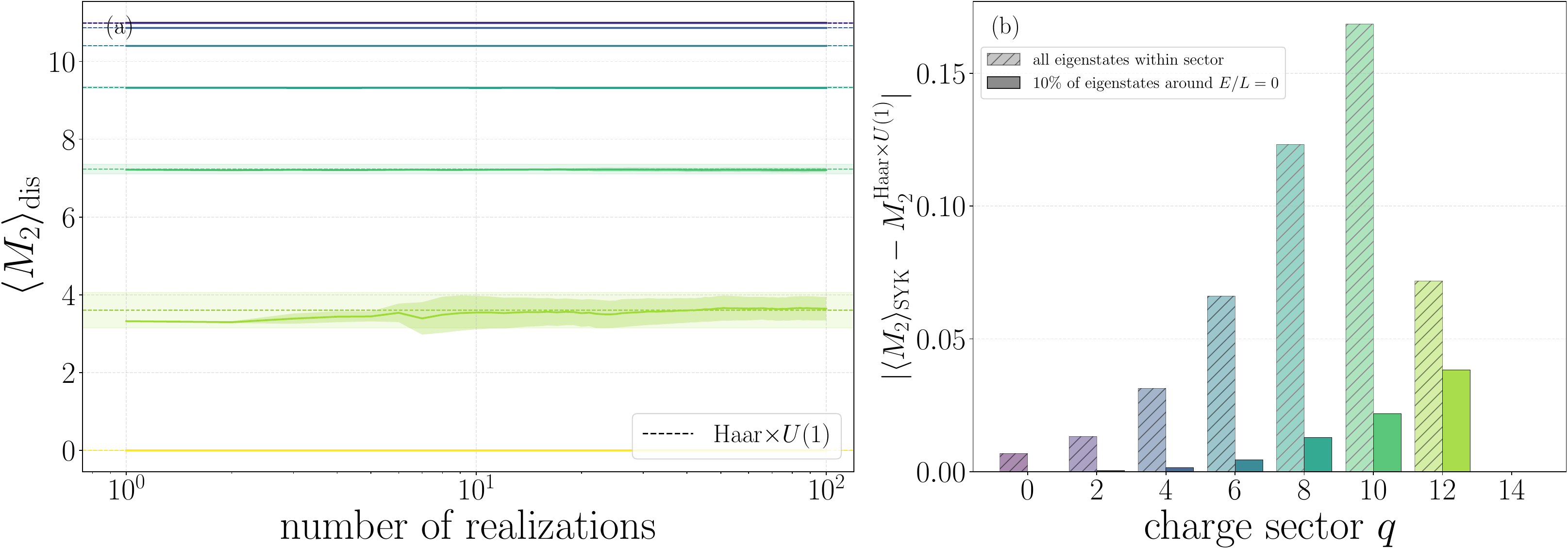}
    \caption{\textit{(a)}~Average stabilizer entropy and corresponding variance across different disorder realizations, showing convergence toward the constrained Haar$\,\times\, U(1)$ random-state ensemble. The convergence is fast, consistent with the self-averaging property of the stabilizer entropy verified in Fig.~\ref{fig:cSYKself}. Here we consider eigenstates around zero energy density, retaining $10\%$ of the spectrum. \textit{(b)}~Absolute difference between the disorder-averaged cSYK stabilizer entropy and the stabilizer entropy of the constrained Haar$\,\times\, U(1)$ ensemble, confirming the predictions and results presented in the main text. Panels share the color palette.}
    \label{fig:cSYKconvergence}
\end{figure}

Finally, fixing the system size to $L = 14$, in Fig.~\ref{fig:cSYKconvergence}(a) we demonstrate the fast convergence toward the ensemble average with the number of disorder realizations across all considered sectors. In panel~(b), we additionally show that the high-energy eigenstates in the bulk (middle of the spectrum) exhibit strong agreement with the Haar$\,\times\, U(1)$ ensemble, whereas this agreement deteriorates when all eigenstates are included. These results are consistent with expectations for the entanglement of mid-spectrum states reported in Ref.~\cite{Jasser_Odavic_Hamma_2025}. Notice that only half of the possible values of the charge sector $q$ are shown, as the results for negative values follow by symmetry. Starting from the central sector $q=0$, the deviation from the expected average value increases as $|q|$ grows. Although at $q=12$ the difference appears to decrease, panel (a) shows that this point is associated with the largest standard deviation, indicating that the overall trend is still increasing.

\clearpage
\bibliography{refs_arXiv_v2}